\documentclass[final,1p,times]{elsarticle}

\usepackage{amssymb}
\usepackage{amsthm}
\usepackage{amsmath}

\usepackage{lineno}

\usepackage{listings}
\usepackage{courier}
\usepackage{url}
\usepackage{subcaption}
\usepackage{floatpag}
\usepackage{proof}

\journal{Science of Computer Programming}

\begin{document}

\begin{frontmatter}



\title{The Optics of Language-Integrated Query\tnoteref{t1}}
\tnotetext[t1]{This work is partially supported by a Doctorate Industry Program
grant to Habla Computing SL, from the Spanish Ministry of Economy, Industry and
Competitiveness.}

\author[1,2]{J. L\'opez-Gonz\'alez\corref{cor1}}
\cortext[cor1]{Corresponding author}
\ead{jesus.lopez@urjc.es}

\author[1,2]{Juan M. Serrano}
\ead{juanmanuel.serrano@urjc.es}

\address[1]{Universidad Rey Juan Carlos, Calle Tulip\'an, s/n, 28933 M\'ostoles, Spain}
\address[2]{Habla Computing SL, Avda. Gregorio Peces Barba, 28918 Legan\'es, Spain}

\title{}
 

\author{}

\address{}

\begin{abstract}
  Monadic comprehensions reign over the realm of language-integrated query
  (LINQ), and for good reasons. Indeed, comprehensions are tightly integrated
  with general purpose programming languages and close enough to common query
  languages, such as SQL, to guarantee their translation into effective
  queries. Comprehensions also support features for writing reusable and
  composable queries, such as the handling of nested data and the use of
  functional abstractions. 

  In parallel to these developments, optics have emerged in recent years as the
  technology of choice to write programs that manipulate complex data structures
  with nested components. Optic abstractions are easily composable and, in
  principle, permit both data access and updates.  This paper attempts to
  exploit the notion of optic for LINQ as a higher-level language that
  complements comprehension-based approaches. 

  In order to do this, we lift a restricted subset of optics, namely getters,
  affine folds and folds, into a full-blown DSL. The type system of the
  resulting {\em language of optics}, that we have named {\em Optica}, distills
  their compositional properties, whereas its denotational semantics is given by
  standard optics.  This formal specification of the concept of optic enables
  the definition of non-standard optic representations beyond van Laarhoven,
  profunctor optics, etc. In particular, the paper demonstrates that a
  restricted subset of XQuery can be understood as an optic representation; it
  introduces Triplets, a non-standard semantic domain to normalize optic
  expressions and facilitate the generation of SQL queries; and it describes how
  to generate comprehension-based queries from optic expressions, thus showing
  that both approaches can coexist. 

  Despite the limited expressiveness of optics in relation to comprehensions,
  results are encouraging enough to anticipate the convenience and feasibility
  of extending existing comprehension-based libraries for LINQ in the functional
  ecosystem, with optic capabilities. In order to show this potential, the paper
  also describes S-Optica, a Scala implementation of Optica using the
  tagless-final approach.
\end{abstract}


\begin{highlights}
\item Getter, Affine Fold and Fold optics are lifted into Optica, a query
  language for LINQ
\item XQuery and SQL queries derived from non-standard optic
  representations
\item Optics as a higher-level interface over comprehension-based
  query languages
\item Typed tagless-final encoding in Scala of the Optica type system and
  semantics 
\end{highlights}


\end{frontmatter}


\newtheorem{proposition}{Proposition}
\newtheorem*{conditions}{Preconditions}

\newdefinition{definition}{Definition}
\newdefinition{remark}{Remark}

\newcommand\Fork{\mathbin{\mathord{*}\mathord{*}\mathord{*}}}
\newcommand\Ext{\mathbin{\mathord{+}\mathord{:}\mathord{=}}}

\lstset{
    language=Scala, 
    basicstyle=\scriptsize\ttfamily,
    literate= {>>>}{{$\ggg$}}1 
              {***}{{$\Fork$}}2 
              {\_gt}{{$_{\mathit{gt}}$}}1
              {\_af}{{$_{\mathit{af}}$}}1
              {\_fl}{{$_{\mathit{fl}}$}}1
              {rho}{{$\rho$}}1
              {lambda}{{$\lambda$}}1
              {=>}{{$\Rightarrow$}}1
              {<-}{{$\leftarrow$}}1
              {≃}{{$\simeq$}}1
}

\newcommand{\sqlinline}[1]{\lstinline[columns=fixed,language=SQL]{#1}}


\section{Introduction}
\label{sec:Introduction}

The research field of language-integrated query (LINQ
\cite{DBLP:conf/icdt/TannenBW92,DBLP:conf/sigmod/MeijerBB06,cheney2013practical})
aims at alleviating the {\em impedance mismatch} problem
\cite{copeland1984making,ireland2009classification} that commonly originates in
software systems where general-purpose programming languages, on the one hand,
and query languages, on the other, need to interoperate. The problem manifests
itself in the form of maintenance, reliability and security problems, which are
essentially due to the mismatches of programming paradigms and data models
endorsed by the interacting languages. In order to tackle this issue, the LINQ
research field favors a {\em domain-specific language} (DSL)-based approach
\cite{Hudak:1996:BDE:242224.242477}. From this perspective, the programmer does
not simply inject query expressions in the general-purpose language as plain
{\em strings}, a practice which is a well-known source of bugs and injection
attack problems; on the contrary, she uses a DSL which ensures that the query is
well-formed, correctly typed, and moreover, that it helps to overcome the
conceptual gap between both the general-purpose and the query language.

Indeed, not every DSL can be given the seal of approval from a
language-integrated perspective. For instance, we may embed SQL in a host
language like Scala ~\cite{odersky2004overview} to attain the stated demand of
type safety and yet, the disparity between the flat and nested computational
models from both languages would not be reduced in the slightest\footnote{ The
entities of a \emph{flat} model are either base types or classes with no nested
collections. Otherwise, we say that the model is \emph{nested}.}. This is,
without a shadow of a doubt, a necessary step in order to generate well-formed
SQL queries. Scala libraries such as Doobie~\cite{doobie19}, which focus on this
specific issue, are worthwhile.  However, to properly bridge the impedance
mismatch gap, we need DSLs at a higher level of abstraction: close enough to
general-purpose languages, yet specific enough to allow for the efficient
generation of queries for a wide range of querying languages
\cite{cheney-shonan-meeting}. Since early in its foundation, the LINQ research
field has exploited monadic comprehensions~\cite{wadler1990comprehending,
wadler1995monads} as its DSL of choice for this purpose. The basic insight was
originally introduced in \cite{176921}, and then developed by the Nested
Relational Calculus (NRC)
\cite{DBLP:journals/sigmod/BunemanLSTW94,DBLP:journals/tcs/BunemanNTW95}, which
provides the foundation of query languages based on comprehensions. NRC subsumes
much work on LINQ theories and systems such as in
Kleisli~\cite{DBLP:journals/jfp/Wong00}, Links \cite{DBLP:conf/dbpl/Cooper09},
Microsoft's LINQ \cite{DBLP:conf/sigmod/MeijerBB06, DBLP:conf/ml/Syme06},
Database Supported Haskell (DSH) \cite{DSHHaskell}, T-LINQ
\cite{cheney2013practical}, QUE$\Lambda$ \cite{suzuki2016finally} and SQUR
\cite{DBLP:conf/aplas/KiselyovK17}.
 
In essence, the purpose of research on LINQ is borrowing the comprehension
syntax that in-memory data structures such as lists, sets, bags, and other bulk
types enjoy in order to express queries at a generic monadic level. To this aim,
bulk types are {\em lifted} into a proper DSL that abstracts away its
characteristic in-memory representation but still allows to express queries
using comprehension syntax. In some cases, as in Kleisli, Links and Microsoft's
LINQ, this lifting mechanism is a primitive part of the general-purpose language
itself. In the case of more conventional, functional programming (FP) languages,
such as Scala, F\#, OCaml, Haskell, etc., the DSL may be {\em embedded} in the
host language using one of the several techniques that FP offers for this
purpose: typed tagless-final \cite{kiselyov2012typed}, generalized algebraic
data types (GADTs)~\cite{xi2003guarded} or quoted domain-specific languages
(QDSLs)~\cite{DBLP:conf/pepm/NajdLSW16}. For instance, quotation is used to
embed the T-LINQ language in F\# \cite{cheney2013practical}, while typed
tagless-final inspires the design of QUE$\Lambda$, in OCaml. Similarly, Quill
\cite{quill19} is a QDSL heavily inspired by T-LINQ, which is embedded in Scala.

To illustrate the use of comprehensions in LINQ, we consider the data structures
in column ``comprehensions'' of Table \ref{tab:intro}, implemented in the Scala
programming language -- our language of choice throughout this
paper\footnote{\ref{app:scala-cheatsheet} offers a brief account of the major
Scala features that are used in this paper.}. According to this model, a couple
consists of two members, first and second, to which we refer by their names;
besides names, each person has also an age. Given a list of couples and a list
of people, we may obtain the names of those partners who occupy the first
position and are under 50 years of age by using a list comprehension, as query
\lstinline{under50_a} shows. Now, using Slick~\cite{slick19} or
Quill~\cite{quill19}, two well-known libraries of the Scala ecosystem, we may
express the very same query in a generic way. To be precise, we call a query
\emph{generic} if it can be efficiently run (e.g. via appropriate compilation)
against data stores of different sorts: in-memory, external relational
databases, non-relational stores such as XML/JSON files, etc. For instance, the
query \lstinline{under50_b} in Table \ref{tab:intro} is a generic query
implemented in Quill. Being generic, this query may be compiled to different
targets according to the mappings between Scala case classes and database
schemas that the Quill framework supports (as of this writing Cassandra's CQL
\cite{cql19} and SQL). For instance, the resulting SQL expression generated for
this query by Quill would be as follows:

\begin{lstlisting}[language=SQL] 
SELECT w.name 
FROM Couple AS c INNER JOIN Person AS w ON c.fst = w.name 
WHERE w.age < 50 
\end{lstlisting}

\begin{table}[]
\begin{tabular}{c|l|l|}
\cline{2-3}
\multicolumn{1}{l|}{} & \multicolumn{1}{c|}{\textit{comprehension}} &
\multicolumn{1}{c|}{\textit{optics}} \\ \hline
\multicolumn{1}{|c|}{\rotatebox[origin=c]{90}{\textit{model }}} & 
\begin{lstlisting}
class Couple(fst: String, snd: String)
class Person(name: String, age: Int)
\end{lstlisting} &
\begin{lstlisting}
class Couple(fst: Person, snd: Person)
class Person(name: String, age: Int)
\end{lstlisting} \\ \hline
\multicolumn{1}{|c|}{\rotatebox[origin=c]{90}{\textit{immutable}}} & 
\begin{lstlisting}
def under50_a(
    couples: List[Couple],
    people:  List[Person]): List[String] =
  for { 
    c <- couples
    w <- people 
    if c.fst == w.name && w.age < 50 
  } yield w.name
\end{lstlisting} &
\begin{lstlisting}
val under50Fl: Fold[List[Couple], String] =
  couples >>> fst >>> filtered (age < 50) >>> name

val under50_c: List[Couple] => List[String] =
  under50Fl.getAll
\end{lstlisting} 
\\ \hline
\multicolumn{1}{|c|}{\rotatebox[origin=c]{90}{\textit{generic}}} & 
\begin{lstlisting}
val under50_b = 
  quote { 
    for { 
      c <- query[Couple]
      w <- query[Person]
      if c.fst == w.name && w.age < 50 
    } yield w.name
  } 
\end{lstlisting} &
\multicolumn{1}{c|}{?} \\ \hline
\end{tabular}
\caption{Towards Optic-Based LINQ.}
\label{tab:intro}
\end{table}

We have illustrated the use of comprehensions with a simple example of the
so-called flat-flat query, i.e. a query that receives and returns flat types.
These queries are significant because relational databases cannot handle nested
results. For instance, we cannot write an SQL query that returns rows in which a
field contains a list of values. This demonstrates a significant mismatch
between SQL and programming languages, where nested data models are customary.
Moreover, comprehension queries, being founded on NRC, can perfectly well handle
nested data, which might lead us to think about wasted expressiveness. In fact,
the opposite is true: several conservativity results show that we can certainly
use nested data as intermediate values in comprehensions
\cite{DBLP:journals/jcss/Wong96}, even in the presence of parameterized queries
(i.e. using lambda expressions) \cite{DBLP:conf/dbpl/Cooper09} and still be able
to generate normalized queries which do not use nested data in any way. We can
even accommodate flat-nested queries through several flat-flat normalized
queries by using techniques for query
shredding~\cite{DBLP:conf/sigmod/CheneyLW14}. In sum, comprehensions are
exceptionally good from a LINQ perspective: well integrated to a wide range of
programming languages and close enough to relational databases in order to
generate effective queries.

In spite of this, we find three major problems or inconveniences in the current
use of comprehensions for LINQ. First, comprehensions can only express retrieval
queries but updates are equally important. This is acknowledged as an open
problem in the LINQ field \cite{cheney-shonan-meeting}. Second, the use of
nested data and functional abstraction in comprehension-based languages such as
Links/T-LINQ undoubtedly helps in obtaining more compositional queries
\cite{cheney2013practical}. However, this is done at the expense of a complex
re-writing machinery, specially in the case of QDSLs like T-LINQ. Alternative
approaches based on normalization-by-evaluation
\cite{DBLP:conf/aplas/KiselyovK17} ameliorate this problem, but the support for
compositional queries is nevertheless limited. Basically, this is due to the
fact that comprehensions are nearer to the point-wise notation exemplified by
the relational calculus than to pure relational algebra and have to deal with
variable (re)naming, freshness and scope. Functionally, both formalisms are
equally expressive, but the point-free combinators of relational algebra are
arguably more {\em flexible} \cite{DBLP:journals/pacmpl/GibbonsHHW18}. This
flexibility naturally derives in more modular queries, which directly impacts
non-functional concerns such as reuse and change tolerance
\cite{Parnas:1972:CUD:361598.361623,hughes1989functional}. Third, there are
potential querying infrastructures which are essentially hierarchical rather
than relational, such as NoSQL document-oriented databases, which build upon
nested data sources in JSON, XML or YAML. The translation of queries at the
programming language level into these infrastructures may benefit from a more
primitive, algebraic querying model, which is hierarchical by nature.

This paper attempts to show that {\em optics}~\cite{pickering2017profunctor} may
play this role in the realm of language-integrated query, and it supports a pure
algebraic approach to LINQ that may potentially be extended for dealing with
updates. Indeed, optics, also known as \emph{functional references}, are
abstractions that select parts which are contextualized within a whole, which
provide methods to access {\em and/or} update the values that they are selecting.
Since the first appearance of \emph{lens}~\cite{foster2007combinators}, arguably
the most prominent optic, a rich catalog of optics has
emerged~\cite{grenrus2018glassery}. They can compose with each other, with a
very few exceptions, so that they can seamlessly produce quite complex
transformations over immutable nested data structures. Indeed, they have become
an essential companion for the functional programmer, as evidenced by the
growing popularity of optic libraries in the functional
ecosystem~\cite{kmett2019lens,truffaut2019monocle}. In sum, we may say that
optics are the de-facto standard to manipulating nested data in a point-free,
algebraic style; they are as ubiquitous as comprehensions in functional
programming languages and, more importantly, the most common variants are
explicitly designed to handle both read and write accessors.

How do we use optics as a higher-level language to express generic
queries? How do these optic-based queries relate to generic comprehension
queries? How do we translate optic expressions to SQL/NoSQL query languages
directly? To answer to these questions, we may follow the same strategy that is
illustrated in Table~\ref{tab:intro} for comprehension queries: by using
lenses~\cite{foster2007combinators}, traversals~\cite{oconnor2011functor},
folds, and other optic abstractions, we can query and update immutable data
structures quite naturally; why not borrow this very same syntax to express
generic queries that can be interpreted over different target storage systems?
For instance, in column ``optics'' of Table~\ref{tab:intro}, an alternative
nested model is defined for the couples example, where keys in
\lstinline{Couple} are replaced by full-blown \lstinline{Person} values. Then,
query \lstinline{under50_c}, derived from optic \lstinline{under50Fl}, which in
turn is composed from several optics (the fold \lstinline{couples}, and getters
\lstinline{her}, \lstinline{age} and \lstinline{name}), offers an alternative
formulation to the comprehension query \lstinline{under50_a}. What we are
looking for is a generic optic-based query, analogous to the comprehension-based
query \lstinline{under50_b}. In essence, we need to lift optics into a
full-fledged DSL.

\paragraph{\bf Contributions and outline}

This paper sets out to define the {\em language of optics} which we have dubbed
{\em Optica}. We aim at showing that Optica may play the role of an effective
query language for LINQ, alone and in combination with comprehension queries,
albeit at the expense of limited expressiveness. In general, we aim to prove
that optics offer a fruitful abstraction for LINQ, and restrict our attention to
proving the feasibility of this approach on a selected subset of optic
abstractions and domain examples. In particular, these are our contributions:

\begin{itemize}

\item A review of concrete optics from the mindset of LINQ. We show how to
    exploit a subset of standard optic abstractions and their combinators in
    order to express compositional queries (Sect.~\ref{sec:Programming}). We
    focus exclusively on \emph{read-only} optics, i.e. those which select parts
    from the whole but do not write back in the data structure, namely, {\em
    getters}, {\em affine folds} and {\em folds}. This allows us to focus on a
    tractable subset of optics, and to prepare the ground to tackle more
    ambitious problems in future work, such as the modeling of updates in LINQ.

\item A {\em formal specification} of read-only optics in terms of Optica, a
    full-blown DSL. The syntax and type system of the language formalize their
    compositional and querying features in an abstract way. Its denotational
    semantics is given by concrete optics themselves (Sect.\ \ref{sec:Optica}).
    We show how to implement generic queries over abstract optic models by using
    Optica in a declarative way, once and for all.

\item The abstract specification of read-only optics provided by Optica enables
    the definition of alternative, non-standard optic representations. We
    provide three major Optica interpretations which attempt to show the
    capabilities of Optica as a general query language for LINQ:

  \begin{itemize}
  \item An XQuery interpretation, that allows us to directly translate Optica
    queries into XQuery~\cite{wadler2002xquery} expressions
    (Sect.~\ref{sec:XQuery}). This aims at showing the adequacy of Optica for
    dealing with common data sources of NoSQL document-oriented databases.
    
  \item A SQL intepretation, that generates SQL queries from Optica expressions
    (Sect.~\ref{sec:SQL}). This non-standard denotational semantics builds upon
    {\em Triplets}, a semantic domain which normalizes the optic expression in
    order to facilitate its direct translation to SQL. The proposed semantics
    works similarly to the normalization-by-evaluation approach of SQUR
    \cite{DBLP:conf/aplas/KiselyovK17}. The major difference is that SQUR
    consists of a relational calculus whereas we work over optics, which are
    more akin to relational algebra.
  
  \item A T-LINQ interpretation, that generates comprehension queries. This
    non-standard semantics is aimed at showing how to use Optica as a
    higher-level language for nested data in conjunction with
    comprehension-based languages (Sect.~\ref{sec:T-LINQ}).
  \end{itemize}
    
\item S-Optica, an embedded DSL implementation of Optica in Scala using the
  tagless-final approach (Sect.~\ref{sec:s-optica}). This implementation is
  intended as a proof-of-concept for illustrating how to implement the formal
  Optica type system and semantics in a common, general purpose programming
  language of the software industry. It also aims at providing an example of the
  tagless final approach, as well as serving as a reference for extending
  existing libraries for LINQ with optic capabilities.
\end{itemize}
   
As can be seen, Sects.~\ref{sec:Programming}-\ref{sec:s-optica} contain the bulk
of the paper. Section~\ref{sec:Discussion} discusses related work and
limitations of the approach. Finally, section~\ref{sec:Conclusions} concludes
and points towards current and future work. The Scala library that accompanies
this paper is publicly available on a Github repository~\cite{habla2019scp19}.

\section{Querying with Optics}
\label{sec:Programming}

This section introduces three different \emph{kinds} of read-only optics:
\emph{getters}, \emph{affine folds} and \emph{folds}\footnote{We ignore other
read-only optics such as \emph{fold1}, since they do not add value in the
particular examples that we have selected for this paper.} together with their
main combinators, where we use Scala as the vehicle to implement
them\footnote{\ref{app:scala-cheatsheet} provides a Scala cheat sheet that
    describes the most fundamental Scala abstractions in the context of this
work.}. In essence, read-only optics are just views without updates, and hence
they are not subject to the familiar optic laws~\cite{grenrus2018glassery}.
They are not as widespread as their siblings with updating capabilities (namely,
\emph{lenses}, \emph{affine traversals} and \emph{traversals}), given that
selecting nested fields from immutable data structures is usually a trivial
task. Nonetheless, they exhibit the same compositional features and patterns as
the rest of optics, and will thus allow us to illustrate the general declarative
querying style advocated by them. The abstractions and examples that we put
forward in this section will be used throughout the paper.

\subsection{Getters, Affine Folds and Folds}
\label{sub:Programming:Getters}

First of all, it is worth noting that we choose the {\em concrete} optic
representation, where the notions of \emph{whole} and \emph{part} are clear, in
order to make definitions easier to understand. There are other representations,
such as \emph{van Laarhoven}~\cite{oconnor2011functor} or
\emph{profunctor}~\cite{pickering2017profunctor}, that implement optic
composition in a remarkably elegant way, whose signatures, however, are not as
easy to approach for an outsider\footnote{As evidenced by the jokes around this
topic in the functional programming community
\url{https://pbs.twimg.com/media/CypY7B1W8AAvqwl.jpg}}.

\begin{definition}[Getter] A getter consists of a function that selects a single
    part from the whole\footnote{A concrete \emph{lens} is basically a getter
    plus a function to update the whole from a new version of the part.}. We
    encode it in Scala as follows:
    \begin{lstlisting}
case class Getter[S, A](get: S => A)
    \end{lstlisting}
    The type parameters \emph{S} and \emph{A} will consistently serve as the
    whole and the part, respectively, throughout the different optic
    definitions.
\end{definition}

There are several getter combinators that will be used frequently in the text
that have been collected in the \emph{companion object} for \lstinline{Getter},
which is shown in Fig.~\ref{fig:getter}. The \lstinline{andThen} method combines
getters that are selecting nested values in order to produce a new getter that selects
a deeply nested value. The getter \lstinline{id} is the neutral component under the
\lstinline{andThen} composition, where whole and part do coincide. The
\lstinline{fork} combinator is required if we wish to put different parts
together. The \lstinline{like} combinator selects a constant part which is taken
as parameter, where the whole is ignored. The remaining combinators essentially
lift arithmetic operations into functions that take getters selecting operands
as parameters and produce a getter that selects the operation result. 

\begin{remark}
  We assume \lstinline{>>>} and \lstinline{***} as infix versions of
  \lstinline{andThen} and \lstinline{fork}, respectively, where the last symbol
  has precedence over the first. Similarly, we will overload the operators
  \lstinline{===}, \lstinline{>} and~\lstinline{-} as infix versions of
  \lstinline{equal}, \lstinline{greaterThan} and \lstinline{subtract},
  respectively. Last, we will use the postfix expression \lstinline{p.not} as an
  alias for \lstinline{not(p)}.
\end{remark}

\begin{figure}
  \begin{lstlisting}
object Getter {

  def id[A]: Getter[A, A] = Getter(a => a)

  def andThen[S, A, B](u: Getter[S, A], d: Getter[A, B]): Getter[S, B] =
    Getter(s => d.get(u.get(s)))

  def fork[S, A, B](l: Getter[S, A], r: Getter[S, B]): Getter[S, (A, B)] =
    Getter(s => (l.get(s), r.get(s)))

  def like[S, A](a: A): Getter[S, A] = Getter(const(a))

  def not[S](b: Getter[S, Boolean]): Getter[S, Boolean] = b >>> Getter(!_)

  def equal[S, A: Equal](x: Getter[S, A], y: Getter[S, A]): Getter[S, Boolean] =
    Getter(s => x.get(s) === y.get(s))

  def greaterThan[S](x: Getter[S, Int], y: Getter[S, Int]): Getter[S, Boolean] =
    Getter(s => x.get(s) > y.get(s))

  def subtract[S](x: Getter[S, Int], y: Getter[S, Int]): Getter[S, Int] =
    Getter(s => x.get(s) - y.get(s))
}
  \end{lstlisting}
  \caption{Getter Combinators.}
  \label{fig:getter}
\end{figure}

\begin{remark}
  Fork-like optic composition (we will also refer to it as \emph{horizontal
  composition}\footnote{And consequently, we will refer to \lstinline{andThen}
  as \emph{vertical composition}.}) is not widespread in the folklore. Indeed,
  it is not possible to implement it in a safe way for most optics. For example,
  an analogous implementation for composing lenses horizontally would violate
  lens laws~\cite{fischer2015clear} when both lenses select the very same part.
\end{remark}

\begin{definition}[AffineFold] An affine fold consists of a function that
    selects at most one part from the whole. We encode it as follows:
    \begin{lstlisting}
case class AffineFold[S, A](preview: S => Option[A]) 
    \end{lstlisting}
    We could see this optic as a simplification of an \emph{affine traversal},
    where we omit the updating part.
\end{definition}

Once again, we have packaged several affine fold combinators in
Fig.~\ref{fig:affine}. The identity affine fold simply selects the whole value
and wraps it in a \lstinline{Some} case. The \lstinline{andThen} combinator
selects the innermost value just in case both optics \lstinline{u} and
\lstinline{d} denote defined values. Otherwise, it will select nothing. We
implement this functionality in terms of the \lstinline{Option} monad using
for-comprehension syntax. Finally, we consider \emph{filtered} as an interesting
builder of affine folds, which declares the same types for whole and part. It
just discards the value (returning \lstinline{None}) in case it is actually
pointing to something and the input predicate does not hold for it.

\begin{remark}
    \label{remark:predicate}
    It is worth emphasizing that the predicate that \lstinline{filtered} takes
    as parameter is a getter itself. This is unusual in folklore libraries,
    where a plain lambda expression is taken as an argument instead. However,
    predicates can be perfectly understood as queries (getters, in particular).
    We will exploit this idea in the next section to avoid introducing lambda
    terms in the Optica DSL.
\end{remark}

\begin{remark}
    \label{rem:heterogeneous}
    One of the major benefits of optics is that they compose
    heterogeneously\footnote{With a very few exceptions, which are beyond the
    scope of this paper.}; in other words, it is possible to combine getters,
    affine folds and folds among them. To put it simply, we can turn getters
    into affine folds and we can turn affine folds into folds. An example of
    such casting is shown in Fig.~\ref{fig:affine} (\lstinline{to_af}), where
    we find it implemented as an implicit converter. Thereby, the Scala compiler
    applies the conversion automatically when it detects a getter where an
    affine fold is expected instead.
\end{remark}

\begin{figure}
  \begin{lstlisting}
object AffineFold {

  def id[A]: AffineFold[A, A] = AffineFold(a => Some(a))

  def andThen[A, B, C](u: AffineFold[A, B], 
                       d: AffineFold[B, C]): AffineFold[A, C] =
    AffineFold(s =>
      for {
        b <- u.preview(s)
        c <- d.preview(b)
      } yield c)

  def filtered[S](p: Getter[S, Boolean]): AffineFold[S, S] =
    AffineFold(s => if (p.get(s)) Some(s) else None)

  implicit def to_af[S, A](g: Getter[S, A]): AffineFold[S, A] =
    AffineFold(s => Some(g.get(s)))
}
  \end{lstlisting}
  \caption{Affine Fold Combinators.}
  \label{fig:affine}
\end{figure}

\begin{definition}[Fold] A fold consists of an optic that selects a (possibly
    empty) sequence of parts from the whole.
    \begin{lstlisting}
case class Fold[S, A](getAll: S => List[A])
    \end{lstlisting}
    We could see this optic as a simplification of a
    \emph{traversal}~\cite{oconnor2011functor}, where we omit the updating
    part. 
\end{definition}

As usual, we have packaged the fold primitives in the corresponding companion
object, which can be found in Fig.~\ref{fig:fold}. The implementation of
\lstinline{id} and \lstinline{andThen} is basically the same as the one we
showed for affine folds, the difference being that we work with the
\lstinline{List} monad instead of the \lstinline{Option} one\footnote{Similarly,
we can have getters following the same pattern by using the \lstinline{Id}
monad, but we avoid doing this for brevity.}. The \lstinline{nonEmpty} method
takes a fold as a parameter and produces a getter that checks whether the number
of selected parts is equal to zero. The remaining combinators
(\lstinline{empty}, \lstinline{all}, \lstinline{any} and \lstinline{elem}) are
just derived definitions, which are implemented in terms of other combinators,
where we assume that object-oriented \emph{dot} syntax is available. For
instance, \lstinline{nonEmpty(fl)} becomes \lstinline{fl.nonEmpty} and
\lstinline{all(fl)(p)} becomes \lstinline{fl.all(p)}. 

The implementation of \lstinline{elem} might deserve further explanation. Since
we favor getters over plain functions as predicates (as stated in
Remark~\ref{remark:predicate}), we need to use optic abstractions and
combinators to build them. The following is a derivation where we start with an
implementation that we consider natural ---which requires lambda expressions---
and we end up with the standing implementation ---where lambda expressions are
removed.
\begin{lstlisting}
  fl.any(Getter(s => s === a))
≃ [definition of `id` getter]
  fl.any(Getter(s => id.get(s) === a)
≃ [definition of `like` getter]
  fl.any(Getter(s => id.get(s) === like(a).get(s))
≃ [definition of `equal` getter]
  fl.any(id === like(a))
\end{lstlisting}
Note that \lstinline{===} is overloaded. In fact, the occurrence of this
method in the last line corresponds to the \lstinline{equal} combinator from
getters, while the rest refer to the standard comparison method from the
\lstinline{Equal} type class.

\begin{figure}
  \begin{lstlisting}
object Fold {

  def id[A]: Fold[A, A] = Fold(a => List(a))

  def andThen[A, B, C](u: Fold[A, B], d: Fold[B, C]): Fold[A, C] =
    Fold(s =>
      for {
        b <- u.getAll(s)
        c <- d.getAll(b)
      } yield c)

  def nonEmpty[S, A](fl: Fold[S, A]): Getter[S, Boolean] = 
    Getter(fl.getAll(_).nonEmpty) /* List.nonEmpty */

  def empty[S, A](fl: Fold[S, A]): Getter[S, Boolean] = 
    fl.nonEmpty.not

  def all[S, A](fl: Fold[S, A])(p: Getter[A, Boolean]): Getter[S, Boolean] = 
    (fl >>> filtered(p.not)).empty

  def any[S, A](fl: Fold[S, A])(p: Getter[A, Boolean]): Getter[S, Boolean] = 
    fl.all(p.not).not
    
  def elem[S, A: Equal](fl: Fold[S, A])(a: A): Getter[S, Boolean] = 
    fl.any(id === like(a))

  implicit def to_fl[S, A](a: AffineFold[S, A]): Fold[S, A] =
    Fold(s => a.preview(s).toList)
}
  \end{lstlisting}
  \caption{Fold Combinators.}
  \label{fig:fold}
\end{figure}

\begin{remark}
    As we have seen throughout this section, read-only optics are essentially
    functions that select parts from the whole, but we have introduced them as
    separated definitions. This distinction between functions and optics turns
    out to be central in this work, since Optica expressions denoting one or the
    other can be evaluated in very different ways, as we will see through
    Sects.~\ref{sec:Optica}-\ref{sec:T-LINQ}.
\end{remark}

\subsection{Composing Optics and Running Queries}
\label{sub:Programming:Real-World}

Once we have seen several standard combinators and some interesting features
from optics, we will exercise them to illustrate the querying style and common
patterns advocated by optics. For this task, we have selected two examples from
\cite{cheney2013practical}\footnote{The first example has been slightly updated
in order to adapt it to today's society.}, which will be used throughout the
paper.

\subsubsection{Couples Example}
\label{subsub:couple-example}

This example extends the one which was introduced in Table~\ref{tab:intro}.
Remember that it consists of a simple relation of couples, where the name and
age of each person forming them is also supplied:
\begin{lstlisting}
type Couples = List[Couple]
case class Couple(fst: Person, snd: Person)
case class Person(name: String, age: Int)
\end{lstlisting}
The associated data structures are defined following a {\em nested}, rather than
a {\em relational} approach, i.e. couples contain a full person value, rather
than a person key. This distinction becomes essential in Sect.~\ref{sec:T-LINQ}.
Once we have defined the model, we provide \lstinline{CoupleModel}-specific
optics to select relevant parts from the domain.
\begin{lstlisting}
object CoupleModel {
  val couples: Fold[Couples, Couple] = Fold(identity)
  val fst: Getter[Couple, Person] = Getter(_.fst)
  val snd: Getter[Couple, Person] = Getter(_.snd)
  val name: Getter[Person, String] = Getter(_.name)
  val age: Getter[Person, Int] = Getter(_.age)
}
\end{lstlisting}
Basically, and for this particular example, we supply a getter for each field,
where whole and part correspond to data and field types, respectively. The Scala
\emph{placeholder syntax} is used in these definitions. There is also a simple
fold that we can use to select each couple from \lstinline{Couples}, that we see
as the root type in the nested model.

\begin{remark}
    The examples that are presented in this paper do not include affine folds as
    part of the domain models, but they could be helpful to model optional
    values. For instance, we could use them to consider an optional
    \emph{address} field associated to each person.
\end{remark}

Now, we can use the standard optics defined in the previous section and the
specific optics defined for this domain to compose new ones. For instance, the
following fold selects the name and age difference from all those couples where
the first member is older than the second one.
\begin{lstlisting}
val differencesFl: Fold[Couples, (String, Int)] = 
  couples >>> filtered((fst >>> age) > (snd >>> age)) 
          >>> (fst >>> name) *** ((fst >>> age) - (snd >>> age))
\end{lstlisting}
Firstly, we use \lstinline{couples} as an entry point and we use
\lstinline{filtered} to remove the couples in which the age of the first member
is not greater than the age of the second one. Right after filtering, we select
the name of the first member and we put it together with the age difference, by
means of \lstinline{***}, to determine the subparts that the optic is selecting.

Once we have defined the fold, we need to generate the query that selects the
corresponding information from the immutable structure, i.e.\ a function that
takes the couples as argument and returns the matching values. For this task,
we simply use \lstinline{getAll}.
\begin{lstlisting}
val differences: Couples => List[(String, Int)] =
  differencesFl.getAll
\end{lstlisting}
If we feed this query with the same data that was used in the original
example~\cite{cheney2013practical}, we should expect the same result.
\begin{lstlisting}
val data: Couples = List(
  Couple(Person("Alex", 60), Person("Bert", 55)), 
  Couple(Person("Cora", 33), Person("Demi", 31)),
  Couple(Person("Eric", 21), Person("Fred", 60)))

val res: List[(String, Int)] = differences(data) 
// res: List[(String, Int)] = List((Alex,5), (Cora,2))
\end{lstlisting}
The comment in the last line of the snippet shows the value that we get in
\lstinline{res} when we run the query with the original data. As expected, it
indicates that Alex and Cora are older than their mates by 5 and 2 years,
respectively.

\subsubsection{Organization Example}
\label{subsub:organization-example}

We move on to the next example, where our model is an organization which is
formed by employees. In addition, each employee has a set of tasks that she is
able to perform.
\begin{lstlisting}
type Org = List[Department]
case class Department(dpt: String, employees: List[Employee])
case class Employee(emp: String, tasks: List[Task])
case class Task(tsk: String)
\end{lstlisting}

Once again, we supply \lstinline{OrgModel}-specific optics to select relevant
parts from the domain:
\begin{lstlisting}
object OrgModel {
  val departments: Fold[Org, Department] = Fold(identity)
  val dpt: Getter[Department, String] = Getter(_.dpt)
  val employees: Fold[Department, Employee] = Fold(_.employees)
  val emp: Getter[Employee, String] = Getter(_.emp)
  val tasks: Fold[Employee, Task] = Fold(_.tasks)
  val tsk: Getter[Task, String] = Getter(_.tsk)
}
\end{lstlisting}
In this case, we find several fields containing lists, thus, we provide folds
instead of getters to deal with sequences of parts. Now, we compose a fold to
select the name of those departments where all employees know how to {\em
abstract}.
\begin{lstlisting}
def expertiseFl: Fold[Org, String] =
  departments >>> filtered(employees.all(tasks.elem("abstract"))) >>> dpt
\end{lstlisting}
This expression refers first to all departments, and then it filters the ones where
all employees contain the task \lstinline{"abstract"}. Finally, it selects their
textual identifier (\lstinline{dpt}). Once the fold is defined, we produce the
query to obtain the selected departments:
\begin{lstlisting}
def expertise: Org => List[String] =
  expertiseFl.getAll
\end{lstlisting}
Once more, we feed the query with the original organization's data.
\begin{lstlisting}
val data: Org = List(
  Department("Product", List(
    Employee("Alex", List(Task("build"))),
    Employee("Bert", List(Task("build"))))),
  Department("Quality", List.empty),
  Department("Research", List(
    Employee("Cora", List(Task("abstract"), Task("build"), Task("design"))),
    Employee("Demi", List(Task("abstract"), Task("design"))),
    Employee("Eric", List(Task("abstract"), Task("call"), Task("design"))))),
  Department("Sales", List(
    Employee("Fred", List(Task("call"))))))

val res: List[String] = expertise(data)
// res: List[String] = List(Quality, Research)
\end{lstlisting}
The resulting value shows that the departments of \emph{Quality} and
\emph{Research} are the only ones where all employees are able to
\emph{abstract}.

The general pattern should be clear now. Firstly, we define the involved data
types in the model and supply specific optics to select their parts. Secondly,
we use these optics and the standard ones to express more sophisticated
selectors in a modular and elegant way. Finally, we run the optic method with an
initial whole to produce the expected query. As it can be seen, the approach is
eminently declarative: the aspects of composing the desired optic and running it
are completely decoupled.

\begin{remark}
    \label{remark:actions}
    We will refer to \lstinline{get}, \lstinline{preview} and \lstinline{getAll}
    as the {\em queries} derived from their corresponding optic types. Read-only
    optics just supply basic reading queries, in a one-to-one mapping.  Thereby,
    the separation of concerns between optic expressions and generated queries
    is not as clear as in other optics. For instance, lenses broaden their
    catalog of derived queries with queries to read, replace or even modify the
    part that the lens is selecting. We further discuss the implications of this
    insight in Sect.~\ref{sec:Discussion}.
\end{remark}

\section{Optica Core}
\label{sec:Optica}

The last section has introduced optics using their so-called {\em concrete}
representation, but the same querying style is actually supported by other
isomorphic representations, such as van Laarhoven or profunctor optics. This
section aims at {\em specifying} the concepts of getters, affine folds and
folds, in a generic way, independently of any particular {\em representation}.
The resulting formalization is a domain-specific language that we have named {\em
  Optica}, which directly supports the optic querying style, and potentially
allows for non-standard optic representations that generate queries in XQuery or
SQL, for instance, as the next sections will show.

This section will first introduce the syntax and type system of Optica, where
standard primitives and combinators are declared. Secondly, we will show how to
provide a generic version of the models and queries that we have seen in
Sect.~\ref{sub:Programming:Real-World}. Finally, we will present the standard
semantics that we can use as an interpretation to deal with immutable data
structures: they recover concrete optics and queries, as introduced in the last
section.

\subsection{Syntax and Type System}
\label{sub:Syntax}

\begin{figure}
\small
\begin{align*}
&\text{Base types} & b &::=\ \mathbb{N} \mid \mathbb{B} \mid \mathbb{S} \\
&\text{Model types} & t & ::=\ b \mid (t, t)\\
&\text{Optic Types} & s &::=\ \mathit{getter}\ t\ t \mid \mathit{affine}\
t\ t \mid \mathit{fold}\ t\ t \\
&\text{Query Types} & u &::=\ t \rightarrow t \mid t\ \rightarrow \mathit{option}\ t \mid t \rightarrow \mathit{list}\ t\\\\
&\text{Constants} & c & \quad (\text{of base type}) \\
&\text{Optic Expressions} & e &::=\ \mathit{id}_{\mathit{gt}} \mid \mathit{id}_{\mathit{af}} \mid
\mathit{id}_{\mathit{fl}} \\
&&&\quad \mid e \ggg_{\mathit{gt}} e \mid e \ggg_{\mathit{af}} e\mid e \ggg_{\mathit{fl}} e \\ 
&&&\quad \mid \mathit{like}\ c \mid \mathit{not}\ e \mid e > e \mid e == e \mid
e - e \mid e\ \Fork\ e \\
&&&\quad \mid \mathit{filtered}\ e \mid \mathit{nonEmpty}\ e \\
&&&\quad \mid \mathit{to}_{\mathit{af}}\ e \mid \mathit{to}_{\mathit{fl}}\ e \\
&\text{Query Expressions} & q &::=\ \mathit{get}\ e \mid \mathit{preview}\ e
\mid \mathit{getAll}\ e
\end{align*}
\caption{Optica syntax}
\label{fig:optica-syntax}
\end{figure}

We introduce the syntax of Optica in Fig.~\ref{fig:optica-syntax}. The upper
part contains the model types (natural numbers, boolean, string and product),
optic types (getters, affine folds and folds) and query types (selection
functions).  The second part shows the set of expressions that form the
language, which are defined in close correspondence with their concrete
counterparts, introduced in the previous section. Despite that, there are
several aspects which deserve further explanation:

\begin{itemize}
    \item Constants are not valid query expressions on their own. As we will see
        later, we use $\mathit{like}$ as the mechanism to represent literals in
        the language. By doing so, we can reuse optic combinators for constants,
        improving the language composability.
      \item The formal syntax avoids the object-oriented \emph{dot} notation,
        idiomatic in Scala, and favours the prefix notation, as is usual practice in
        related work.
    \item The methods \lstinline{all}, \lstinline{any}, \lstinline{elem} and
        \lstinline{empty} are not included as syntax primitives. Instead, they
        are introduced as derived definitions, as can be seen in
        Fig.~\ref{fig:derived-defs}.
      \item At present, query expressions are atomic, i.e. it is not possible to
        compose several of them.
\end{itemize}

\begin{figure}
    \small
\begin{align*}
&\mathbf{def}\ \mathit{empty}\ \mathit{fl}=&&\mathbf{def}\ \mathit{all}\ \mathit{fl}\ \mathit{p}=&\\
&\quad \mathit{nonEmpty}\ \mathit{fl} \ggg \mathit{not}&
&\quad \mathit{empty}\ (\mathit{fl} \ggg\ \mathit{filtered}\ (\mathit{not}\ \mathit{p}))&\\\\
&\mathbf{def}\ \mathit{any}\ \mathit{fl}\ \mathit{p}=&&\mathbf{def}\
    \mathit{elem}\ \mathit{fl}\ \mathit{u}=&\\
&\quad \mathit{not}\ (\mathit{all}\ \mathit{fl}\ (\mathit{not}\ p))&
&\quad \mathit{any}\ \mathit{fl}\ (\mathit{id} == \mathit{like\ a})&
\end{align*}
\caption{Optica derived definitions}
\label{fig:derived-defs}
\end{figure}

\begin{figure}[]
    \centering
    \small
\begin{tabular}{ll}

\infer[id_{\mathit{gt}}]{id_{\mathit{gt}}: \mathit{getter}\ \alpha\ \alpha}{}
&
\infer[\ggg_{\mathit{gt}}]{g_1\ \ggg_{\mathit{gt}} g_2:\ \mathit{getter}\ \alpha\ \gamma}{
    g_1:\ \mathit{getter}\ \alpha\ \beta
    & g_2:\ \mathit{getter}\ \beta\ \gamma
} \\\\

\infer[\Fork]{g_1\ \Fork\ g_2:\ \mathit{getter}\ \alpha\ (\beta,\ \gamma)}{
    g_1:\ \mathit{getter}\ \alpha\ \beta
    & g_2:\ \mathit{getter}\ \alpha\ \gamma
}
&
\infer[\mathit{like}]{\mathit{like}\ b:\ \mathit{getter}\ \alpha\ \beta}{
    b:\ \beta
& \beta \in \text{base types}
} \\\\

\infer[\mathit{not}]{\mathit{not}\ g:\ \mathit{getter}\ \alpha\ \mathbb{B}}{
    g:\ \mathit{getter}\ \alpha\ \mathbb{B}
}
&
\infer[>]{g_1 > g_2:\ \mathit{getter}\ \alpha\ \mathbb{B}}{
    g_1:\ \mathit{getter}\ \alpha\ \mathbb{N}
    & g_2:\ \mathit{getter}\ \alpha\ \mathbb{N}
} \\\\

\infer[==]{g_1 == g_2:\ \mathit{getter}\ \alpha\ \mathbb{B}}{
    g_1:\ \mathit{getter}\ \alpha\ \beta
    & g_2:\ \mathit{getter}\ \alpha\ \beta
}
&
\infer[-]{g_1 - g_2:\ \mathit{getter}\ \alpha\ \mathbb{N}}{
    g_1:\ \mathit{getter}\ \alpha\ \mathbb{N}
    & g_2:\ \mathit{getter}\ \alpha\ \mathbb{N}
} \\\\\\\\

\infer[id_{\mathit{af}}]{id_{\mathit{af}}: \mathit{affine}\ \alpha\ \alpha}{}
&
\infer[\ggg_{\mathit{af}}]{a_1\ \ggg_{\mathit{af}} a_2:\ \mathit{affine}\ \alpha\ \gamma}{
    a_1:\ \mathit{affine}\ \alpha\ \beta
    & a_2:\ \mathit{affine}\ \beta\ \gamma
} \\\\

\infer[\mathit{filtered}]{\mathit{filtered}\ p:\ \mathit{affine}\ \alpha\ \alpha}{
    p:\ \mathit{getter}\ \alpha\ \mathbb{B}
}

&

\infer[\mathit{to}_{\mathit{af}}]{\mathit{to}_{\mathit{af}}\ g: \mathit{affine}\ \alpha\ \beta}{
    g:\ \mathit{getter}\ \alpha\ \beta
} \\\\\\\\

\infer[id_{\mathit{fl}}]{id_{\mathit{fl}}: \mathit{fold}\ \alpha\ \alpha}{}
&
\infer[\ggg_{\mathit{fl}}]{f_1\ \ggg_{\mathit{fl}} f_2:\ \mathit{fold}\ \alpha\ \gamma}{
    f_1:\ \mathit{fold}\ \alpha\ \beta
    & f_2:\ \mathit{fold}\ \beta\ \gamma
} \\\\

\infer[\mathit{nonEmpty}]{\mathit{nonEmpty}\ f:\ \mathit{getter}\ \alpha\ \mathbb{B}}{
    f:\ \mathit{fold}\ \alpha\ \beta
} &

\infer[\mathit{to}_{\mathit{fl}}]{\mathit{to}_{\mathit{fl}}\ a: \mathit{fold}\ \alpha\ \beta}{
    a:\ \mathit{affine}\ \alpha\ \beta
} \\\\\\\\

\infer[\mathit{get}]{\mathit{get}\ g:\ \alpha \rightarrow \beta}{
    g:\ \mathit{getter}\ \alpha\ \beta
} 
&
\infer[\mathit{preview}]{\mathit{preview}\ a:\ \alpha \rightarrow option\ \beta}{
    a:\ \mathit{affine}\ \alpha\ \beta
} \\\\

\infer[\mathit{getAll}]{\mathit{getAll}\ f:\ \alpha \rightarrow list\ \beta}{
    f:\ \mathit{fold}\ \alpha\ \beta
}
\end{tabular}
\caption{Optica type system}
\label{fig:typing-rules}
\end{figure}

The type system is presented in Fig.~\ref{fig:typing-rules}, where $\alpha$,
$\beta$ and $\gamma$ represent model types (see Fig.~\ref{fig:optica-syntax}).
Unlike T-LINQ~\cite{cheney-shonan-meeting} or
QUE$\Lambda$~\cite{DBLP:conf/aplas/KiselyovK17}, Optica does not introduce terms
for variables. Thereby, its type rules are slightly simplified, since they omit
the characteristic `$\Gamma \vdash$' prefix. They are structured into four
groups, corresponding to getters, affine folds, folds, and their derived
queries. The only optic constructors are $\mathit{id}_*$ and $\mathit{like}$,
which allow us to form new optic expressions from scratch. The remaining
combinators should be straightforward, since they exactly correspond with the
ones introduced in the companion objects for getters, affine folds and folds of
Sect.~\ref{sec:Programming}\footnote{With the exception of those combinators,
    like {\em any}, {\em all}, {\em elem} and {\em empty}, which can be defined
compositionally in terms of other combinators and do not need to access the
internal optic representation.}. In regard to queries, note that their type
rules do not proceed from the companion objects, but from the case class
definitions of concrete optics themselves. Their formalization leads to
introducing functions as a new semantic domain for Optica. However, note that
the part of the language corresponding to optics is purely first-order, i.e. no
lambdas are needed in order to create optic expressions. 

\subsection{Core Extensions and Generic Queries}
\label{sub:Optica:Generic}

As we have seen in Sect.~\ref{sec:Programming}, we defined domain optics to
model the couple and organization examples. Now, we want to do the same, but in
a general way. To do so, we need to extend the core language from
Sect.~\ref{sub:Syntax} with new primitives, specific to the particular domain.
We present them, along with their associated type rules, in
Fig.~\ref{fig:couple-syntax}. As can be seen, it introduces the entity types
($\mathit{Couple}$ and $\mathit{Person}$) and a term for each optic introduced
in Sect.~\ref{subsub:couple-example}. The type rules just determine the type
associated to each term, i.e. the kind, whole and part associated to each optic.

\begin{figure}[h]
\small
\centering
\begin{align*}
&\text{Entity Types} & t &\Ext\ \mathit{Couples} \mid \mathit{Couple} \mid \mathit{Person} \\
&\text{Optic Expressions} & e &\Ext\ \mathit{couples} \mid \mathit{fst} \mid
\mathit{snd} \mid \mathit{name} \mid \mathit{age}\\
\end{align*}
\begin{tabular}{ll}

\infer[\mathit{couples}]{\mathit{couples}: \mathit{fold}\ \mathit{Couples}\ \mathit{Couple}}{}
&
\infer[\mathit{fst}]{\mathit{fst}: \mathit{getter}\ \mathit{Couple}\ \mathit{Person}}{}\\\\

\infer[\mathit{snd}]{\mathit{snd}: \mathit{getter}\ \mathit{Couple}\ \mathit{Person}}{}
&
\infer[\mathit{name}]{\mathit{name}: \mathit{getter}\ \mathit{Person}\ \mathbb{S}}{}\\\\

\infer[\mathit{age}]{\mathit{age}: \mathit{getter}\ \mathit{Person}\ \mathbb{N}}{}
\end{tabular}
\caption{Couple syntax and type system}
\label{fig:couple-syntax}
\end{figure}

Once we have defined the Optica language primitives (where we place the standard
optics and combinators) and the domain extension (where we find the structure of
the domain data model in terms of specific optics), we should be able to provide
generic domain queries. We adapt \lstinline{differences}
(Sect.~\ref{subsub:couple-example}) as follows:
\begin{definition} \label{def:differences}
    The generic versions for $\mathit{differencesFl}$ (optic
    expression) and $\mathit{differences}$ (query expression) are implemented as
    follows, in terms of the Optica and couple-specific primitives.
\small
\begin{flalign*}
&\mathbf{def}\ \mathit{differencesFl}=&\\
&\quad \mathit{couples} \ggg \mathit{to}_{\mathit{fl}}\ (\mathit{filtered}\
((\mathit{fst} \ggg
\mathit{age}) > (\mathit{snd} \ggg \mathit{age})) \ggg&\\
&\quad \mathit{to}_{\mathit{af}}\ ((\mathit{fst} \ggg \mathit{name}) \Fork
((\mathit{fst} \ggg \mathit{age}) -
(\mathit{snd} \ggg \mathit{age}))))&
\end{flalign*}
\begin{flalign*}
&\mathbf{def}\ \mathit{differences}=&\\
&\quad \mathit{getAll}\ \mathit{differencesFl}&
\end{flalign*}
\end{definition}
The implementation of the generic versions of \lstinline{differencesFl} and
\lstinline{differences} are basically the same as the ones we introduced in
Sect.~\ref{subsub:couple-example} ---modulo the differences that we listed in
Sect.~\ref{sub:Syntax} and the fact that the invocations to casting methods such
as $\mathit{to}_{\mathit{af}}$ and $\mathit{to}_{\mathit{fl}}$ are made
explicit.

\begin{figure}
\small
\centering
\begin{align*}
&\text{Entity Types} & t &\Ext\ \mathit{Org} \mid \mathit{Department} \mid
\mathit{Employee} \mid \mathit{Task} \\
&\text{Optic Expressions} & e &\Ext\ \mathit{departments} \mid \mathit{dpt} \mid
\mathit{employees} \mid \mathit{emp} \mid \mathit{tasks} \mid \mathit{tsk}\\
\end{align*}
\begin{tabular}{ll}

\infer[\mathit{departments}]{\mathit{departments}: \mathit{fold}\ \mathit{Org}\ \mathit{Department}}{}
&
\infer[\mathit{dpt}]{\mathit{dpt}: \mathit{getter}\ \mathit{Department}\ \mathbb{S}}{}\\\\

\infer[\mathit{employees}]{\mathit{employee}: \mathit{fold}\ \mathit{Department}\ \mathit{Employee}}{}
&
\infer[\mathit{emp}]{\mathit{emp}: \mathit{getter}\ \mathit{Employee}\ \mathbb{S}}{}\\\\

\infer[\mathit{tasks}]{\mathit{tasks}: \mathit{getter}\ \mathit{Employee}\ \mathit{Task}}{}
&
\infer[\mathit{tsk}]{\mathit{tsk}: \mathit{getter}\ \mathit{Task}\ \mathbb{S}}{}\\\\
\end{tabular}

\caption{Organization syntax and type system}
\label{fig:org-syntax}
\end{figure}

We can carry out the same exercise for the organization example
(Sect.~\ref{subsub:organization-example}). Once again, we introduce a language
extension containing the entity types and terms associated to this example
(Fig.~\ref{fig:org-syntax}). Once we do that, we are able to introduce a generic
counterpart for the \lstinline{expertise} query.
\begin{definition} \label{def:expertise}
    The generic versions for $\mathit{expertiseFl}$ (optic expression) and
    $\mathit{expertise}$ (query expression) are implemented as follows, in terms
    of the Optica and organization-specific primitives.
\small
\begin{flalign*}
&\mathbf{def}\ \mathit{expertiseFl}\ =&\\
&\quad \mathit{departments} \ggg \mathit{to}_{\mathit{fl}}\ (\mathit{filtered}\ (\mathit{all}\
\mathit{employees}\ (\mathit{elem}\ (\mathit{tasks} \ggg
\mathit{to}_{\mathit{fl}}\ (\mathit{to}_{\mathit{af}}\ \mathit{tsk}))\ `\mathit{abstract}\textrm')) \ggg \mathit{to}_{\mathit{af}}\
\mathit{dpt})
\end{flalign*}
\begin{flalign*}
&\mathbf{def}\ \mathit{expertise}\ =&\\
&\quad \mathit{getAll}\ \mathit{expertiseFl}\ &
\end{flalign*}
\end{definition}
At this point, we have defined generic queries which are not coupled to any
particular querying infrastructure. In the rest of the paper, we will show how
to reuse such queries for generating in-memory, XQuery, SQL and
comprehension-based queries.

\subsection{Standard Semantics}
\label{sub:Optica:Standard}

\begin{figure}
\small
\begin{align*}
&\mathcal{T}[\mathbb{N}] &=& \quad \texttt{Int} \\
&\mathcal{T}[\mathbb{B}] &=& \quad \texttt{Boolean} \\
&\mathcal{T}[\mathbb{S}] &=& \quad \texttt{String} \\
&\mathcal{T}[(\alpha, \beta)] &=& \quad \texttt{(}\mathcal{T}[\alpha],\ \mathcal{T}[\beta]\texttt{)} \\
&\mathcal{T}[\mathit{getter}\ \alpha\ \beta] &=& \quad \texttt{Getter}[\mathcal{T}[\alpha],\ \mathcal{T}[\beta]]\\
&\mathcal{T}[\mathit{affine}\ \alpha\ \beta] &=& \quad \texttt{Affine}[\mathcal{T}[\alpha],\ \mathcal{T}[\beta]]\\
&\mathcal{T}[\mathit{fold}\ \alpha\ \beta]   &=& \quad \texttt{Fold}[\mathcal{T}[\alpha],\ \mathcal{T}[\beta]]\\
&\mathcal{T}[\alpha \rightarrow\ \beta]                 &=& \quad \mathcal{T}[\alpha] \Rightarrow\ \mathcal{T}[\beta]\\
&\mathcal{T}[\alpha \rightarrow \mathit{option}\ \beta] &=& \quad \mathcal{T}[\alpha] \Rightarrow\ \texttt{Option}[\mathcal{T}[\beta]]\\
&\mathcal{T}[\alpha \rightarrow \mathit{list}\ \beta]   &=& \quad \mathcal{T}[\alpha] \Rightarrow\ \texttt{List}[\mathcal{T}[\beta]]
\end{align*}

\caption{Optica semantic domains}
\label{fig:semantic-domains}
\end{figure}

\begin{figure}
\small
\begin{align*}
&\mathcal{E}[\mathit{id}_{\mathit{gt}}:\ \mathit{getter}\ \alpha\ \alpha]&=& \quad
\texttt{Getter.id} \\
&\mathcal{E}[g\ \ggg_{\mathit{gt}} h: \mathit{getter}\ \alpha\ \gamma]&=& \quad 
\texttt{Getter.andThen}(\mathcal{E}[g: \mathit{getter}\ \alpha\ \beta],\
\mathcal{E}[h: \mathit{getter}\ \beta\ \gamma]) \\
&\mathcal{E}[g\ \Fork\ h:\ \mathit{getter}\ \alpha\ (\beta,\ \gamma)]&=& \quad 
\texttt{Getter.fork}(\mathcal{E}[g: \mathit{getter}\ \alpha\ \beta],\
\mathcal{E}[h: \mathit{getter}\ \alpha\ \gamma]) \\
&\mathcal{E}[\mathit{like}\ b:\ \mathit{getter}\ \alpha\ \beta]&=& \quad
\texttt{Getter.like}(\mathcal{E}[b: \beta]) \\
&\mathcal{E}[\mathit{not}\ g:\ \mathit{getter}\ \alpha\ \mathbb{B}]&=& \quad
\texttt{Getter.not}(\mathcal{E}[g: \mathit{getter}\ \alpha\ \mathbb{B}]) \\
&\mathcal{E}[g \oplus h:\ \mathit{getter}\ \alpha\ \delta]&=& \quad
\texttt{Getter.}\oplus(\mathcal{E}[g: \mathit{getter}\ \alpha\ \beta],\
\mathcal{E}[h: \mathit{getter}\ \alpha\ \gamma]) \\\\\
&\mathcal{E}[\mathit{id}_{\mathit{af}}:\ \mathit{affine}\ \alpha\ \alpha]&=& \quad
\texttt{AffineFold.id} \\
&\mathcal{E}[g\ \ggg_{\mathit{af}} h: \mathit{affine}\ \alpha\ \gamma]&=& \quad
\texttt{AffineFold.andThen}(\mathcal{E}[g: \mathit{affine}\ \alpha\ \beta],\
\mathcal{E}[h: \mathit{affine}\ \beta\ \gamma]) \\
&\mathcal{E}[\mathit{filtered}\ p:\ \mathit{affine}\ \alpha\ \alpha]&=& \quad 
\texttt{AffineFold.filtered}(\mathcal{E}[p: \mathit{getter}\ \alpha\ \mathbb{B}]) \\
&\mathcal{E}[\mathit{to}_{\mathit{af}}\ g:\ \mathit{affine}\ \alpha\ \beta]&=& \quad
\texttt{AffineFold.to}_{\mathit{af}}(\mathcal{E}[g: \mathit{getter}\ \alpha\
\beta]) \\\\
&\mathcal{E}[\mathit{id}_{\mathit{fl}}:\ \mathit{fold}\ \alpha\ \alpha]&=& \quad
\texttt{Fold.id} \\
&\mathcal{E}[g\ \ggg_{\mathit{fl}} h: \mathit{fold}\ \alpha\ \gamma]&=& \quad 
\texttt{Fold.andThen}(\mathcal{E}[g: \mathit{fold}\ \alpha\ \beta],\
\mathcal{E}[h: \mathit{fold}\ \beta\ \gamma]) \\
&\mathcal{E}[\mathit{nonEmpty}\ g:\ \mathit{getter}\ \alpha\ \mathbb{B}]&=& \quad
\texttt{Fold.nonEmpty}(\mathcal{E}[g:\ \mathit{fold}\ \alpha\ \beta]) \\
&\mathcal{E}[\mathit{to}_{\mathit{fl}}\ a:\ \mathit{fold}\ \alpha\ \beta]&=& \quad
\texttt{Fold.to}_{\mathit{fl}}(\mathcal{E}[g: \mathit{affine}\ \alpha\ \beta]) \\\\
&\mathcal{E}[\mathit{get}\ g:\ \alpha \rightarrow \beta] &=& \quad 
\mathcal{E}[g: \mathit{getter}\ \alpha\ \beta]\texttt{.get}\\
&\mathcal{E}[\mathit{preview}\ g:\ \alpha \rightarrow \mathit{option}\ \beta] &=& \quad
\mathcal{E}[g: \mathit{affine}\ \alpha\ \beta]\texttt{.preview}\\
&\mathcal{E}[\mathit{getAll}\ g:\ \alpha \rightarrow \mathit{list}\ \beta]
&=& \quad 
\mathcal{E}[g: \mathit{fold}\ \alpha\ \beta]\texttt{.getAll}
\end{align*}

\caption{Optica standard semantics}
\label{fig:standard-semantics}
\end{figure}

In defining a new language, it is common practice to start with its syntax and
type system, and then proceed to define its semantics. In our case, we have
proceeded in reverse: we started with the intended semantics (optics and
queries) and created an abstract syntax and type system which mimic its
structure. Therefore, what is new in this section is how to formalize the
connection between the syntax and type system of Optica and concrete optics, its
intended semantics. For this task, we provide semantic functions $\mathcal{T}$
(Fig.~\ref{fig:semantic-domains}) and $\mathcal{E}$
(Fig.~\ref{fig:standard-semantics}). The first maps Optica types to their
corresponding semantic domains. The second maps an expression of type $t$ to an
element of the semantic domain $\mathcal{T}(t)$. As can be seen, $\mathcal{T}$
just maps types to their Scala counterparts\footnote{Scala does not include a
standard type for natural numbers. Instead of supplying them on our own, we
prefer to choose the standard \lstinline{Int} type for simplicity.}. Given this
scenario, the implementation of $\mathcal{E}$ turns out to be trivial. In fact,
we just translate the Optica expressions into their Scala analogues from
Sect.~\ref{sec:Programming}. We use $\oplus$ to unify the different binary
combinators (\lstinline{>}, \lstinline{-}, etc.).

\begin{figure}
\small
\begin{center}
\begin{tabular}{lcl}
$\mathcal{T}[Couples]$ &=& $\texttt{Couples}$ \\
$\mathcal{T}[Couple]$ &=& $\texttt{Couple}$ \\
$\mathcal{T}[Person]$ &=& $\texttt{Person}$ \\\\
$\mathcal{E}[\mathit{couples}:\ \mathit{fold}\ \mathit{Couples}\ \mathit{Couple}]$ & $=$ & $\texttt{CoupleModel.couples}$ \\
$\mathcal{E}[fst: \mathit{getter}\ \mathit{Couple}\ \mathit{Person}]$ & $=$ &
$\texttt{CoupleModel.fst}$ \\
$\mathcal{E}[snd: \mathit{getter}\ \mathit{Couple}\ \mathit{Person}]$ & $=$ &
$\texttt{CoupleModel.snd}$ \\
$\mathcal{E}[name: \mathit{getter}\ \mathit{Person}\ \mathbb{S}]$ & $=$ & $\texttt{CoupleModel.name}$ \\
$\mathcal{E}[age: \mathit{getter}\ \mathit{Person}\ \mathbb{N}]$ & $=$ & $\texttt{CoupleModel.age}$
\end{tabular}
\end{center}
\caption{Semantic domains and standard semantics for couples extension}
\label{fig:couple-standard-semantics}
\end{figure}

We also need to take into account the evaluation of the extended versions of the
language, where terms specific to each example are introduced. For instance,
Fig.~\ref{fig:couple-standard-semantics} shows the semantic domains and the
evaluation of each term from the couples example extension. It is also trivial,
since it just maps the domain-specific terms to the concrete optics from
Sect.~\ref{subsub:couple-example}. The corresponding instance for the
organization extension follows the very same pattern and we omit it for brevity.
Once we have defined the standard semantics for all terms, we should be able to
translate generic queries into plain functions, by means of $\mathcal{E}$. We
evaluate $\mathit{differences}$ (Def.~\ref{def:differences}) as follows:
\par\nobreak\vspace{-0.4cm}
{\small
\begin{flalign*}
&\mathbf{def}\ \mathit{differencesR}: \texttt{Couples} \Rightarrow
\texttt{List[(String, Int)]}=&\\
&\quad \mathcal{E}[\mathit{differences}: \mathit{Couples}
\rightarrow \mathit{list}\ (\mathbb{S}, \mathbb{N})]
\end{flalign*}
}%
As can be seen, the resulting value is a Scala function that works with
immutable data structures. Finally, $\mathit{expertise}$
(Def.~\ref{def:expertise}) is evaluated in this way:
\par\nobreak\vspace{-0.4cm}
{\small
\begin{flalign*}
&\mathbf{def}\ \mathit{expertiseR}: \texttt{Org} \Rightarrow \texttt{List[String]}=&\\
&\quad \mathcal{E}[\mathit{expertise}: \mathit{Org} \rightarrow
\mathit{list}\ \mathbb{S}]
\end{flalign*}
}%
It recovers a Scala function which selects the corresponding department names.
These functions are exactly the same as their counterparts from
Sect.~\ref{sec:Programming}.

\section{XQuery}
\label{sec:XQuery}

So far, we have seen that optics allow us to manipulate immutable data
structures in a modular and elegant way, and that concrete optics can be lifted
into the Optica language, a full-blown DSL. The standard semantics of Optica is
given in terms of concrete optics; however, this is not significant from
the point of view of language-integrated query. The state of real applications
is mostly handled through SQL and NoSQL databases, web services, etc.; therefore, this
section and the following will show how to reuse Optica expressions in order to
generate queries for external data sources beyond in-memory data structures.
In particular, this section shows how getters, affine folds and folds from the
Optica DSL can actually be given a non-standard representation in terms of
XQuery expressions. Prior to that, we will manually adapt the
\lstinline{differences} and \lstinline{expertise} queries and corresponding
models into the XML/XQuery setting~\cite{wadler2002xquery} in an idiomatic way.
This will serve as a point of reference to implement the aforementioned
semantics. In this sense, there are several assumptions that we need to make in
order to map optics and XML models, which are described subsequently.

\subsection{XML/XQuery Background}
\label{sec:xmlxquery-background}

\newcommand{\xmlinline}[1]{\lstinline[columns=fixed,language=XML]{#1}}

\begin{figure}
  \centering
  \begin{lstlisting}[language=XML, basicstyle=\footnotesize\ttfamily]
<xml>
    <couple>
        <fst><name>Alex</name><age>60</age></fst>
        <snd><name>Bert</name><age>55</age></snd>
    </couple>
    <couple>
        <fst><name>Cora</name><age>33</age></fst>
        <snd><name>Demi</name><age>31</age></snd>
    </couple>
    <couple>
        <fst><name>Eric</name><age>21</age></fst>
        <snd><name>Fred</name><age>60</age></snd>
    </couple>
</xml>
  \end{lstlisting}
  \caption{Couples data represented as XML.}
  \label{fig:couple-xml}
\end{figure}

Accommodating objects into XML types is not a trivial
task~\cite{lammel2006revealing}. Figure~\ref{fig:couple-xml} shows a possible
way of encoding the state of the couples example in an XML document. It contains
a root element \xmlinline{xml} where couples hang from as \xmlinline{couple}
elements, which in turn contain subelements for the first (\xmlinline{fst}) and
second (\xmlinline{snd}) members that form the couple. Finally, \xmlinline{name}
and \xmlinline{age} are simple tags that contain primitive attributes.

Usually, an XML document is accompanied by an XSD schema, which is essential to
validate the information that we place in the document. The schema associated to
the couple document can be found in \ref{app:XML:Couple}. Among other things, it
prevents us from defining people without a \xmlinline{name} element, placing non
numerical values as \xmlinline{age} values, and defining several \xmlinline{fst}
elements inside a couple. Later, we will see that it is important to take this
schema into account while implementing queries.

Now, we would like to produce an XQuery expression, analogous to
\lstinline{differences}. It should be able to collect the name and age
difference of all people who occupy the first position in the couple and are
older than their mates. Since we do not want to calculate a single value for
this query, like a number or a boolean, the results should be presented as a
sequence of nodes, i.e.\ the output is also an XML tree. For example, this is
the output of the query we are looking for:
\begin{lstlisting}[language=XML, basicstyle=\scriptsize\ttfamily]
<xml>
    <tuple>
        <one><name>Alex</name></one>
        <two>5</two>
    </tuple>
    <tuple>
        <one><name>Cora</name></one>
        <two>2</two>
    </tuple>
</xml>
\end{lstlisting}
We pair values by means of a contrived \xmlinline{tuple} element, which contains
\xmlinline{one} and \xmlinline{two} projection subelements, where data is
finally stored. Once we know the output that we want to produce, we show the
XQuery expression that we could use to generate it\footnote{We broke the query
into several lines for readability, but note that certain interpreters require a
single line in order to produce a valid result.}.
\begin{lstlisting}[language=XML, basicstyle=\scriptsize\ttfamily]
/xml/couple[fst/age > snd/age]/<tuple>
                                 <one>{fst/name}</one>
                                 <two>{fst/age - snd/age}</two>
                               </tuple>
\end{lstlisting}
We describe its main components in the following paragraphs:
\begin{itemize}
\item One of the most fundamental queries is \xmlinline{/}, which grants access
    to the so-called \emph{document} node, which can be seen as the entry point
    in the document. Since XML documents are essentially nested data structures,
    XQuery provides concise syntax to access nested elements. For example,
    \xmlinline{/xml/couple} selects all elements \xmlinline{couple} that are
    hanging from an element \xmlinline{xml} which in turn should be accessible
    from the document node.

\item XQuery does provide filters to enrich queries, which are placed inside
    square brackets. For example, \xmlinline{[fst/age > snd/age]} is a filter
    that we apply over \xmlinline{/xml/couple} to discard the couples in which
    the age of the fist member is not greater than the age of the second one.
    The operator \xmlinline{>} is able to extract the inner value of these
    elements and interpret them as numbers. In fact, this should be safe, if we
    take into account the XML schema.

\item XQuery supports XML interpolation to enrich its results. It serves us to
    provide the structure that we need in order to put pairs of values together.
    It is worth mentioning that this is the only feature from XQuery which is
    not available in XPath, among the ones we use in this work.
\end{itemize}

\begin{figure}
  \centering
  \begin{lstlisting}[language=XML, basicstyle=\footnotesize\ttfamily]
<xml>
    <department>
        <dpt>Product</dpt>
        <employee>
            <emp>Alex</emp>
            <task><tsk>build</tsk></task>
        </employee>
        <employee>
            <emp>Bert</emp>
            <task><tsk>build</tsk></task>
        </employee>
    </department>
    <department>
        <dpt>Quality</dpt>
    </department>
    <department>
        <dpt>Research</dpt>
        <employee>
            <emp>Cora</emp>
            <task><tsk>abstract</tsk></task>
            <task><tsk>build</tsk></task>
            <task><tsk>design</tsk></task>
        </employee>
        <employee>
            <emp>Demi</emp>
            <task><tsk>abstract</tsk></task>
            <task><tsk>design</tsk></task>
        </employee>
        <employee>
            <emp>Eric</emp>
            <task><tsk>abstract</tsk></task>
            <task><tsk>call</tsk></task>
            <task><tsk>design</tsk></task>
        </employee>
    </department>
    <department>
        <dpt>Sales</dpt>
        <employee>
            <emp>Fred</emp>
            <task><tsk>call</tsk></task>
        </employee>
    </department>
</xml>
  \end{lstlisting}
  \caption{Organization deployed as XML.}
  \label{fig:org-xml}
\end{figure}

Now, we could adapt the organization example, along with the
\lstinline{expertise} query. Figure~\ref{fig:org-xml} shows the XML document
where we adapt the information from the original example. This document is valid
according to the schema that we have placed in \ref{app:XML:Organization}. As we
already showed, \lstinline{expertise} returns the name of the departments where
all employees are able to \emph{abstract}. Once again, we need to return a node
sequence since we could find many departments matching the criteria. Thereby,
the output that the query should produce might be:
\begin{lstlisting}[language=XML, basicstyle=\scriptsize\ttfamily]
<xml>
    <dpt>Quality</dpt>
    <dpt>Research</dpt>
</xml>
\end{lstlisting}

Producing such a query, analogous to \lstinline{expertise}, is by no means
straightforward since there is no standard XQuery method to check if all
elements that are hanging from a certain context do satisfy a predicate.
Fortunately, we can implement the desired behavior in terms of simpler
primitives, as we show in the following query:
\begin{lstlisting}[language=XML, basicstyle=\scriptsize\ttfamily]
/xml/department[not(employee[not(task[tsk = "abstract"])])]/dpt
\end{lstlisting}
The query produces the expected results, although it is difficult to read due
primarily to the combination of filters and negations. This query uses new
XQuery features, that we describe next:
\begin{itemize}
\item There are several invocations to the \xmlinline{not} function. This is
    just the negation function that we could find in many programming languages,
    but it adds extra functionality beyond negating booleans. Namely, it also
    produces \xmlinline{true} if the argument corresponds to a non-empty
    sequence of elements, and \xmlinline{false} if the argument corresponds to
    an empty one.
\item We find a new operator \xmlinline{=}, which corresponds to equality. The
    first operand in the equality is a tag, and thus its value is extracted to
    do the comparison. The second operand is a string literal.  Beyond strings,
    XQuery provides literals for other basic types, like numbers or booleans.
\end{itemize}

Finally, we would like to introduce a new element that we have deliberately
ignored so far since it was not present in the queries. It is the \emph{self
axis} and it refers to the current context. In XQuery, it is represented as a
\xmlinline{.} (dot). This self notion is neutral under nested access. For
example, \xmlinline{./couple/./fst/.} is equivalent to \xmlinline{couple/fst}.
We will need this self notion afterwards, in order to implement the non-standard
semantics.

\subsection{XQuery Non-standard Semantics}
\label{sub:xquery_non-standard}

We come back to our objective of turning Optica expressions into XQuery
expressions. For this task, we will use $\mathcal{E}^{xml}$ as the semantic
function that assigns well-typed Optica expression their denotations. Prior to
that, we need to choose $\mathcal{T}^{xml}$, which maps Optica types to semantic
domains for this infrastructure. Since we aim at generating XQuery expressions,
it seems reasonable to use $\mathit{XQuery}$ as the semantic domain for query
types. We also need to identify the semantic domain for optic types.  Although
this might sound contrived at this point, we adopt the very same semantic domain
as the one that we have embraced for queries. Therefore, we
define~$\mathcal{T}^{xml}$ as follows:
\par\nobreak\vspace{-0.2cm}
{\small
    \[\mathcal{T}^{xml}[t] = \mathit{XQuery}\]
}%
In fact, regardless of the input type, it will always evaluate to an XQuery
expression. Remark~\ref{rem:xquery} will shed some light on this decision. The
rest of the section revolves around the details of~$\mathcal{E}^{xml}$ and
discusses the results.

\subsubsection{Evaluating domain primitives} 

Before presenting the implementation of $\mathcal{E}^{xml}$, there are several
assumptions about the adaptation of the Optic models into XML schemas that need
to be made, where we basically adopt the same conventions that we have seen in
Sect.~\ref{sec:xmlxquery-background}. Firstly, we will assume that all
information is hanging from an \xmlinline{xml} element, which acts as the root
of the XML document. Secondly, we will assume that every optic corresponds to an
XML element, where the optic kind determines the cardinality. Finally, optics
that select base types are adapted as \emph{simple type} elements containing a
value with the corresponding base type; optics that select domain entity types
are adapted as elements with \emph{complex type}, since they nest other elements
in turn. Each of the previous conventions are assumed in the XSD schemas that
can be found in the appendix.

\begin{figure}
\small
\begin{center}
\begin{tabular}{lcl}
$\mathcal{E}^{xml}[\_]$&::&$\mathit{XQuery}$\\\\
$\mathcal{E}^{xml}[\mathit{couples}:\ \mathit{fold}\ \mathit{Couples}\ \mathit{Couple}]$&=&$\mathit{couple}$\\
$\mathcal{E}^{xml}[fst: \mathit{getter}\ \mathit{Couple}\ \mathit{Person}]$&=&$\mathit{fst}$\\
$\mathcal{E}^{xml}[snd: \mathit{getter}\ \mathit{Couple}\ \mathit{Person}]$&=&$\mathit{snd}$\\
$\mathcal{E}^{xml}[name: \mathit{getter}\ \mathit{Person}\ \mathbb{S}]$&=&$\mathit{name}$\\
$\mathcal{E}^{xml}[age: \mathit{getter}\ \mathit{Person}\ \mathbb{N}]$&=&$\mathit{age}$
\end{tabular}
\end{center}
\caption{XQuery non-standard semantics for couples extension}
\label{fig:couple-xquery-semantics}
\end{figure}

Now, we have all the ingredients that we need to provide the implementation of
the XQuery evaluator. Given its simplicity, we will start with the evaluation of
the extended terms for the couples example that we have collected in
Fig.~\ref{fig:couple-xquery-semantics}. As we have said in the previous
paragraph, optics correspond to XML elements, and thus we represent them as mere
element selection. Indeed, optic names are good candidates as tag names.
However, we need to adjust the plural names of folds into the singular, like in
$\mathit{couple}$, since this information is supplied as individual elements.
The evaluation for the organization model should be straightforward now and does
not add any value; therefore, we do not show it.

\subsubsection{Evaluating core primitives} 

\begin{figure}
\small
\begin{align*}
&\mathcal{E}^{xml}[\_]&::&\quad \mathit{XQuery} \\\\
&\mathcal{E}^{xml}[\mathit{id}_{\mathit{gt}}:\ \mathit{getter}\ \alpha\ \alpha]&=& \quad . \\
&\mathcal{E}^{xml}[g\ \ggg_{\mathit{gt}} h: \mathit{getter}\ \alpha\ \gamma]&=& \quad
\mathcal{E}^{xml}[g: \mathit{getter}\ \alpha\ \beta]/\mathcal{E}^{xml}[h:
\mathit{getter}\ \beta\ \gamma] \\
&\mathcal{E}^{xml}[g\ \Fork\ h:\ \mathit{getter}\ \alpha\ (\beta,\ \gamma)]&=& \quad 
<\mathit{tuple}>\\
&&&\qquad<\mathit{fst}>\mathcal{E}^{xml}[g: \mathit{getter}\ \alpha\
\beta]</\mathit{fst}>\\
&&&\qquad<\mathit{snd}>\mathcal{E}^{xml}[h: \mathit{getter}\ \alpha\
\gamma]</\mathit{snd}>\\
&&&\quad</\mathit{tuple}>\\
&\mathcal{E}^{xml}[\mathit{like}\ b:\ \mathit{getter}\ \alpha\ \beta]&=& \quad
\_b\_ \\
&\mathcal{E}^{xml}[\mathit{not}\ g:\ \mathit{getter}\ \alpha\ \mathbb{B}]&=&
\quad \mathbf{not}(\mathcal{E}^{xml}[g:\ \mathit{getter}\ \alpha\ \mathbb{B}]) \\
&\mathcal{E}^{xml}[g \oplus h:\ \mathit{getter}\ \alpha\ \delta]&=& \quad 
(\mathcal{E}^{xml}[g:\ \mathit{getter}\ \alpha\ \beta] \oplus
\mathcal{E}^{xml}[h:\ \mathit{getter}\ \alpha\ \gamma]) \\\\
&\mathcal{E}^{xml}[\mathit{id}_{\mathit{af}}:\ \mathit{affine}\ \alpha\ \alpha]&=& \quad
. \\
&\mathcal{E}^{xml}[g\ \ggg_{\mathit{af}} h: \mathit{affine}\ \alpha\ \gamma]&=& \quad 
\mathcal{E}^{xml}[g: \mathit{affine}\ \alpha\ \beta]/\mathcal{E}^{xml}[h:
\mathit{affine}\ \beta\ \gamma] \\
&\mathcal{E}^{xml}[\mathit{filtered}\ p:\ \mathit{affine}\ \alpha\ \alpha]&=&
\quad .\Big{[}\mathcal{E}^{xml}[p:\ \mathit{affine}\ \alpha\ \mathbb{B}]\Big{]} \\
&\mathcal{E}^{xml}[\mathit{to}_{\mathit{af}}\ g:\ \mathit{affine}\ \alpha\ \beta]&=& \quad
\mathcal{E}^{xml}[g: \mathit{getter}\ \alpha\ \beta] \\\\
&\mathcal{E}^{xml}[\mathit{id}_{\mathit{fl}}:\ \mathit{fold}\ \alpha\ \alpha]&=& \quad . \\
&\mathcal{E}^{xml}[g\ \ggg_{\mathit{fl}} h: \mathit{fold}\ \alpha\ \gamma]&=& \quad 
\mathcal{E}^{xml}[g: \mathit{fold}\ \alpha\ \beta]/\mathcal{E}^{xml}[h:
\mathit{fold}\ \beta\ \gamma] \\
&\mathcal{E}^{xml}[\mathit{nonEmpty}\ g:\ \mathit{getter}\ \alpha\
\mathbb{B}]&=& \quad \mathbf{exists}(\mathcal{E}^{xml}[g:\ \mathit{fold}\
\alpha\ \beta]) \\
&\mathcal{E}^{xml}[\mathit{to}_{\mathit{fl}}\ a:\ \mathit{fold}\ \alpha\ \beta]&=& \quad
\mathcal{E}^{xml}[a: \mathit{affine}\ \alpha\ \beta] \\\\
&\mathcal{E}^{xml}[\mathit{get}\ g:\ \alpha \rightarrow \beta] &=& \quad
/\mathit{xml}/\mathcal{E}^{xml}[g:\ \mathit{getter}\ \alpha\ \beta] \\
&\mathcal{E}^{xml}[\mathit{preview}\ g:\ \alpha \rightarrow option\ \beta] &=& \quad 
/\mathit{xml}/\mathcal{E}^{xml}[g:\ \mathit{affine}\ \alpha\ \beta] \\
&\mathcal{E}^{xml}[\mathit{getAll}\ g:\ \alpha \rightarrow list\ \beta]
&=& \quad /\mathit{xml}/\mathcal{E}^{xml}[g:\ \mathit{fold}\ \alpha\ \beta]
\end{align*}

\caption{XQuery non-standard semantics}
\label{fig:xquery-non-standard-semantics}
\end{figure}

The evaluations for the core combinators are collected in
Fig.~\ref{fig:xquery-non-standard-semantics}. We start with the combinators for
getters. Firstly, $\ggg_{\mathit{gt}}$ is translated as nested access, where the
evaluations of the composing expressions $g$ and $h$ are tied together. For
$\Fork$ we use the XML interpolation, where the evaluation of the composing
expressions is placed in the corresponding projection elements. Finally,
$\mathit{id}_{\mathit{gt}}$ is interpreted as a self reference ($.$), which is
neutral under composition. Now, we move on to standard getter constructions,
beginning with $\mathit{like}$. Since it produces constant optics, whose part
does not depend on the surrounding whole, we decide to map them to XQuery
literals\footnote{The notation $\_b\_$ just indicates the adaptation of $b$ into
an XQuery literal.}. Next, we can see that $\mathit{not}$ is interpreted as the
function \xmlinline{not}, and binary combinators, which are unified by the
symbol $\oplus$, are interpreted as the corresponding XQuery operations.

Moving on to affine folds, we find that the composition and identity primitives
share the same implementation as the ones we have seen for getters. This
situation ---which also occurs in fold combinators--- demonstrates that we do
not make a difference between semantic domains in the interpretation. In fact,
if we understand $\mathit{XQuery}$ as a representation of an affine fold, it is
natural that we can also use it as a representation of a getter, and the
implementation of $\mathit{to}_{\mathit{af}}$ confirms this intuition. This
module also contains $\mathit{filtered}$. Since we have a filtering mechanism
available in XQuery, we simply interpret this primitive into square brackets
($[]$), passing the semantics of the predicate getter as an argument to it.

Finally, we present the fold related method $\mathit{nonEmpty}$. In this
particular case, we need to adapt any fold into a getter that selects a boolean.
Luckily, XQuery provides a function \xmlinline{exists} which turns XQuery
expressions into booleans. It does it by checking that the result produced by
the query is not empty. It might have been noticed that \xmlinline{exists} was
not even mentioned in the background section. In fact, it was not necessary
since the \xmlinline{not} function does the trick by turning an expression
denoting a sequence into a boolean. In particular, \xmlinline{not(exists(sq))}
(where \xmlinline{sq} denotes a sequence of elements) is equivalent to
\xmlinline{not(sq)}. However, while evaluating, we do not know whether the
\xmlinline{exists} invocation denoted by $\mathit{nonEmpty}$ will be consumed by
a function like \xmlinline{not} (denoted by another expression), and thus we
need to invoke \xmlinline{exists} explicitly\footnote{There are different
implementation techniques in the literature that we could use to optimize this
kind of situation, but we ignore them for brevity.}.

As final notes, we must say that interpreting optic expressions like
\lstinline{differencesFl} (Def.~\ref{def:differences}) or
\lstinline{expertiseFl} (Def.~\ref{def:expertise}) leads to relative queries,
i.e.\ queries that do not start with \xmlinline{/} and which are relative to the
current context. Those queries are valid XQuery expressions, but they will not
produce any results if we run them against the XML document which contains the
whole hierarchy. Fortunately, we could easily compose such relative
queries with the ones generated by external models to produce queries over more
complex domains. Leaving this possibility aside, the next section shows the
final refinement that we need to perform in order to obtain the expected XQuery
expressions.

\subsubsection{Target queries and results} 

The evaluation of query expressions from Optica can be found at the bottom of
Fig.~\ref{fig:xquery-non-standard-semantics}. Since both optic and query types
denote an XQuery expression, the semantics of query expressions is almost
direct.  The only caveat is that $\mathit{get}$, $\mathit{preview}$ and
$\mathit{getAll}$ prepend the \xmlinline{/xml} fragment to the relative query
obtained in the optic representation.  This is just a consequence of one of the
assumptions that we made when adopting XML, where we have stated that an
\xmlinline{<xml/>} root element was necessary by convention. Thereby, we take
the opportunity to prepend it here.

At this point, with the required evaluations at hand, we should be able to
recover the target queries. As a result,
$\mathcal{E}^{xml}[\mathit{differences}]$ provides the following XQuery
expression:
\begin{lstlisting}[language=XML]
/xml/couple[fst/age > snd/age]/<tuple>
                                   <one>{fst/name}</one>
                                   <two>{fst/age - snd/age}</two>
                               </tuple>
\end{lstlisting}
The resulting query is exactly the same as the one that we have introduced in
Sect.~\ref{sec:xmlxquery-background}. We also supply the output provided by
$\mathcal{E}^{xml}[\mathit{expertise}]$:
\begin{lstlisting}[language=XML]
/xml/department[not(exists(employee[not(exists(task/tsk[. = "abstract"]))]))]/dpt
\end{lstlisting}
We can see a self-reference (dot) when comparing the task with the literal
`$\mathit{abstract}$', which is a consequence of the particular implementation
of $\mathit{elem}$ that we have presented in Sect.~\ref{sub:Syntax} (we use the
identity getter to refer to the predicate parameter while invoking
$\mathit{any}$). Besides, we find redundant invocations to
\xmlinline{exists}\footnote{Once again, these invocations could be removed from
    the resulting query by means of annotations, as in~\cite{suzuki2016finally},
    but we wanted to keep the interpretation compositional in order to make it
simpler.}. If we ignore these minor differences, the query is essentially the
same as the one that we presented at the beginning of this section, and
therefore it produces the very same output. The accompanying implementation in
Scala of the XQuery interpreter contains tests that show the right behavior of
these queries, where we use
\emph{BaseX}\footnote{\url{http://basex.org/basex/xquery/}} as the XQuery engine
to access XML documents.

\begin{remark}
    \label{rem:xquery}
    As we have seen throughout this section, both optic and query types are
    evaluated into the same semantic domain $\mathit{XQuery}$. Indeed, if we
    leave interpolation facilities aside, this is essentially an interpretation
    into XPath, which is just a language to select parts from an XML document,
    just like optics select parts from immutable data structures. In this sense,
    it is only natural that XQuery can behave as a non-standard optic
    representation. 

\end{remark}

\section{SQL}
\label{sec:SQL}

SQL is a query language for relational data sources which greatly differs from
the hierarchical nature of both XML and optic models. Nevertheless, this section will
show that we can generate SQL statements from Optica expressions. Firstly, we
manually adapt the couple and organization examples into the SQL setting to
better understand the kind of queries that we want to produce. Then, we will
present the SQL non-standard semantics of Optica and the assumptions that we build
upon in order to automatically generate analogous queries to the ones that we have
obtained manually.

\subsection{SQL Background}
\label{sub:SQL:Background}

As opposed to XML, relational databases are organized around flat data sources.
As a consequence, we face the object-relational impedance
mismatch~\cite{ireland2009classification} when trying to accommodate the object
models underlying optics into the relational setting. Fortunately, there are
patterns that we can embrace to approach this task, like the \emph{Foreign Key
Aggregation} or the \emph{Foreign Key Association}
patterns~\cite{keller1997mapping}. We take them as a reference and propose the
following tables to adapt the couples example model that we introduced
in~\ref{subsub:couple-example}:
\begin{lstlisting}[columns=fixed,language=SQL]
CREATE TABLE Person (
  name varchar(255) PRIMARY KEY,
  age int NOT NULL
);

CREATE TABLE Couple (
  fst varchar(255) NOT NULL,
  snd varchar(255) NOT NULL,
  FOREIGN KEY (fst) REFERENCES Person(name),
  FOREIGN KEY (snd) REFERENCES Person(name)
);
\end{lstlisting}
As can be seen, case classes are adapted as tables and their attributes are
adapted as columns. Once again, as we have seen in the XQuery interpretation, it
is necessary to distinguish between attributes which contain base types and
attributes containing other entities. In fact, attributes that refer to entities
require pointers to establish the precise connections between the adapted
tables, following the Foreign Key Aggregation pattern. We assume
figures~\ref{subfig:Person} and ~\ref{subfig:Couple} as the initial state for
these tables, where the columns in \sqlinline{Couple} are clearly selecting
names from \sqlinline{Person}.

\begin{figure}
  \centering
  \begin{subfigure}[b]{0.15\linewidth}
      \footnotesize
      \begin{tabular}{ll}
          \hline
          \textbf{name} & \textbf{age} \\ \hline
          Alex          & 60           \\
          Bert          & 55           \\
          Cora          & 33           \\
          Demi          & 31           \\
          Eric          & 21           \\
          Fred          & 60           \\ \hline
      \end{tabular}
      \caption{\sqlinline{Person}}
      \label{subfig:Person}
  \end{subfigure}\hspace{0.1\textwidth} 
  \begin{subfigure}[b]{0.15\linewidth}
      \footnotesize
      \begin{tabular}{ll}
          \hline
          \textbf{fst} & \textbf{snd} \\ \hline
          Alex         & Bert         \\
          Cora         & Demi         \\
          Eric         & Fred         \\ \hline
      \end{tabular}
      \caption{\sqlinline{Couple}}
      \label{subfig:Couple}
  \end{subfigure}\hspace{0.1\textwidth} 
  \begin{subfigure}[b]{0.16\linewidth}
      \footnotesize
      \begin{tabular}{ll}
          \hline
          Alex         & 5             \\
          Cora         & 2             \\ \hline
      \end{tabular}
      \caption{\sqlinline{Differences}}
      \label{subfig:Differences}
  \end{subfigure}
  \caption{Data for the couples example.}
  \label{fig:couple-tables}
\end{figure}

Previously, we have seen that the adaptation of \lstinline{differences} in the
XML setting produced XML as output. We are now dealing with SQL tables, where
the output of a statement is a table itself. Thereby, we would expect
Fig.~\ref{subfig:Differences} as the result of executing the adaptation of
\lstinline{differences}. In particular, we could produce such output with the
following query:
\begin{lstlisting}[columns=fixed,language=SQL]
SELECT w.name, w.age - m.age
FROM Couple c INNER JOIN Person w ON c.fst = w.name
              INNER JOIN Person m ON c.snd = m.name
WHERE w.age > m.age;
\end{lstlisting}
This statement is clearly separated in three major sections. First, we describe
\sqlinline{FROM}, which builds the raw table that the other parts use to gather
information from. This table is created by joining the table \sqlinline{Couple}
with two occurrences of table \sqlinline{Person}, thereby incorporating the
information from the couple members \sqlinline{fst} and \sqlinline{snd}.
Variables \sqlinline{c}, \sqlinline{w} and \sqlinline{m} allow us to refer to
these three tables. Second, the \sqlinline{WHERE} clause introduces filters that
are applied over the compound table to discard the rows that do not match the
criteria: those where the age of the first member is not greater than the age of
the second one. Last, the \sqlinline{SELECT} clause indicates the columns that
we are interested in: the name of the first member and the age difference.

Now, we move on to the organization example. First of all, we create tables for
departments, employees and tasks following the same adaptation pattern:
\begin{lstlisting}[language=SQL]
CREATE TABLE Department (
  dpt varchar(255) PRIMARY KEY
);

CREATE TABLE Employee (
  emp varchar(255) PRIMARY KEY,
  dpt varchar(255) NOT NULL,
  FOREIGN KEY (dpt) REFERENCES Department(dpt)
);

CREATE TABLE Task (
  tsk varchar(255) NOT NULL,
  emp varchar(255) NOT NULL,
  FOREIGN KEY (emp) REFERENCES Employee(emp)
);
\end{lstlisting}
All components in the previous statements should be familiar at this point, but
there is an important change in the way we configure foreign keys. As we have seen
in the couples example, getters selecting entities were mapped into a column
containing a foreign key. However, the organization example contains multivalued
attributes, like \lstinline{employees} or \lstinline{tasks}, that should not be
adapted as a single column. For this situation we adopt the Foreign Key
Association pattern. We assume that these tables have been populated with the
data in figures~\ref{subfig:Department}, \ref{subfig:Employee} and
\ref{subfig:Task}.

\begin{figure}
    \centering
    \begin{subfigure}[b]{0.15\linewidth}
        \footnotesize
        \begin{tabular}{l}
            \hline
            \textbf{dpt} \\ \hline
            Product      \\
            Quality      \\
            Research     \\
            Sales        \\ \hline
        \end{tabular}
        \caption{\sqlinline{Department}}
        \label{subfig:Department}
    \end{subfigure}\hspace{0.09\textwidth} 
    \begin{subfigure}[b]{0.2\linewidth}
        \footnotesize
        \begin{tabular}{ll}
            \hline
            \textbf{emp} & \textbf{dpt} \\ \hline
            Alex         & Product      \\
            Bert         & Product      \\
            Cora         & Research     \\
            Demi         & Research     \\
            Eric         & Research     \\
            Fred         & Sales        \\ \hline
        \end{tabular}
        \caption{\sqlinline{Employee}}
        \label{subfig:Employee}
    \end{subfigure}\hspace{0.09\textwidth} 
    \begin{subfigure}[b]{0.2\linewidth}
        \footnotesize
        \begin{tabular}{ll}
            \hline
            \textbf{tsk} & \textbf{emp} \\ \hline
            build        & Alex         \\
            build        & Bert         \\
            abstract     & Cora         \\
            build        & Cora         \\
            design       & Cora         \\
            abstract     & Demi         \\
            design       & Demi         \\
            abstract     & Eric         \\
            call         & Eric         \\
            design       & Eric         \\
            call         & Fred         \\ \hline
        \end{tabular}
        \caption{\sqlinline{Task}}
        \label{subfig:Task}
    \end{subfigure}\hspace{0.09\textwidth} 
    \begin{subfigure}[b]{0.15\linewidth}
        \footnotesize
        \begin{tabular}{ll}
            \hline
            Quality      \\
            Research     \\ \hline
        \end{tabular}
        \caption{\sqlinline{Expertise}}
        \label{subfig:Expertise}
    \end{subfigure}
    \caption{Data for the organization example.}
    \label{fig:org-tables}
\end{figure}

As we have seen before, \emph{Quality} and \emph{Research} are the departments where
all employees are able to abstract; therefore, the adaptation of
\lstinline{expertise} should produce Fig.~\ref{subfig:Expertise} as a
result. We propose the following query to generate it:
\begin{lstlisting}[language=SQL]
SELECT d.dpt 
FROM Department AS d 
WHERE NOT(EXISTS(SELECT e.* 
                 FROM Employee AS e 
                 WHERE NOT(EXISTS(SELECT t.* 
                                  FROM Task AS t 
                                  WHERE (t.tsk = "abstract") AND (e.emp = t.emp))) 
                       AND (d.dpt = e.dpt)));
\end{lstlisting}
Reading this query is by no means trivial. Fortunately, it shares the same
pattern as the query that we have seen while adapting \lstinline{expertise} in the XQuery
setting.  In fact, \sqlinline{EXISTS} is a function that returns true as long
as the nested statement produces non-empty results. If we combine it with
\sqlinline{NOT} to negate predicates, we can check if all rows satisfy a
condition. Beyond the noise generated by this pattern, there are additional
filters which manifest relations between nested and outer variables that
introduce even more complexity in the picture. 

\subsection{SQL Non-standard Semantics}

Peculiarities of SQL have been known for a long time
now~\cite{date1984critique}. As Date states in connection with certain aspects
of SQL, ``there is so much confusion in this area that it is difficult to
criticize it coherently''. Part of the problem resides in that the formal
definition of SQL was produced \emph{after the fact}, where many academic
considerations were neglected. Consequently, the language does have its weak
points, where the lack of orthogonality becomes a central issue. Although many
deficiencies have been remedied in the last decades, the obtrusive syntax of the
\sqlinline{SELECT} statement remains a problem. For example, despite the fact
that relational algebra combinators may appear in any order, the rigid structure
of \sqlinline{SELECT} statements might demand the programmer to recast a
relational algebra expression that is considered natural (like \sqlinline{UNION
(tabexp1, tabexp2)}) into a semantically equivalent form, compliant with the SQL
standard (like \sqlinline{(SELECT ... FROM ... WHERE ...) UNION (SELECT ... FROM
... WHERE ...)}). Fortunately, \cite{DBLP:conf/dbpl/Cooper09} supplies a list of
syntactic rules which we can use to rewrite any expression from an ordinary
impure functional programming language into its SQL form. Optica expressions
share with relational algebra the purely compositional character of algebraic
expressions; hence, they also require a set of transformations before being able
to be translated to SQL queries. These transformations will not be carried out
on the optic expression itself, but through a new semantic domain which plays
the role of an intermediate expression that can be directly translated to SQL. 

\begin{figure}[hbt]
\begin{center}
\begin{tabular}{lcl}
$\mathcal{T}^{sql}[\mathit{getter}\ \alpha\ \beta]$ &=& $\mathit{Triplet} \rightarrow \mathit{Triplet}$ \\
$\mathcal{T}^{sql}[\mathit{affine}\ \alpha\ \beta]$ &=& $\mathit{Triplet} \rightarrow \mathit{Triplet}$ \\
$\mathcal{T}^{sql}[\mathit{fold}\ \alpha\ \beta]$ &=& $\mathit{Triplet} \rightarrow \mathit{Triplet}$ \\
$\mathcal{T}^{sql}[\alpha \rightarrow \mathit{list}\ \beta]$ &=& 
$(\mathbb{S} \rightarrow \mathbb{S}) \rightarrow \mathit{SQL}$ \\
$\mathcal{T}^{sql}[\mathbb{N}]$ &=& $\mathit{Fragment}$ \\
$\mathcal{T}^{sql}[\mathbb{B}]$ &=& $\mathit{Fragment}$ \\
$\mathcal{T}^{sql}[\mathbb{S}]$ &=& $\mathit{Fragment}$
\end{tabular}
\end{center}
\caption{SQL semantic domains}
\label{fig:sql-semantic-domains}
\end{figure}

Accordingly, the new semantic domains defined by the semantic function
$\mathcal{T}^{sql}$ are shown in Fig.~\ref{fig:sql-semantic-domains}. Firstly,
all optic types are mapped to a $\mathit{Triplet}$ endofunction. Triplets are
the intermediate expressions which lie between optic and SQL expressions, whose
major purpose is to reconcile the main disagreements among them. Secondly, since
we aim at generating SQL statements, the semantic domain associated to query
types is, as expected, an SQL expression. However, it is required to supply a
function ($\mathbb{S} \rightarrow \mathbb{S}$) that maps relational table names
to the column name which corresponds to the primary key ---information that is
not contemplated by the optic model--- in order to produce SQL statements. It is
important to remark that the types of the queries $\mathit{get}$ and
$\mathit{preview}$ are ignored here. Later on we will explain why this
partiality is needed. Finally, base types are mapped to \emph{triplet
fragments}, i.e. their evaluation will be used to form triplets.

For the rest of the section we will proceed as usual, introducing the semantic
function $\mathcal{E}^{sql}$, which is responsible for evaluating domain, optic
and query terms, and we will conclude discussing the results. Prior to that,
we find it essential to describe the details about the intermediate
structure $\mathit{Triplet}$.

\subsubsection{$\mathit{Triplet}$: motivation and details}

As we have just seen, a SQL select statement exhibits a remarkable separation of
concerns, where selection and filtering, although sharing syntax, belong to
different query clauses. This separation requires a unifying mechanism to refer
to the very same item from both clauses. SQL solves this problem by means of
variables declared in the \sqlinline{FROM} clause, which are accessible from the
\sqlinline{SELECT} and \sqlinline{WHERE} scopes.

This way of representing queries in SQL contrasts with its optic counterpart. In
Optica, the aspects of selection and filtering may appear anywhere within the
expression. Moreover, the information required by these components does not need
to be collected in a single \sqlinline{FROM} component, but specified on demand.
In optic expressions, there is no need for variables either, since it is the
context where two optics appear that determines whether they are selecting the
same item or not. For example, consider the following optic
expression\footnote{This query is only a less direct way of implementing query
  \lstinline{under50} from Sect.~\ref{sec:Introduction}.}, where we find two
occurrences of $\mathit{fst}$: 
\par\nobreak\vspace{-0.2cm}
{\small
\[
\mathit{couples} \ggg \mathit{filtered}\ (\mathit{fst} \ggg \mathit{age} <
50) \ggg \mathit{fst} \ggg \mathit{name}
\]
}%
Despite having one of them surrounded by $\mathit{filtered}$, we can see that
both of them are selecting the very same person. Furthermore, note that the
information required by the filtering expression (the age of the first member)
is collected within the predicate scope and not shared globally. 

%

\begin{figure}
\centering
\includegraphics[width=\linewidth]{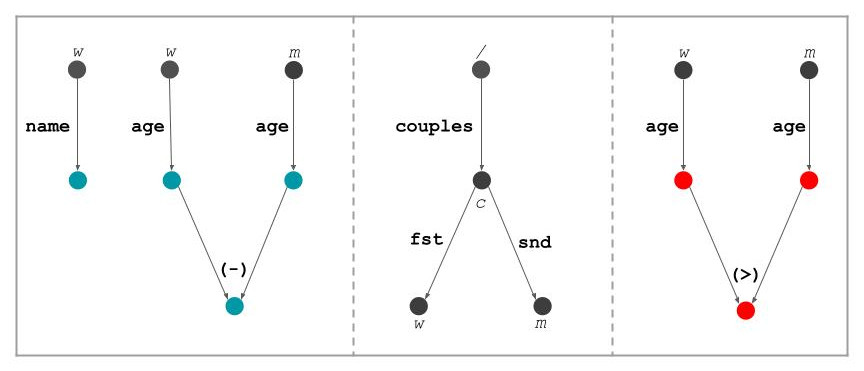}
\caption{Triplet generated for $\mathit{differencesFl}$.}
\label{fig:decoupling}
\end{figure}

$\mathit{Triplet}$ is the data structure that we use as an intermediary to
alleviate the aforementioned disagreements. Its main objective is to segregate
the three different aspects, which are evident in a \sqlinline{SELECT}
statement, from an Optica expression. In particular, a triplet is made of three
components which correspond to the \sqlinline{SELECT}, \sqlinline{FROM} and
\sqlinline{WHERE} clauses, respectively. We present an informal view of this
concept in Fig.~\ref{fig:decoupling}, where we represent the triplet associated
to the expression $\mathit{differencesFl}$ (Def.~\ref{def:differences}). A
triplet may be considered as a \emph{structured} optic whose actual focus is
determined through three components:
\begin{itemize}

\item The middle component determines the {\em potential} focus of the optic. In
    particular, Fig.~\ref{fig:decoupling} shows this component as a
    \emph{trie}\footnote{\url{https://en.wikipedia.org/wiki/Trie}} whose edges
    are optics focusing on entity types (not base types). Its elements are
    sequences of optics that represent a vertical composition, e.g. the sequence
    made of the primitive fold \lstinline{couples} and getter \lstinline{fst}
    represents the fold \lstinline{couples >>> fst}. The figure labels each node
    with a distinct name that refers to the entities of that unique path. In
    this example we potentially refer to the list of couples ($c$) and two lists
    of people: its first ($w$) and second ($m$) members. The nodes of the trie,
    colored in black, and its associated names can be reused in the left and
    right components.

\item The right component further {\em constrains} the potential collections of
    entities identified by the entity trie, by imposing conditions over them. In
    the example, there is just one condition that restricts the collection of
    couples (and, consequently, its dependent collections of people) to those
    where her age ($w$) is greater than his ($m$). These conditions are
    represented in terms of directed graphs whose edges are optics or binary
    combinators, like $>$, which make two different paths converge. Note that
    red nodes form restriction graphs.

\item Last, the left component defines the \emph{actual} selection of the
    overall optic by selecting in sequence certain collections from the entity
    trie, and possibly by further refining them through additional optic
    expressions selecting base values. In the example, we select her name and
    the age difference of the couple (which will be greater than 0) according to
    the constraints which were imposed by the right component. Selections are
    represented using the same graphs as in the constrain component, but nodes
    forming them are colored in blue.

\end{itemize}

\begin{figure}
\small
\begin{center}
\begin{tabular}{lcl}
$t$ &::=& $(s,\ f,\ w)$\\
$s$ &::=& $(e,\ e,\ \ldots,\ e)$\\
$f$ &::=& $/ \mid \mathit{insert}\ \hat{p}\ f$\\
$w$ &::=& $\{e,\ e,\ \ldots,\ e\}$\\
$e$ &::=& $\mathit{like}\ c \mid \mathit{not}\ e \mid e > e \mid e == e \mid e - e \mid \hat{p} \mid \hat{p}.\mathit{optic} \mid \mathit{nonEmpty}\ t$\\
$\hat{p}$ &::=& $(\mathit{optic},\ \mathit{optic},\ \ldots,\ \mathit{optic})$
\end{tabular}
\end{center}
\caption{Triplet syntax}
\label{fig:triplet-syntax}
\end{figure}

We formalize the notion of triplet in Fig.~\ref{fig:triplet-syntax} through its
associated syntax. As we have pointed out previously, the middle component is
just a trie whose keys are primitive optic expressions focusing on entities.
Thus, the elements stored in the trie are sequences of such expressions, which
we will refer to as \emph{paths} ($\hat{p}$)\footnote{We will use \emph{hats},
as in $\hat{p}$, to emphasize the terms which correspond to paths.}.  Entity
tries may be the empty trie, $/$, or the result of inserting a new path,
$insert\ \hat{p}\ f$. The left and right components, $s$ and $w$, are a sequence
and a set of expressions ($e$), respectively. Repeated restrictions in $w$ are
redundant and their ordering is irrelevant ---that is why a set is chosen.
Expressions $e$ are very similar to those from Optica ($\mathit{like}$,
$\mathit{not}$, $\mathit{>}$, etc.), but there are a few major changes that
deserve further explanation. Essentially, expressions do not include vertical
composition as such; instead, if the vertical composition selects an entity, it
is simply represented through a path from the entity trie. Otherwise, if it
selects a base type, it is represented as the projection of an attribute from a
path, as in $\mathit{\hat{p}.optic}$. For instance, the Optica expression
\lstinline{couples >>> fst} would denote the path $(\mathit{couples},\
\mathit{fst})$, while the expression \lstinline{couples >>> fst >>> name} would
denote the projection $(\mathit{couples},\ \mathit{fst}).\mathit{name}$.
Horizontal composition is also unneeded since the left component is able to
collect a sequence of single selections. Finally, expressions also contain a
$\mathit{nonEmpty}$ term, where we keep a snapshot of a triplet that is later
used to produce nested queries.  Section~\ref{sub:sql-actions-queries}, where we
formalize the precise correspondence between triplets and SQL, will show that a
$\mathit{nonEmpty}$ term is eventually translated into an \sqlinline{EXISTS}
operator.

\begin{figure}
  \vspace*{-4cm}
  \hspace*{-2.5cm}
  \centering
  \includegraphics[width=1.4\linewidth]{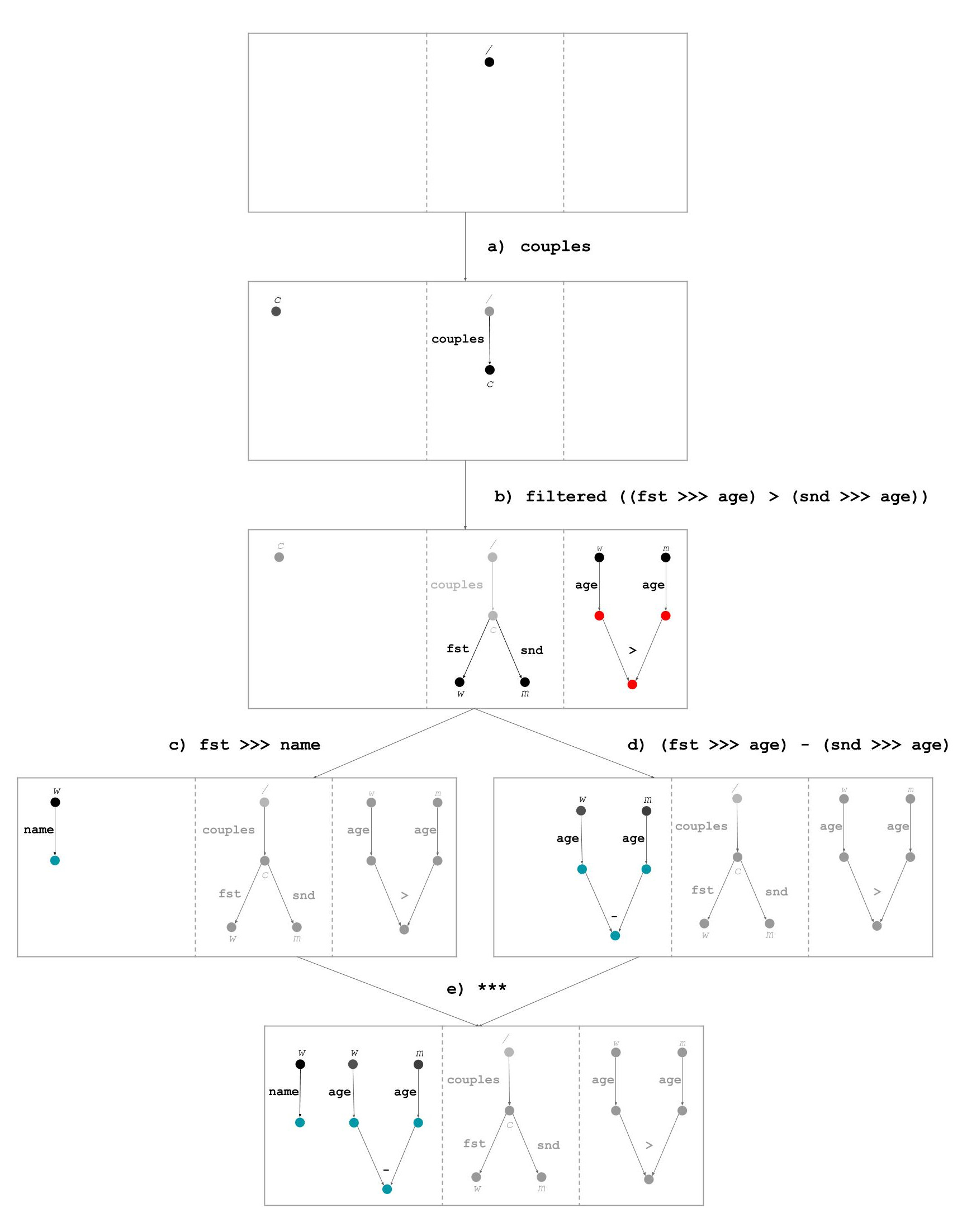}
  \caption{Triplet evolution for $\mathit{differencesFl}$}
  \label{fig:evolution}
  \thisfloatpagestyle{empty}
\end{figure}

At this point, we might consider using $\mathit{Triplet}$ as the chosen optic
representation. However, composing the different triplets generated by optic
subexpressions turns out to be a clumsy task. Instead, we would like to use a
representation with better compositional guarantees. In this sense, it is more
convenient to use a triplet endofunction so that each subexpression can describe
the precise transformation that it performs over the input triplet when it is
composed through vertical composition\footnote{This is reminiscent of the
    functional representation of difference lists, where concatenation is
    implemented in terms of plain composition, and the list is recovered by
    passing the empty list as input. In this case, the analogues of the empty
    list and concatenation are the empty triplet (Def.~\ref{def:empty-triplet})
and vertical composition.}. This is how we obtain $\mathit{Triplet} \rightarrow
\mathit{Triplet}$, the chosen semantic domain for optic types. We illustrate the
idea behind this function in Fig.~\ref{fig:evolution}, which shows the evolution
of the resulting triplet for the $\mathit{differencesFl}$ query, starting from
the empty triplet. The arcs in this figure are labelled by the optic
subexpressions that identify the applied transformations. As expected, the last
triplet in the chain corresponds to the structure that we presented in
Fig.~\ref{fig:decoupling}. We will detail these steps throughout the next
sections while presenting the $\mathcal{E}^{sql}$ definition.

\subsubsection{Evaluating domain primitives}

\begin{figure}
\small
\begin{center}
\begin{tabular}{lcl}
$\mathcal{E}^{sql}[\_\ : \mathit{op}\ \alpha\ \beta \quad \text{where}\ \mathit{op} \in \{\mathit{getter}, \mathit{affine}, \mathit{fold}\}]$ &::& $\mathit{Triplet}\ \rightarrow \mathit{Triplet}$\\\\
$\mathcal{E}^{sql}[\mathit{couples}:\ \mathit{fold}\ \mathit{Couples}\ \mathit{Couple}]$ &=& $\mathit{entity}\ \mathit{couples}$\\
$\mathcal{E}^{sql}[fst: \mathit{getter}\ \mathit{Couple}\ \mathit{Person}]$&=&$\mathit{entity}\ \mathit{fst}$\\
$\mathcal{E}^{sql}[snd: \mathit{getter}\ \mathit{Couple}\ \mathit{Person}]$&=&$\mathit{entity}\ \mathit{snd}$\\
$\mathcal{E}^{sql}[name: \mathit{getter}\ \mathit{Person}\ \mathbb{S}]$ &=& $\mathit{base}\ \mathit{name}$\\
$\mathcal{E}^{sql}[age: \mathit{getter}\ \mathit{Person}\ \mathbb{N}]$ &=& $\mathit{base}\ \mathit{age}$\\\\
$\mathit{base}\ b$ &=& $((\hat{x}),\ f,\ w) \mapsto ((\hat{x}.b),\ f,\ w) \quad \text{where}$ \\
&& $\hat{x}\ \text{is an element of}\ f$\\
$\mathit{entity}\ \mathit{e}$ &=& $((\hat{x}),\ f,\ w) \mapsto ((\hat{y}),\ \mathit{f2},\ w) \quad \text{where}$ \\
&& $\hat{y} = \hat{x} \smallfrown (\mathit{e}) \quad \text{and}$ \\
&& $\mathit{f2} = \mathit{insert}\ \hat{y}\ f \quad \text{and}$ \\
&& $\hat{x}\ \text{is an element of}\ f$
\end{tabular}
\end{center}

\caption{Triplet non-standard semantics for couples extension}
\label{fig:domain-triplet-semantics}
\end{figure}

This section provides the semantics of primitive optics from the domain syntax,
such as $\mathit{couples}$, $\mathit{fst}$, etc., in terms of the precise
transformations that they carry out over an input $\mathit{Triplet}$. Their
formalization can be found in Fig.~\ref{fig:domain-triplet-semantics}, using the
couples domain for illustration purposes. Note that $\smallfrown$ represents the
concatenation of sequences. Before explaining this formalization, we will
describe the occurrences of domain primitives in the particular example shown in
Fig.~\ref{fig:evolution} as well as in Fig.~\ref{fig:filtered}, where
Step~\emph{b)} is shown in detail.

\begin{itemize}

    \item Step~\emph{a)} shows the changes introduced by the term
        $\mathit{couples}$. This is a very special case since it takes the
        initial triplet as input. As can be seen, the new changes consist of
        introducing the new path in the trie and selecting it in the left
        component. Bear in mind that we can only introduce optics selecting domain
        entities in the trie, like $\mathit{couples}$, that selects a sequence
        of $\mathit{Couple}$ entities.

      \item Step~\emph{b)} contains more domain terms in the predicate. In
        particular, Step~\emph{b1)} (Fig.~\ref{fig:filtered}) shows the changes
        introduced by $\mathit{fst}$ when it is applied to a triplet that
        focuses on couples. Since this optic focuses on entities and the input
        triplet is not empty, the result is a triplet that extends the entity
        trie by appending the new optic to the couples path, changing its
        focus (i.e. the left part of the triplet) to the new path $w$.

      \item Step~\emph{b3)} represents the changes introduced by $\mathit{age}$.
        In this case, we deal with an optic selecting a base type $\mathbb{N}$.
        Thereby, it cannot be introduced in the trie. Instead, we refine the
        focus of the input triplet which becomes a projection to the new optic.

\end{itemize} 

Fortunately, the behaviour of the first and second cases can be factorized, as
long as we contemplate the following definition for the empty triplet:
\begin{definition}
    \label{def:empty-triplet}
    We formalize the \emph{empty triplet} as the one that contains a single
    selection $\hat{()}$, an empty trie and an empty set of restrictions.
    \[\mathit{empty} = ((\hat{()}), /, \varnothing) \] The empty sequence $()$
    is used in tries to refer to its root, and thereby, we use it as the initial
    path in the left component.
\end{definition}

However, our formalization must take into account the distinction between optics
selecting entities and optics selecting base types.
Figure~\ref{fig:domain-triplet-semantics} introduces functions $\mathit{base}$
and $\mathit{entity}$ for this task, where $\mapsto$ just represents the
standard ``maps to'' notation from functions. They take the optic expression as
parameter and they produce triplet endofunctions as result. The rest of the
implementation should be straightforward, given the previous explanations. We do
not show the evaluation of the organization terms since they follow the very
same pattern, exploiting $\mathit{entity}$ and $\mathit{base}$.

\subsubsection{Evaluating core primitives}

This section specifies the triplet transformations that are associated to the
Optica core combinators which can be found in
Fig.~\ref{fig:core-triplet-non-standard}. Before delving into the semantics of
the getter, affine fold and fold combinators, we will introduce the next
definitions, which will be useful for ensuring the consistency of the
formalization:

\begin{definition}
  Given $e: \mathit{optic}\ \beta\ \gamma$, where $\mathit{optic} \in
  \{\mathit{getter},\mathit{affine},\mathit{fold}\}$, a triplet $t$ \emph{is a
    valid input for} $e$ if one of the following conditions holds:
{\small
\begin{center}
\begin{tabular}{cl}
  (1) & $t = \mathit{empty}$  \\
  (2) & $t = \mathcal{E}^{sql}[\mathit{e_2}: \mathit{optic}\ \alpha\ \beta]\
  \mathit{t_2}\text{, for some}\ \mathit{e_2},\ \mathit{t_2}\ \text{such that}\
  \mathit{t_2}\ \text{is a valid input for}\ \mathit{e_2}$ 
\end{tabular}
\end{center}
}%
Basically, an input triplet is valid for a given optic $e$ if it is the empty
triplet (Def.~\ref{def:empty-triplet}), or if it is the result obtained from
evaluating an optic expression $e_2$ with a valid input, where the `part' type
of optic $e_2$ coincides with the `whole' type from $e$.
\end{definition}

\begin{definition}
    \label{def:singleton-model-type}
    A \emph{singleton model type} is either a base type or a domain type, i.e.
    it is the result of discarding product types from Optica model types.
\end{definition}
 
\begin{proposition}
    \label{pro:single}
    Let $e: \mathit{optic}\ \alpha\ \beta$, where $\mathit{optic} \in
    \{\mathit{getter},\mathit{affine},\mathit{fold}\}$, $\beta \in
    \text{singleton model type}$, and $t$ a valid input for $e$; then: 
{\small
\begin{center}
$((s), \_, \_) = \mathcal{E}^{sql}[e: \mathit{optic}\ \alpha\ \beta]\ t$
\end{center}
}%

\end{proposition}

The proposition states that, given a valid input, the result from evaluating an
optic that selects a singleton model type always returns a single selection $s$.
This can be easily proven by induction since all combinators producing optics
that select singleton model types do generate a single expression as left
component, according to Fig.~\ref{fig:core-triplet-non-standard}. In fact, this
proposition turns out to be necessary to consider that the implementations of
$\mathit{base}$ and $\mathit{entity}$ (Fig.~\ref{fig:domain-triplet-semantics})
are well-defined.

\begin{figure}
\small
\begin{align*}
& \mathcal{E}^{sql}[\_\ : \mathit{op}\ \alpha\ \beta\ \text{where}\ \mathit{op}
    \in \{\mathit{getter}, \mathit{affine}, \mathit{fold}\}] &::& \quad \mathit{Triplet}\ \rightarrow \mathit{Triplet}\\\\
&\mathcal{E}^{sql}[\mathit{id}_{\mathit{gt}}:\ \mathit{getter}\ \alpha\ \alpha]&=& \quad t \mapsto t \\
&\mathcal{E}^{sql}[g \ggg_{\mathit{gt}} h: \mathit{getter}\ \alpha\ \gamma]&=& \quad
\mathcal{E}^{sql}[h:\ \mathit{getter}\
\beta\ \gamma] \cdot \mathcal{E}^{sql}[g:\ \mathit{getter}\ \alpha\ \beta] \\
&\mathcal{E}^{sql}[g\ \Fork\ h:\ \mathit{getter}\ \alpha\ (\beta,\ \gamma)]&=&
\quad t \mapsto (s_1
\smallfrown \mathit{s_2},\ \mathit{f_1}\ \triangledown\ \mathit{f_2},\
\mathit{w_1} \cup \mathit{w_2}) \quad \text{where} \\
&&& \quad (\mathit{s_1},\ \mathit{f_1},\ \mathit{w_1}) = \mathcal{E}^{sql}[g:\
\mathit{getter}\ \alpha\ \beta]\ t \quad \text{and} \\
&&& \quad (\mathit{s_2},\ \mathit{f_2},\ \mathit{w_2}) = \mathcal{E}^{sql}[h:\ \mathit{getter}\ \alpha\ \gamma]\ t \\
&\mathcal{E}^{sql}[\mathit{like}\ b:\ \mathit{getter}\ \alpha\ \beta]&=& \quad (\_,\ f,\ w)
\mapsto ((\mathit{like}\ \_b\_),\ f,\ w)\\
&\mathcal{E}^{sql}[\mathit{not}\ g:\ \mathit{getter}\ \alpha\ \mathbb{B}]&=& \quad (((b),\ f,\ w)
\mapsto ((\mathit{not}\ b),\ f,\ w))) \cdot \mathcal{E}^{sql}[g:\ \mathit{getter}\
\alpha\ \mathbb{B}] \\
&\mathcal{E}^{sql}[g\ \oplus h:\ \mathit{getter}\ \alpha\ \delta]&=& \quad t \mapsto
((\mathit{b_1} \oplus \mathit{b_2}),\ \mathit{f_1}\ \triangledown\ \mathit{f_2},\
\mathit{w_1} \cup \mathit{w_2}) \quad \text{where} \\
&&& \quad ((\mathit{b_1}),\ \mathit{f_1},\ \mathit{w_1}) = \mathcal{E}^{sql}[g:
\mathit{getter}\ \alpha\ \beta]\ t \quad \text{and} \\
&&& \quad ((\mathit{b_2}),\ \mathit{f_2},\ \mathit{w_2}) = \mathcal{E}^{sql}[f:\
\mathit{getter}\ \alpha\ \gamma]\ t\\\\
&\mathcal{E}^{sql}[\mathit{id}_{\mathit{af}}:\ \mathit{affine}\ \alpha\ \alpha]&=& \quad t \mapsto t \\
&\mathcal{E}^{sql}[g \ggg_{\mathit{af}} h: \mathit{affine}\ \alpha\ \gamma]&=& \quad
\mathcal{E}^{sql}[h:\ \mathit{affine}\
\beta\ \gamma] \cdot \mathcal{E}^{sql}[g:\ \mathit{affine}\ \alpha\ \beta] \\
&\mathcal{E}^{sql}[\mathit{filtered}\ g:\ \mathit{affine}\ \alpha\ \alpha]&=& \quad (s,\ f,\ w)
\mapsto (s,\ \mathit{f_1},\ \{b\} \cup w) \quad \text{where} \\
&&& \quad ((b),\ \mathit{f_1},\ \varnothing) = \mathcal{E}^{sql}[g:\ \mathit{getter}\ \alpha\
\mathbb{B}]\ (s,\ f,\ \varnothing) \\
&\mathcal{E}^{sql}[\mathit{to}_{\mathit{af}}\ g:\ \mathit{affine}\ \alpha\ \beta]&=& \quad
\mathcal{E}^{sql}[g: \mathit{getter}\ \alpha\ \beta] \\\\
&\mathcal{E}^{sql}[\mathit{id}_{\mathit{fl}}:\ \mathit{fold}\ \alpha\ \alpha]&=& \quad t \mapsto t \\
&\mathcal{E}^{sql}[g \ggg_{\mathit{fl}} h: \mathit{fold}\ \alpha\ \gamma]&=& \quad
\mathcal{E}^{sql}[h:\ \mathit{fold}\
\beta\ \gamma] \cdot \mathcal{E}^{sql}[g:\ \mathit{fold}\ \alpha\ \beta] \\
&\mathcal{E}^{sql}[\mathit{nonEmpty}\ g:\ \mathit{getter}\ \alpha\ \mathbb{B}]&=&
\quad (s,\ f,\ w) \mapsto ((nonEmpty\ (\mathcal{E}^{sql}[g:\ \mathit{fold}\ \alpha\
\beta]\ (s,\ f,\ \varnothing))),\ f,\ w) \\
&\mathcal{E}^{sql}[\mathit{to}_{\mathit{fl}}\ a:\ \mathit{fold}\ \alpha\ \beta]&=& \quad
\mathcal{E}^{sql}[a: \mathit{affine}\ \alpha\ \beta]
\end{align*}

\caption{Triplet non-standard semantics}
\label{fig:core-triplet-non-standard}
\end{figure}

\begin{figure}
  \vspace*{-4cm}
  \hspace*{-2.5cm}
  \centering
  \includegraphics[width=1.4\linewidth]{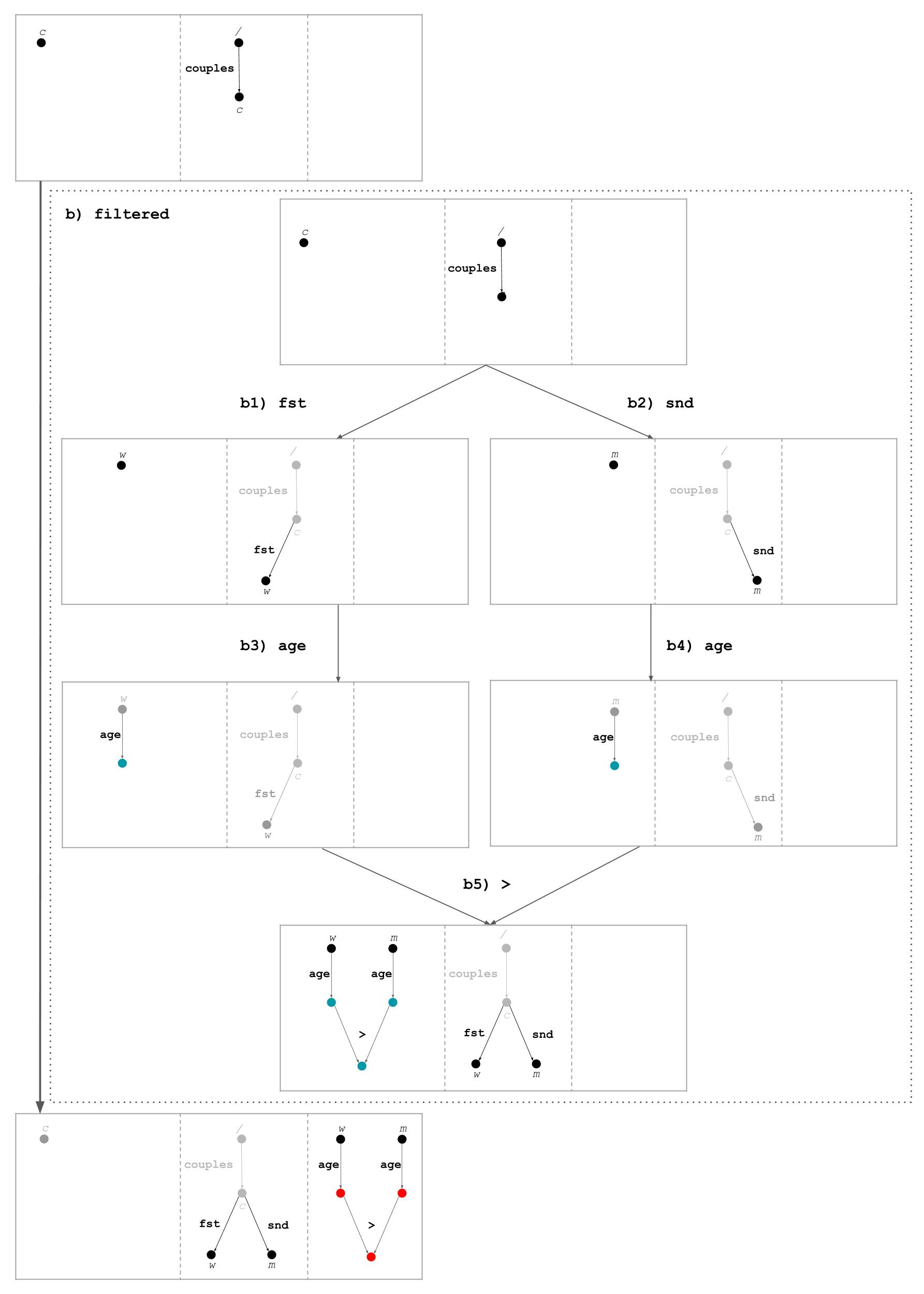}
  \caption{Filtered step in detail}
  \label{fig:filtered}
  \thisfloatpagestyle{empty}
\end{figure}

\paragraph{Getters}

First, we describe the implementation of $\ggg_{\mathit{gt}}$. As
Fig.~\ref{fig:evolution} suggests, vertical composition should be evaluated as
the chaining of transformations, i.e. as function composition. Consequently,
\lstinline{id_gt} is implemented as the identity function, meaning no
transformation at all. 

Step~\emph{e)} (Fig.~\ref{fig:evolution}) shows an example of horizontal
composition ($\Fork$), where a pair of diverging triplets are somehow combined.
In this special case, the changes are only reflected in the left component since
the middle and right components are exactly the same in both triplets. The
evaluation of $\Fork$ supplies the input triplet $t$ as an argument to the
evaluations of $g$ and $h$, which results in a pair of diverging triplets, as
those in the illustration. To carry out the combination of them, we concatenate
the selections $\mathit{s_1}$ and $\mathit{s_2}$, we merge the tries
$\mathit{f_1}$ and $\mathit{f_2}$ ($\triangledown$), and we make the union of
the sets of restrictions $\mathit{w_1}$ and $\mathit{w_2}$. Both the union of
sets and the merging of tries are idempotent operations.

Next, we find \lstinline{like} and \lstinline{not} as examples of unary standard
combinators which just update the left component of the input triplet. The
former ignores the previous selection and replaces it by the constant value. The
latter transforms the triplet by applying the operation over the current
selection. The evaluation demands a single expression as input, where we rely on
Prop.~\ref{pro:single}. Moreover, the Optica type system guarantees that such an
expression represents an optic selecting a boolean.

Finally, Step~\emph{b5)} (Fig.~\ref{fig:filtered}) represents a binary
combinator. The situation is very similar to that of
$\Fork$\footnote{Incidentally, this step is better to illustrate the result of
merging two tries.}. However, instead of concatenating the selections, their
single components are fused into the corresponding expression. The evaluation of
this term assumes that the triplets which derive from the evaluations of $g$ and
$h$ contain singleton selections. Once again, we rely on Prop.~\ref{pro:single}
since all the binary combinators that we can find in Optica take base types as
operands.

\paragraph{Affine Folds}

As in the XQuery evaluation, composition and identity primitives are exactly the
same as those we have just presented for getters. In addition,
$\mathit{to}_{\mathit{af}}$ only returns the evaluation of its argument. The same concept
will apply to folds. Consequently, there only remains $\mathit{filtered}$. As we
have seen before, Step~\emph{b)} (Fig.~\ref{fig:evolution}) represents this
combinator, which was further detailed in Fig.~\ref{fig:filtered} given its
complexity. This figure shows an inner box that describes the triplet evolution
specified by the predicate, which starts from the same input triplet as the
$\mathit{filtered}$ whole expression. The rest of the evolution inside the box
should be straightforward now. However, it has yet to be explained how to move
from the result of Step~\emph{b5)} to the final result of Step~\emph{b)}.
Informally, what happens in this example is that the selection of the whole
expression does by no means change, which seems meaningful since the filter
expression should not change the focus; the left component of the inner
expression represents the predicate, which becomes a new constraint in the right
component of the resulting triplet; last, the middle component remains unchanged
in this particular case.

The evaluation of $\mathit{filtered}$ in
Fig.~\ref{fig:core-triplet-non-standard} formalizes the previous intuitions.
Firstly, the overall input triplet is passed as argument to the evaluation of
the predicate. Its right component is reset to the empty set and the triplet
generated by the predicate is expected to contain an empty set of restrictions,
since getters are unable to update the restriction component of the triplet. In
the resulting triplet, the selection $s$ passes as is, while the restriction
that was selected in the predicate is appended to the existing ones in $w$.
Finally, note that the new entity tree results from the inner triplet,
$\mathit{f_1}$, since new paths may have been created internally.

\paragraph{Folds}

Lastly, we present the interpretation of $\mathit{nonEmpty}$, which introduces a
significant difference in comparison with the rest of combinators: it takes a
fold as parameter. The evaluation of folds is problematic since they lead to the
introduction of nested queries in this infrastructure, as we will see later.
This is the reason why we use the $\mathit{nonempty}$ term from
Fig.~\ref{fig:triplet-syntax} here, which basically stores the triplet resulting
from the evaluation of the fold over the input triplet (after resetting its
constraints, since they will not be necessary). This resulting triplet will
typically extend the entity trie with new paths, but these will not be
propagated into the overall resulting triplet since they are considered private
to the inner scope. As a result, the entity trie of the nonempty expression, as
well as its constraints, are kept unchanged in the overall triplet.

\begin{remark}
    \label{pro:total}
    An Optica expression is always translatable into a triplet endofunction, as
    evidenced by the total implementation of $\mathcal{E}^{\mathit{sql}}$, where
    Prop.~\ref{pro:single} has proven essential. In fact, this evaluation just
    consists on moving things around to adapt Optica expressions to the triplet
    configuration. Unfortunately, translating triplets into SQL statements is a
    partial process, as described in the following section.
\end{remark}

\subsubsection{Target queries and results}
\label{sub:sql-actions-queries}

\begin{figure}[h]
  \centering
  \includegraphics[width=\linewidth]{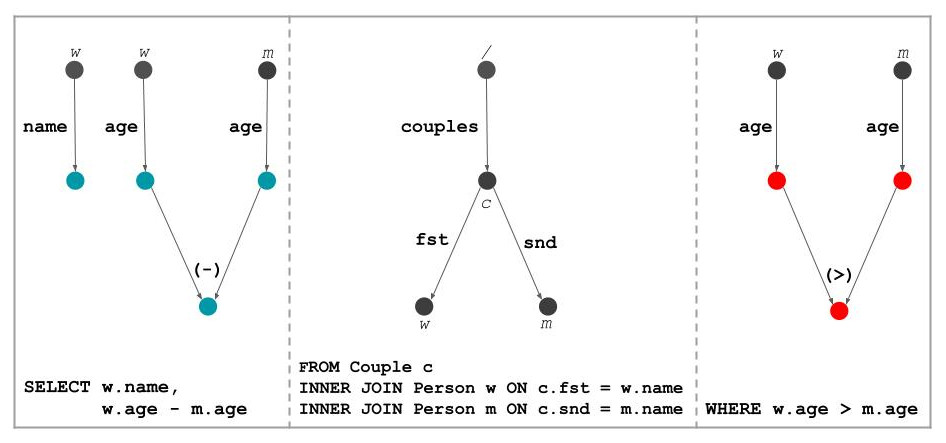}
  \caption{From triplet to SQL}
  \label{fig:sql}
\end{figure}

We have designed triplets to be easily translatable into \sqlinline{SELECT}
statements. This is clearly evidenced in Fig.~\ref{fig:sql}, where we compare
the triplet generated for $\mathit{differencesFl}$ (Def.~\ref{def:differences})
and the expected SQL query that we presented in the background
(Sect.~\ref{sub:SQL:Background}) for the same example. While the translation of
the expressions in the left and right components is straightforward, the
generation of the \sqlinline{FROM} clause from the middle component requires
further explanation. We present the formal translation of Optica query
expressions into SQL in Fig.~\ref{fig:from-triplet-to-sql}. What first calls our
attention is the absence of translations for $\mathit{get}$ and
$\mathit{preview}$. In fact, it is only possible to produce a SQL statement from
$\mathit{getAll}$. As suggested in Prop.~\ref{pro:total}, the translation of
triplets into SQL statements is a partial process. 

\begin{conditions}

    We describe the precise conditions that an Optica query should satisfy in
    order to produce a valid SQL statement\footnote{It is important to note that
    an error should be raised if any of them is not satisfied.}:
    \begin{enumerate}

        \item The optic selected type, i.e. its `part', is a flat type. For
            instance, $\mathit{couples}$ is not translatable into SQL, since it
            selects $\mathit{Couple}$ as part, which contains nested references
            to the entity $\mathit{Person}$. The expression $\mathit{couples}
            \ggg \mathit{fst}$ is valid, since $\mathit{Person}$ does not
            contain further nested data structures: name and age are plain
            values.

        \item The expression cannot contain a fold selecting a base type. For
            example, $\mathit{departments} \ggg \mathit{employees} \ggg
            \mathit{tasks}$ is valid since all the involved folds do select
            entity types.

        \item The original kind (ignoring castings) of the leftmost expression
            forming a query has to be a \emph{fold}. For example,
            $\mathit{couples} \ggg \mathit{fst} \ggg \mathit{name}$ is
            translatable into SQL (it starts with the $\mathit{couples}$ fold)
            while $\mathit{fst} \ggg \mathit{name}$ is not (it starts with the
            $\mathit{fst}$ getter).  Thereby, $\mathit{get}$ and
            $\mathit{preview}$ are omitted, since getter or affine fold
            expressions do not satisfy such condition.

    \end{enumerate} We will further motivate each limitation in the following
    paragraphs, where the whole process of generating SQL statements from
    triplets is described. 

\end{conditions}

Since we aim at turning triplets into SQL expressions, the very first step is to
produce a triplet. We achieve this by evaluating the fold expression that
accompanies $\mathit{getAll}$ and supplying the empty triplet
(Def.~\ref{def:empty-triplet}) to the resulting function. Then, we need to
refine the entity trie of the obtained triplet by assigning {\em fresh} names
for each of its paths (which the evaluation function $\mathcal{E}^{sql}$ does
not generate). Last, we pass the refined triplet as argument to the actual
translator ($\mathit{sql}$). Besides the triplet, note that this function
receives an additional argument, $\hat{local}$, that specifies the path which
actually determines the scope of the query to be generated. This is necessary
since the {\em sql} function will be used to translate both the whole SQL query,
and the inner queries of $\mathit{nonEmpty}$ expressions. In this very first
invocation, we aim at translating the whole triplet; thus, we pass as
$\hat{local}$ the $\hat{top}$ of the entity trie, which represents the common
prefix of every path of the trie. 

The {\em sql} function delegates the generation of each clause of the whole
\sqlinline{SELECT} statement into the corresponding functions $\mathit{select}$,
$\mathit{from}$ and $\mathit{where}$. Moreover, it calls an additional function,
$\mathit{where_+}$, whose purpose will be motivated later on. The results
obtained from each function are concatenated to form the final query. Note that
parentheses and brackets are discarded in the result, they are simply introduced
to delimit the arguments supplied to each function. In particular, an invocation
surrounded by brackets informs that the invocation may be omitted, taking into
account the accompanying conditions. We describe the generation process of each
clause in the next paragraphs, where we will make frequent use of the following
additional definitions:

\begin{center}
\begin{tabular}{rl}
  $\rho.\hat{top}$ &  \em The key which starts every path of the entity trie, if any \\
  $\rho_1.local(\rho_2)$ &\em  The local path of $\rho_2$ which is extended by $\rho_1$, if any \\
  $\rho(\hat{p})$ &\em  The name assigned to the given path in the refined trie\\
  $\hat{p}.last$ & \em  The key which finishes the given path \\
  $\hat{p}.up$ &\em  The second to last key of the given path \\
  $optic.name$ & \em The name of the given optic primitive \\
  $optic.kind$ & \em  The kind of optic: getter, affine fold or fold \\
  $optic.whole$ & \em The type of the whole entity that the optic inspects \\
  $optic.part$ & \em  The type of focus to which the optic points to (an entity or base type)\\
  $pk(type)$ & \em The primary key of the relational table associated to the given type\\
\end{tabular}
\end{center}

With a little abuse of notation, we will omit the $last$ attribute in path
expressions, as in $\hat{p}.name$, instead of writing the more verbose
$\hat{p}.last.name$.

\paragraph{Select clause} The $\mathit{select}$ function generates the
\sqlinline{SELECT} clause by separating the result of translating each
expression with commas. We describe the translation of the different types of
expressions in the following lines:
\begin{itemize}

    \item The translation of a path $\hat{x}$ simply refers to all the columns
        of the table corresponding to that path, which was assigned by the
        $\mathit{fresh}$ function. Also note that the path must refer to an
        entity with no further nested entities (Precondition 1). Otherwise, the
        query output would not contain all the data required to reassemble the
        entity, i.e.  this work does not support \emph{query
        shredding}~\cite{DBLP:conf/sigmod/CheneyLW14} yet.

    \item The translation of a projection $\hat{x}.\mathit{base}$ is basically
        the same, but we find an interesting restriction here. SQL does not
        support multivalued columns, and therefore we cannot use a fold to
        project values (Precondition 2).

    \item The translation of $\mathit{nonEmpty}$ is given in terms of
        \sqlinline{EXISTS}, which contains a nested SQL expression. Thereby, we
        invoke the $\mathit{sql}$ generator recursively. Before doing this, we
        need first to generate fresh names for the trie of the {\em nonEmpty}
        expression and to merge it with the outer entity trie\footnote{The
        combinator $\vartriangleleft$ merges tries, keeping the names from the
    left when it finds conflicting paths.}. Second, we need to calculate the
    right $\hat{local}$ path and to pass it to the {\em sql} generator.

    \item The evaluation of the rest of expressions should be straightforward
        since they just adapt operators and literals into their SQL form.

\end{itemize}

\begin{remark}
    None of the optic models associated to the guiding examples include affine
    folds in their definitions. In the particular case of the SQL
    interpretation, such optics are assumed as fields which may contain a
    \sqlinline{NULL} value, i.e. as \emph{nullable} table columns.
\end{remark}

\paragraph{Where clause} We continue with the \sqlinline{WHERE} clause given its
similarity with the \sqlinline{SELECT} clause, which is generated by the
$\mathit{where}$ and $\mathit{where}_+$ functions. The former is quite similar
to $\mathit{select}$ since it basically delegates the evaluation of the
restriction expressions, although it uses \sqlinline{AND} as delimiter for the
results. The evaluation of expressions is exactly the same as the one that we
have introduced in the previous paragraph. Note that $\mathit{where}$ produces
\sqlinline{WHERE True} if the set of restrictions is empty. Concerning
$\mathit{where}_+$, this function is responsible for appending the precise
connection between nested and outer variables, which were introduced at the very
end of Sect.~\ref{sub:SQL:Background}. We will explain it together with the
discussion of the generation of the \sqlinline{FROM} clause in the next
paragraph.

\paragraph{From clause} Before venturing into the $\mathit{from}$ function,
there are a few conditions that the generator should preserve. Firstly, it is
assumed that $\rho.\hat{\mathit{top}}$ must refer to a $\mathit{fold}$
(Precondition 3), since we need an entry point in the hierarchical tables. This
means that we can only translate expressions that start with a fold, like
$\mathit{differencesFl}$ (Def.~\ref{def:differences}) ---which starts with
$\mathit{couples}$--- or $\mathit{expertiseFl}$ (Def.~\ref{def:expertise})
---which starts with $\mathit{departments}$.  Secondly, the invocation to
$\mathit{from}$ is omitted if $\hat{\mathit{local}}$ is not defined, since this
indicates that the current query is not introducing new variables, and therefore
no \sqlinline{FROM} clause is required.

As expected, the $\mathit{from}$ function prepares the \sqlinline{FROM} clause.
It selects the `part' type from $\hat{\mathit{local}}$ as the initial table.
Then, it produces an \sqlinline{INNER JOIN} expression for each element hanging
from it.  This is the reason why tries contain nothing more than entities, since
they correspond to relational tables. In general, the complexity associated to
these definitions is due to the choice and implementation of the corresponding
Foreign Key patterns (Sect.~\ref{sub:SQL:Background}).

\begin{figure}
\small
\begin{align*}
&\mathcal{E}^{sql}[\mathit{getAll}\ g:\ \alpha \rightarrow \mathit{list}\ \beta]
&::& \quad (\mathbb{S} \rightarrow \mathbb{S}) \rightarrow \mathit{SQL} \\\\
%
%
&\mathcal{E}^{sql}[\mathit{getAll}\ g:\ \alpha \rightarrow \mathit{list}\ \beta] \quad \mathit{pk}
&=& \quad \mathit{sql}\ (s,\ \rho,\ w)\ \mathit{pk}\ \rho.\hat{top} \quad \text{where} \\
&&& \quad (s,\ f,\ w) = \mathcal{E}^{sql}[g:\ \mathit{fold}\ \alpha\ \beta]\
\mathit{empty} \quad \text{and}\\
&&& \quad \rho = \mathit{fresh}\ f \quad \text{and} \\
&&& \quad \rho.\hat{top}\ \text{is defined} \\\\
%
%
&sql \quad (s,\ \rho,\ w) \quad \mathit{pk} \quad [\hat{local}] &=& \quad (\mathit{select}\ s\ \rho\ \mathit{pk}) \quad
[\mathit{from}\
\rho\ \mathit{pk}\ \hat{local}] \quad (\mathit{where}\ w\ \rho\ \mathit{pk}) \quad [\mathit{where}_+\ \rho\ \mathit{pk}\ \hat{local}]; \quad \text{where}\\ 
&&& \quad \rho.\hat{\mathit{top}}.\mathit{kind} = \mathit{fold} \quad \text{and} \\
&&& \quad \mathit{from}\ \text{invocation is omitted if}\ \hat{\mathit{local}}\ \text{is not defined} \quad \text{and}\\
&&& \quad \mathit{where_+}\ \text{invocation is omitted if}\ \rho.\hat{\mathit{top}} = \hat{\mathit{local}} \\\\
%
%
&\mathit{select} \quad (e_1, e_2, \ldots, e_n) \quad \rho \quad \mathit{pk} &=& \quad
\mathbf{SELECT}\ \mathit{expr}\ e_1\ \rho\ \mathit{pk}\ \boldsymbol{,}\ \mathit{expr}\ e_2\ \rho\ \mathit{pk}\ \boldsymbol{,}\ \ldots\ \boldsymbol{,}\ \mathit{expr}\ e_n\ \rho\ \mathit{pk}\\\\
%
%
&\mathit{expr} \quad \hat{x} \quad \rho \quad \mathit{pk} &=& \quad \rho(\hat{x})\boldsymbol{.*} \quad \text{where} \\
&&& \quad \hat{x}.last.part\ \in \text{flat types}\\
&\mathit{expr} \quad \hat{x}.\mathit{optic} \quad \rho \quad \mathit{pk} &=& \quad
\rho(\hat{x})\boldsymbol{.}\mathit{(optic.name)} \quad \text{where} \\
&&& \quad \mathit{optic.part} \in \text{base types} \quad \text{and} \\
&&& \quad \mathit{optic}.\mathit{kind} \in \{\mathit{getter}, \mathit{affine}\}\\
&\mathit{expr} \quad (t \oplus u) \quad \rho \quad \mathit{pk} &=& \quad \mathit{expr}\ t\ \rho\ \mathit{pk}
\boldsymbol{\oplus} \mathit{expr}\ u\ \rho\ \mathit{pk} \\
&\mathit{expr} \quad (\mathit{not}\ e) \quad \rho \quad \mathit{pk} &=& \quad \mathbf{NOT}\ \boldsymbol{(}\ \mathit{expr}\ e\ \rho\ \mathit{pk}\ \boldsymbol{)} \\
&\mathit{expr} \quad (\mathit{like}\ a) \quad \rho \quad \mathit{pk} &=& \quad a \\
&\mathit{expr} \quad (\mathit{nonEmpty}\ (s,\ f,\ w)) \quad \rho \quad \mathit{pk} &=& \quad
\mathbf{EXISTS}\ \boldsymbol{(}\ \mathit{sql}\ (s,\ \rho', w)\ \mathit{pk}\ \rho.local(\rho')\ \boldsymbol{)} \quad \text{where}\\
&&& \quad \rho' = \rho \vartriangleleft\mathit{fresh}\ f\\\\
%
%
&\mathit{where} \quad \varnothing \quad \rho \quad \_ &=& \quad \mathbf{WHERE}\
\mathit{True}\\
&\mathit{where} \quad \{e_1, e_2, \ldots, e_n\} \quad \rho \quad \mathit{pk} &=& \quad
\mathbf{WHERE}\ \mathit{expr}\
e_1\ \rho\ \mathit{pk}\ \mathbf{AND}\ \mathit{expr}\ e_2\ \rho\ \mathit{pk}\ \mathbf{AND}\ \ldots\ \mathbf{AND} \
\mathit{expr}\ e_n\ \rho\ \mathit{pk}\\\\\
%
%
&\mathit{where}_+ \quad \rho \quad \mathit{pk} \quad \hat{local} &=& \quad
\mathbf{AND}\ \rho(\hat{\mathit{local}})\boldsymbol{.}\mathit{key} \boldsymbol{=}
\rho(\hat{\mathit{local}}.\mathit{up})\boldsymbol{.}\mathit{key} \quad \text{where} \\
&&& \quad \mathit{key} = \mathit{pk}(\hat{\mathit{local}}.\mathit{whole})\\\\
%
%
&\mathit{from} \quad \rho \quad \mathit{pk} \quad \hat{local} &=& \quad \mathbf{FROM}\ \hat{\mathit{local}}.\mathit{part}\
\mathbf{AS}\ \rho(\hat{\mathit{local}})\ \mathit{joins} \quad \text{where}\\
&&&\quad \mathit{joins} = \{\mathit{eqjoin}\ \hat{x}\ \rho\ \mathit{pk} \mid \hat{x} \in \rho,\ \hat{x} = \hat{\mathit{local}} \smallfrown \hat{y} \quad \text{for some} \quad \hat{y} \neq ()\} \\\\
&\mathit{eqjoin} \quad \hat{x} \quad \rho \quad \mathit{pk} &=& \quad
\mathbf{INNER\ JOIN}\ \mathit{\hat{x}.part}\ \mathbf{AS}\ \rho(\hat{x})\
    \mathit{cond}\ \quad \text{where}\\ 
&&& \quad cond = \begin{cases} 
    \mathbf{USING}\ \mathit{pk}(\hat{x}.\mathit{whole}) & \text{if} \quad \hat{x}.\mathit{kind}\ = \mathit{fold}\\
    \mathbf{ON}\ \rho(\hat{x}.\mathit{up})\boldsymbol{.}(\hat{x}.\mathit{up}.\mathit{name}) \boldsymbol{=} \rho(\hat{x})\boldsymbol{.}(\mathit{pk}(\hat{x}.\mathit{part})) & \text{otherwise} 
   \end{cases}
\end{align*}

\caption{SQL generation}
\label{fig:from-triplet-to-sql}
\end{figure}

Once we have implemented $\mathcal{E}^{sql}$, we can use it to translate generic
queries into SQL statements. For $\mathit{differences}$
(Def.~\ref{def:differences}) we get:
{\small
\begin{flalign*}
&\mathbf{def}\ \mathit{differencesSQL}: \mathit{SQL}=&\\
&\quad \mathcal{E}^{sql}[\mathit{differences}: \mathit{Couples}
\rightarrow \mathit{list}\ (\mathbb{S}, \mathbb{N})]\ (\mathit{Person} \leadsto
\mathit{name})
\end{flalign*}
}%
and we adapt $\mathit{expertise}$ (Def.~\ref{def:expertise}) as follows:
{\small
\begin{flalign*}
&\mathbf{def}\ \mathit{expertiseSQL}: \mathit{SQL}=&\\
&\quad \mathcal{E}^{sql}[\mathit{expertise}: \mathit{Org} \rightarrow \mathit{list}\
\mathbb{S}]\ (\mathit{Department} \leadsto \mathit{dpt},\ \mathit{Employee}
\leadsto \mathit{emp})
\end{flalign*}
}%
Unlike other evaluators, $\mathcal{E}^{sql}$ requires the relation of primary
keys for the involved tables as an additional argument, since this information
is not contemplated in the optic model.  We use the notation $(t_0 \leadsto
k_0,\ t_1 \leadsto k_1,\ ...,\ t_n \leadsto k_n)$ to build such argument. For
instance, the primary key associated to the table $\mathit{Person}$ is the
column $\mathit{name}$. If we ignore variable names, the SQL statements which
are generated by the previous definitions are exactly the same as those that
were introduced in Sect.~\ref{sub:SQL:Background}. 

As can be inferred from Fig.~\ref{fig:from-triplet-to-sql}, the evaluation of
$\mathit{getAll}$ always leads to a SQL \sqlinline{SELECT} statement, unless an
error condition is present. The resulting query does not contain nested
subqueries, beyond the ones that emerge in the context of \sqlinline{EXISTS}
(due to the $\mathit{nonEmpty}$ term). The \sqlinline{FROM} clause uses
\sqlinline{INNER JOIN}s as the means to navigate downwards the tables in the
model. Besides the previous elements, the evaluator just produces expressions
with basic functions, operators and literals; no additional SQL features are
required.

Clearly, the SQL semantics are not as neat as those associated the XML
infrastructure (Sect.~\ref{sec:XQuery}), since they require a non-trivial
normalization into triplets prior to generating the SQL statement. Besides, such
generation is partial, and thus triplets must meet certain conditions to
guarantee a correct translation. Fortunately, as we will see in the next
section, we can take a different path towards the generation of SQL where we can
rely on existing work on language-integrated query. Despite this fact, in
Sect.~\ref{sec:Discussion} we will discuss why the triplet normalization is
still relevant.

\section{T-LINQ}
\label{sec:T-LINQ}


This section introduces Optica as a higher level language that we can interpret
into comprehensions. In particular, we generate T-LINQ
\cite{cheney2013practical} queries from optic expressions, essentially following
a similar translation implemented in the Links language in \cite{cheneyEmail}.
By doing this, we demonstrate that the compositional style embraced by optics can be
fruitfully exploited in order to generate comprehension expressions automatically.
Moreover, we open the possibility of delegating the arduous task of generating
SQL statements from optic expressions, described in the previous section, to the
existing translation and normalization techniques of T-LINQ. As usual, we supply
a brief background on the querying language, T-LINQ, and then we show the
non-standard semantics that is needed in order to produce the corresponding
comprehension-based queries.

\begin{figure}[hbt]
\begin{subfigure}{.4\textwidth}
  \centering
  \small
\begin{flalign*}
& \mathbf{type}\ \mathit{NestedOrg} = \mathit{NestedDepartment}\ \mathbf{list}&\\
& \mathbf{type}\ \mathit{NestedDepartment} = &\\
& \quad \{\mathit{dpt}: \mathbf{string};\ \mathit{employees}: \mathit{NestedEmployee}\ \mathbf{list}\}&\\
& \mathbf{type}\ \mathit{NestedEmployee} = &\\
& \quad \{\mathit{emp}: \mathbf{string};\ \mathit{tasks}: \mathit{Task}\ \mathbf{list}\}&\\
& \mathbf{type}\ \mathit{Task} = \{\mathit{tsk}: \mathbf{string}\}&
\end{flalign*}
  \caption{Nested organization}
  \label{fig:nested-org}
\end{subfigure}%
\begin{subfigure}{.6\textwidth}
  \small
  \centering
\begin{flalign*}
& \mathbf{type}\ \mathit{Org} = \{\mathit{departments}: \{\mathit{dpt}: \mathbf{string}\}\ \mathbf{list}; &\\
& \qquad \qquad \quad \mathit{employees}: \{\mathit{dpt}: \mathbf{string};\ \mathit{emp}: \mathbf{string}\}\ \mathbf{list}; &\\
& \qquad \qquad \quad \mathit{tasks}: \{\mathit{emp}: \mathbf{string};\ \mathit{tsk}: \mathbf{string}\}\ \mathbf{list}\} &
\end{flalign*}
  \caption{Flattened organization}
  \label{fig:flat-org}
\end{subfigure}
\caption{Alternative organization models}
\label{fig:model-comparison}
\end{figure}

\subsection{Background}
\label{sub:background-tlinq}

In order to manually adapt the $\mathit{expertise}$ query
(Def.~\ref{def:expertise}) as a T-LINQ expression\footnote{This section omits
the couples example for the sake of brevity. We select $\mathit{expertise}$ over
$\mathit{differences}$ since we consider it to be more challenging.}, we will
first review the difference between a relational (or flattened) model and a
nested model. Figure~\ref{fig:model-comparison} shows the nested ({\em NestOrg})
and flat models ({\em Org}) for the organization example
from~\cite{cheney2013practical}, as T-LINQ records. Note that $\mathit{Org}$
differs from the nested version $\mathit{NestedOrg}$ in the type of their
fields, since it contains textual values which act as foreign keys to refer to
the corresponding entities. In fact, the second version has a strong
correspondence with the SQL tables that we introduced in
Sect~\ref{sub:SQL:Background}. Cheney \emph{et al} show the quoted expression
that adapts the flattened model into the nested one
(Fig.~\ref{fig:from-flat-to-nested}), where $\%\mathit{org}$ splices the
database representation (${<}@ \mathbf{database}(\text{``Org''}) @{>}$). In
particular, the programmer understands such representation as a list of entities
from the relational model; therefore, she can use the widespread notation of
list comprehensions to implement the desired queries, where filtering
($\mathit{if ... then}$) and mapping ($\mathit{yield}$) features are also
available.  Figure~\ref{fig:expertise-tlinq} shows the implementation of the
$\mathit{expertise}$ query in terms of T-LINQ, which builds upon the flattened
model\footnote{T-LINQ does support a compositional style, where analogous
    combinators for $\mathit{all}$, $\mathit{any}$, etc. could be
    supplied~\cite[Sect.  3.2]{cheney2013practical}. Using these combinators and
    the nested version of the organizational model, $\mathit{NestedOrg}$, the
    $\mathit{expertiseTlinq}$ could be written more concisely. Then, and thanks
    to its normalization engine, the query could be rewritten to its equivalent
version over the relational model.}. Later on, we will see that the nested model
becomes essential in the evaluation of Optica expressions, where we will try to
produce an equivalent for $\mathit{expertiseTlinq}$ from the evaluation of
$\mathit{expertise}$.

\begin{figure}[hbt]
\small
\begin{align*}
&\mathbf{def}\ \mathit{nestedOrg} =\ {<}@\\
& \quad \mathbf{for}\ d\ \mathbf{in}\ \%\mathit{org}.\mathit{departments}\ \mathbf{do}\\
& \quad \mathbf{yield}\ \{\mathit{dpt} = d.\mathit{dpt}, \mathit{employees} = \\
& \qquad \qquad \qquad \mathbf{for}\ e\ \mathbf{in}\ \%\mathit{org}.\mathit{employees}\ \mathbf{do}\\
& \qquad \qquad \qquad \mathbf{if}\ d.\mathit{dpt} = e.\mathit{dpt}\ \mathbf{then}\\
& \qquad \qquad \qquad \mathbf{yield}\ \{\mathit{emp} = e.\mathit{emp}, \mathit{employees} = \\
& \qquad \qquad \qquad \qquad \qquad \quad \mathbf{for}\ t\ \mathbf{in}\ \%\mathit{org}.\mathit{tasks}\ \mathbf{do}\\
& \qquad \qquad \qquad \qquad \qquad \quad \mathbf{if}\ e.\mathit{emp} = t.\mathit{emp}\ \mathbf{then}\\
& \qquad \qquad \qquad \qquad \qquad \quad \mathbf{yield}\ \{\mathit{tsk} = t.\mathit{tsk}\}\}\}\ @{>}
\end{align*}
\caption{From flat to nested organization model}
\label{fig:from-flat-to-nested}
\end{figure}

\begin{figure}[hbt]
    \small
\begin{align*}
&\mathbf{def}\ \mathit{expertiseTlinq} =\ {<}@&\\
& \quad \mathbf{for}\ d\ \mathbf{in}\ \%\mathit{org}.\mathit{departments}\ \mathbf{do}&\\
& \quad \mathbf{if}\ \mathbf{not}\ \mathbf{exists}&\\
& \quad \quad \mathbf{for}\ e\ \mathbf{in}\ \%\mathit{org}.\mathit{employees}\ \mathbf{do}&\\
& \quad \quad \mathbf{if}\ d.\mathit{dpt} = e.\mathit{dpt}\ \wedge\
    \mathbf{not}\ \mathbf{exists}\ &\\
& \quad \quad \quad \mathbf{for}\ t\ \mathbf{in}\ \%\mathit{org}.\mathit{tasks}\ \mathbf{do}&\\
& \quad \quad \quad \mathbf{if}\ e.\mathit{emp} = t.\mathit{emp}\ \wedge\
    t.\mathit{tsk} = `abstract\textrm'\ \mathbf{then}\ \mathbf{yield}\ t.\mathit{tsk}&\\
& \quad \quad \mathbf{then}\ \mathbf{yield}\ e.\mathit{emp}&\\
& \quad \mathbf{then}\ \mathbf{yield}\ d.\mathit{dpt}\ @{>}&
\end{align*}
\caption{T-LINQ analogous for $\mathit{expertise}$}
\label{fig:expertise-tlinq}
\end{figure}

\subsection{T-LINQ non-standard Semantics}

As usual, we provide $\mathcal{E}^{tlinq}$ in order to evaluate Optica
expressions into T-LINQ expressions. Prior to this, we need to determine the
semantic domains for this evaluation by means of $\mathcal{T}^{tlinq}$, which is
shown in Fig.~\ref{fig:tlinq-semantic-domains}. As expected, this semantic
function maps Optica types to T-LINQ representation types ($\mathit{Repr}$).  In
particular it just relies on an auxiliary function $\mathcal{T}^{aux}$ and wraps
the resulting type with $\mathit{Expr}$.  The implementation of
$\mathcal{T}^{aux}$ is direct for base types, whereas tuples are represented as
records. Concerning query types, its representation is also straightforward
since functions are directly supported by T-LINQ, although we map
$\mathit{option}$ to $\mathbf{list}$, since such datatype is not contemplated by
T-LINQ. Last, optic types are simply represented by the query type they
generate. The next sections present the semantic domains for domain types, the
implementation of $\mathcal{E}^{tlinq}$ and discusses the final results.

\begin{figure}[hbt]
\small
\begin{center}
\begin{tabular}{lcl}
$\mathcal{T}^{tlinq}[t]$ &=& $\mathit{Expr}{<}\ \mathcal{T}^{aux}[t]\ {>}$ \\\\
$\mathcal{T}^{aux}[\mathbb{N}]$ &=& $\mathbf{int}$ \\
$\mathcal{T}^{aux}[\mathbb{B}]$ &=& $\mathbf{bool}$ \\
$\mathcal{T}^{aux}[\mathbb{S}]$ &=& $\mathbf{string}$ \\
$\mathcal{T}^{aux}[(\alpha, \beta)]$ &=& $\boldsymbol{\{}\_1:\mathcal{T}^{aux}[\alpha]\boldsymbol{,}\ \_2:\mathcal{T}^{aux}[\beta]\boldsymbol{\}}$ \\
$\mathcal{T}^{aux}[\alpha \rightarrow\ \beta]$                 &=& $\mathcal{T}^{aux}[\alpha] \rightarrow\ \mathcal{T}^{aux}[\beta]$\\
$\mathcal{T}^{aux}[\alpha \rightarrow \mathit{option}\ \beta]$ &=& $\mathcal{T}^{aux}[\alpha] \rightarrow\ \mathbf{list}\ \mathcal{T}^{aux}[\beta]$\\
$\mathcal{T}^{aux}[\alpha \rightarrow \mathit{list}\ \beta]$   &=& $\mathcal{T}^{aux}[\alpha] \rightarrow\ \mathbf{list}\ \mathcal{T}^{aux}[\beta]$\\
$\mathcal{T}^{aux}[\mathit{getter}\ \alpha\ \beta]$ &=& $\mathcal{T}^{aux}[\alpha \rightarrow\ \beta]$\\
$\mathcal{T}^{aux}[\mathit{affine}\ \alpha\ \beta]$ &=& $\mathcal{T}^{aux}[\alpha \rightarrow \mathit{option}\ \beta]$\\
$\mathcal{T}^{aux}[\mathit{fold}\ \alpha\ \beta]$   &=& $\mathcal{T}^{aux}[\alpha \rightarrow \mathit{list}\ \beta]$\\
\end{tabular}
\caption{Semantic domains of the T-LINQ evaluation}
\label{fig:tlinq-semantic-domains}
\end{center}
\end{figure}

\subsubsection{Evaluating domain and core primitives}

This section introduces the evaluation of domain and core terms into T-LINQ
expressions. As we have already seen, all domain terms represent optic
expressions, and thus they have to be adapted as functions.
Figure~\ref{fig:couple-tlinq-semantics} shows the semantic domains (by extending
$\mathcal{T}^{aux}$) and the evaluation of the terms in the organization
example. Note how the organization types are mapped to the corresponding nested
(instead of relational) types. This aspect will be relevant later on while
generating the target queries. Back to the evaluation of terms, we can see that
this is essentially a T-LINQ adaptation of the code that we presented in
\lstinline{OrgModel} (Sect.~\ref{subsub:organization-example}) where we used
lambda expressions to build concrete optics. 

\begin{figure}[hbt]
\small
\begin{center}
\begin{tabular}{lcl}
$\mathcal{T}^{aux}[\mathit{Org}]$ &=& $\mathit{NestedOrg}$ \\
$\mathcal{T}^{aux}[\mathit{Department}]$ &=& $\mathit{NestedDepartment}$ \\
$\mathcal{T}^{aux}[\mathit{Employee}]$ &=& $\mathit{NestedEmployee}$ \\
$\mathcal{T}^{aux}[\mathit{Task}]$ &=& $\mathit{Task}$ \\
\\
$\mathcal{E}^{tlinq}[\mathit{departments}:\ \mathit{fold}\ \mathit{Org}\ \mathit{Department}]$ &=& ${<}@\ \mathbf{fun}(\mathit{ds}) \rightarrow \mathit{ds}\ @{>}$\\
$\mathcal{E}^{tlinq}[\mathit{dpt}: \mathit{getter}\ \mathit{Department}\ \mathbb{S}]$ &=& ${<}@\ \mathbf{fun}(d) \rightarrow d.\mathit{dpt}\ @{>}$\\
$\mathcal{E}^{tlinq}[\mathit{employees}: \mathit{fold}\ \mathit{Department}\ \mathit{Employee}]$ &=& ${<}@\ \mathbf{fun}(\mathit{es}) \rightarrow \mathit{es}\ @{>}$\\
$\mathcal{E}^{tlinq}[\mathit{emp}: \mathit{getter}\ \mathit{Employee}\ \mathbb{S}]$ &=& ${<}@\ \mathbf{fun}(e) \rightarrow e.\mathit{emp}\ @{>}$\\
$\mathcal{E}^{tlinq}[\mathit{tasks}: \mathit{fold}\ \mathit{Employee}\ \mathit{Task}]$ &=& ${<}@\ \mathbf{fun}(\mathit{ts}) \rightarrow \mathit{ts}\ @{>}$\\
$\mathcal{E}^{tlinq}[\mathit{tsk}: \mathit{getter}\ \mathit{Task}\ \mathbb{S}]$ &=& ${<}@\ \mathbf{fun}(t) \rightarrow t.\mathit{tsk}\ @{>}$
\end{tabular}
\end{center}
\caption{T-LINQ semantic domains and non-standard semantics for organization
extension}
\label{fig:couple-tlinq-semantics}
\end{figure}

The evaluation of core combinators (Fig.~\ref{fig:tlinq-non-standard-semantics})
also shares a strong resemblance with those that we have seen for their concrete
counterparts in Sect.~\ref{sub:Programming:Getters}. In essence, the difference
lies in the fact that concrete optics build directly upon the type system of
Scala and the T-LINQ interpretation on its own type system. Thus, the
evaluation of $\Fork$ creates a T-LINQ lambda expression using T-LINQ records
instead of using the lambda expressions and products of Scala.  Similarly,
$\ggg_{\mathit{af}}$ and $\ggg_{\mathit{fl}}$ implement composition by using directly the primitives
of T-LINQ, whereas the implementation of this combinator in concrete optics is
based upon the standard Scala implementation.

\begin{figure}
\small
\begin{align*}
&\mathcal{E}^{tlinq}[\mathit{id}_{\mathit{gt}}:\ \mathit{getter}\ \alpha\ \alpha]&=&
\quad {<}@\ \mathbf{fun}(a) \rightarrow a\ @{>}\\
&\mathcal{E}^{tlinq}[g\ \ggg_{\mathit{gt}} h: \mathit{getter}\ \alpha\ \gamma]&=& \quad
{<}@\ \mathbf{fun}(a) \rightarrow \%\mathcal{E}^{tlinq}[h: \mathit{getter}\ \beta\
\gamma]\ (\%\mathcal{E}^{tlinq}[g: \mathit{getter}\ \alpha\ \beta]\ a) \ @{>}\\
&\mathcal{E}^{tlinq}[g\ \Fork\ h:\ \mathit{getter}\ \alpha\ (\beta,\ \gamma)]&=&
\quad {<}@\ \mathbf{fun}(a) \rightarrow \{\_1 = \%\mathcal{E}^{tlinq}[g: \mathit{getter}\
\alpha\ \beta]\ a,\ \_2 = \%\mathcal{E}^{tlinq}[h: \mathit{getter}\ \alpha\ \gamma]\ a\} \ @{>}\\
&\mathcal{E}^{tlinq}[\mathit{like}\ b:\ \mathit{getter}\ \alpha\ \beta]&=& \quad
{<}@\ \mathbf{fun}(a) \rightarrow b \ @{>}\\
&\mathcal{E}^{tlinq}[\mathit{not}\ g:\ \mathit{getter}\ \alpha\ \mathbb{B}]&=&
\quad {<}@\ \mathbf{fun}(a) \rightarrow \mathbf{not}\ (\%\mathcal{E}^{tlinq}[g:\
\mathit{getter}\ \alpha\ \mathbb{B}]\ a) \ @{>}\\
&\mathcal{E}^{tlinq}[g \oplus h:\ \mathit{getter}\ \alpha\ \delta]&=& \quad 
{<}@\ \mathbf{fun}(a) \rightarrow (\%\mathcal{E}^{tlinq}[g:\ \mathit{getter}\ \alpha\
\beta]\ a \quad \oplus \quad
\%\mathcal{E}^{tlinq}[h:\ \mathit{getter}\ \alpha\ \gamma]\ a) \ @{>}\\\\
&\mathcal{E}^{tlinq}[\mathit{id}_{\mathit{af}}:\ \mathit{affine}\ \alpha\ \alpha]&=&
\quad {<}@\ \mathbf{fun}(a) \rightarrow \mathbf{yield}\ a \ @{>}\\
&\mathcal{E}^{tlinq}[g\ \ggg_{\mathit{af}} h: \mathit{affine}\ \alpha\ \gamma]&=& \quad 
{<}@\ \mathbf{fun}(a) \rightarrow \mathbf{for}\ b\ \mathbf{in}\ \%\mathcal{E}^{tlinq}[g:
\mathit{affine}\ \alpha\ \beta]\ a\ \mathbf{do}\\ 
&&&\qquad\qquad\qquad\quad \mathbf{for}\ c\ \mathbf{in}\
\%\mathcal{E}^{tlinq}[h: \mathit{affine}\ \beta\ \gamma]\ b\ \mathbf{yield}\ c \ @{>}\\
&\mathcal{E}^{tlinq}[\mathit{filtered}\ p:\ \mathit{affine}\ \alpha\ \alpha]&=&
\quad {<}@\ \mathbf{fun}(a) \rightarrow \mathbf{if}\ \%\mathcal{E}^{tlinq}[p:\
\mathit{affine}\ \alpha\ \mathbb{B}]\ a\ \mathbf{then}\ \mathbf{yield}\ a\ @{>}\\
&\mathcal{E}^{tlinq}[\mathit{to}_{\mathit{af}}\ g:\ \mathit{affine}\ \alpha\ \beta]&=& \quad
{<}@\ \mathbf{fun}(a) \rightarrow \mathbf{yield}\ (\%\mathcal{E}^{tlinq}[g:
\mathit{getter}\ \alpha\ \beta]\ a) \ @{>}\\\\
&\mathcal{E}^{tlinq}[\mathit{id}_{\mathit{fl}}:\ \mathit{fold}\ \alpha\ \alpha]&=& \quad 
{<}@\ \mathbf{fun}(a) \rightarrow \mathbf{yield}\ a \ @{>}\\
&\mathcal{E}^{tlinq}[g\ \ggg_{\mathit{fl}} h: \mathit{fold}\ \alpha\ \gamma]&=& \quad 
{<}@\ \mathbf{fun}(a) \rightarrow \mathbf{for}\ b\ \mathbf{in}\ \%\mathcal{E}^{tlinq}[g:
\mathit{fold}\ \alpha\ \beta]\ a\ \mathbf{do}\\ 
&&&\qquad\qquad\qquad\quad \mathbf{for}\ c\ \mathbf{in}\
\%\mathcal{E}^{tlinq}[h: \mathit{fold}\ \beta\ \gamma]\ b\ \mathbf{yield}\ c \ @{>}\\
&\mathcal{E}^{tlinq}[\mathit{nonEmpty}\ g:\ \mathit{getter}\ \alpha\
\mathbb{B}]&=& \quad {<}@\ \mathbf{fun}(a) \rightarrow \mathbf{exists}\
(\%\mathcal{E}^{tlinq}[g:\ \mathit{fold}\ \alpha\ \beta]\ a )\ @{>}\\
&\mathcal{E}^{tlinq}[\mathit{to}_{\mathit{fl}}\ a:\ \mathit{fold}\ \alpha\ \beta]&=& \quad
\mathcal{E}^{tlinq}[a: \mathit{affine}\ \alpha\ \beta]
\end{align*}

\caption{T-LINQ non-standard semantics for optic terms}
\label{fig:tlinq-non-standard-semantics}
\end{figure}

\subsubsection{Target queries and results}

\begin{figure}[hbt]
\small
\begin{center}
\begin{tabular}{lcl}
$\mathcal{E}^{tlinq}[\mathit{get}\ g:\ \alpha \rightarrow \beta]$ &=& $\mathcal{E}^{tlinq}[g:\ \mathit{getter}\ \alpha\ \beta]$\\
$\mathcal{E}^{tlinq}[\mathit{preview}\ g:\ \alpha \rightarrow \mathit{option}\ \beta]$ &=& $\mathcal{E}^{tlinq}[g:\ \mathit{affine}\ \alpha\ \beta]$\\
$\mathcal{E}^{tlinq}[\mathit{getAll}\ g:\ \alpha \rightarrow \mathit{list}\ \beta]$ &=& $\mathcal{E}^{tlinq}[g:\ \mathit{fold}\ \alpha\ \beta]$
\end{tabular}
\end{center}
\caption{T-LINQ non-standard semantics for query terms}
\label{fig:tlinq-query-non-standard-semantics}
\end{figure}

The last step towards the generation of final queries is supplying the
non-standard semantics for queries, which are shown in
Fig.~\ref{fig:tlinq-query-non-standard-semantics}. This step is trivial since
they share the very same semantic domain as their associated optics; therefore,
we just need to evaluate their optic argument. However, and in order to produce
the final queries, there is a non-negligible disagreement that we need to
address: the T-LINQ expressions which are generated by $\mathcal{E}^{tlinq}$
refer to entities from the nested model, as introduced by
Fig.~\ref{fig:couple-tlinq-semantics}. To resolve this mismatch, we need to
reconcile the relational model with the nested model, so that we can use
$\mathit{nestedOrg}$ (Fig.~\ref{fig:from-flat-to-nested}) for the task.
Thereby, we just supply the nested data to the T-LINQ lambda expression
generated from the Optica expression:
\par\nobreak\vspace{-0.2cm}
{\small
\[\mathbf{def}\ \mathit{expertiseTlinq} =
    {<}@\ \%\mathcal{E}^{tlinq}[\mathit{expertise}: \mathit{Org} \rightarrow \mathit{list}\ \mathbb{S}]\ \%\mathit{nestedOrg}\ @{>}
\]
}%
This produces an alternative implementation of the query which was presented in
Fig.~\ref{fig:expertise-tlinq}. However, the T-LINQ expression generated by the
new version is much more difficult to read and less efficient, given the
complexity introduced by $\mathit{nestedOrg}$. Fortunately, this is not a
problem, since both queries share the very same normal form, and consequently,
they produce the same SQL statement.

\section{S-Optica: Optica as a Scala library}
\label{sec:s-optica}

This section aims at implementing the Optica DSL in Scala. The resulting library
(which we call S-Optica) is provided as a proof-of-concept of the feasibility of
extending existing libraries for LINQ, especially those based on comprehensions
with optic capabilities. We will show in detail the S-Optica implementation of
the syntax and type system of Optica, as well as its standard semantics. The
reader may want to look into the accompanying sources for more information about
the S-Optica implementation of the interpreters for XQuery, SQL and T-LINQ. The
S-Optica implementation is also intended to serve as an illustration of the
tagless-final style~\cite{kiselyov2012typed}, that we have chosen in order to
implement our DSL.

\subsection{Syntax and type system}

In the tagless-final style, the syntax and type system of a typed DSL is
implemented through a {\em type constructor class}\footnote{We supply a brief
tour of how to encode type (constructor) classes in Scala in
\ref{appsub:encoding-type-classes}.}, which represents the class 
of representations, or possible interpretations, of that DSL. This type class
does not need to be a single, monolithic module, but it is usually decomposed
into different type classes which encode different aspects of the DSL. In our
case, the division of classes has taken into consideration the structure of
optics and combinators that we followed in Sect.~\ref{sec:Programming}, and the
difference between optic and query types as introduced in Sect.~\ref{sub:Syntax}.
Accordingly, Fig.~\ref{fig:symantics} shows the syntax and semantics of the
Optica fragment corresponding to getters, affine folds and folds;
Fig.~\ref{fig:optica} shows the implementation of the fragment of queries, as
well as the overall Optica type class. Some comments on the implementation
follow below:
\begin{itemize}

    \item The primitive combinators of the different types of optics, getters,
        affine folds and folds, are implemented in their respective modules.
        Those which are not primitive, but can be defined in terms of other
        combinators, namely, \lstinline{any}, \lstinline{all}, \lstinline{elem}
        and \lstinline{empty}, are defined in the \lstinline{OpticCom} type
        class. 

    \item The implementation of these derived combinators benefits from the same
        syntactic enhancements that we assumed in
        Sect.~\ref{sub:Programming:Getters}. In fact, their implementation is
        literally the same as that for concrete optics shown in
        Fig.~\ref{fig:fold}. The differences simply lie in their signatures and
        the intended semantics: whereas the implementations of
        Fig.~\ref{fig:fold} only work for concrete optics, the implementations
        of Fig.~\ref{fig:symantics} work for any optic representation
        \lstinline{Repr[_]}. Thus, we may instantiate this class in order to
        work with concrete optics, or any other standard representation such as
        van Laarhoven or profunctor optics; of course, we may also instantiate
        this class in order to work with XQuery, TripletFun or T-LINQ, since
        these are legitimate read-only optic representations as have been shown
        throughout the paper.

    \item We actually used concrete optic types in the signatures of these type
        classes, i.e. the types \lstinline{Getter[_,_]},
        \lstinline{Affine[_,_]}, etc. (to which the signatures refer to) are
        exactly those defined in Sect.~\ref{sub:Programming:Getters}. How can
        these signatures work for any representation, then? The reason is simply
        that these combinators do not receive and return plain concrete optics
        types, but their {\em representations}: the \lstinline{empty} combinator
        does not receive a concrete fold, but anything that {\em counts as} a
        fold optic. Concrete types thus behave mostly as \emph{phantom
        types}~\cite{leijen1999domain, hinze2003fun}, which specify the abstract
        semantic domains of the language and aid in the definition of its type
        system.

    \item The query types of Optica correspond in the tagless-final style to
        {\em observations}~\cite{suzuki2016finally}. These can be understood as
        the standard interpretations that we demand from any representation of
        the implemented DSL. This matches perfectly with the distinction between
        optics and query types: for instance, we will always want a
        \lstinline{getAll} interpretation from a fold program, irrespective of
        the optic representation. In the tagless-final style, we commonly assign
        different type constructors to DSL expressions and observations. Thus,
        in Fig.~\ref{fig:optica} we use \lstinline{Repr[_]} and
        \lstinline{Obs[_]} for optic and query representations, respectively.
        This is actually equivalent to having two different DSLs, one for optics
        and another for queries, into which optics are compiled.

    \item Base types of Optica also enjoy a different representation.  As the
        implementation of the \lstinline{like} combinator shows, base values are
        represented using the very type system of the host language, i.e. Scala.
        Thus, its representation is not \lstinline{Repr[_]} nor
        \lstinline{Obs[_]}, but the identity type constructor. This
        representation for base types is also common practice in tagless-final
        style\footnote{We would run into problems, however, if the target optic
        representation does not also use itself the Scala types for representing
    base types.}.

    \item To avoid overloading the \lstinline{like} method for the different
        base types, \lstinline{Int}, \lstinline{String} and \lstinline{Boolean},
        we use the GADT \lstinline{Base}, whose object instances are marked as
        implicits, thereby enabling the context bound syntax in Scala. The
        \lstinline{Base} GADT is also declared in those signatures that depends
        on the \lstinline{like} combinator, namely, \lstinline{elem}, and the
        combinator \lstinline{equal}\footnote{This constraint could be slightly
            lifted since we may want to compare not only base, but model types
            in general (cf. Fig.~\ref{fig:optica-syntax}).}.

\end{itemize}

\begin{figure}
  \begin{lstlisting}
trait GetterCom[Repr[_]] {
  def id_gt[S]: Repr[Getter[S, S]]
  def andThen_gt[S, A, B](u: Repr[Getter[S, A]],
                         d: Repr[Getter[A, B]]): Repr[Getter[S, B]]
  def fork_gt[S, A, B](l: Repr[Getter[S, A]],
                      r: Repr[Getter[S, B]]): Repr[Getter[S, (A, B)]]
  def like[S, A: Base](a: A): Repr[Getter[S, A]]
  def not[S](b: Repr[Getter[S, Boolean]]): Repr[Getter[S, Boolean]]
  def equal[S, A: Base](x: Repr[Getter[S, A]],
                        y: Repr[Getter[S, A]]): Repr[Getter[S, Boolean]]
  def greaterThan[S](x: Repr[Getter[S, Int]],
                     y: Repr[Getter[S, Int]]): Repr[Getter[S, Boolean]]
  def subtract[S](x: Repr[Getter[S, Int]],
                  y: Repr[Getter[S, Int]]): Repr[Getter[S, Int]]
}

trait AffineFoldCom[Repr[_]] {
  def id_af[S]: Repr[AffineFold[S, S]]
  def andThen_af[S, A, B](u: Repr[AffineFold[S, A]],
                         d: Repr[AffineFold[A, B]]): Repr[AffineFold[S, B]]
  def filtered[S](p: Repr[Getter[S, Boolean]]): Repr[AffineFold[S, S]]
  def to_af[S, A](gt: Repr[Getter[S, A]]): Repr[AffineFold[S, A]]
}

trait FoldCom[Repr[_]] {
  def id_fl[S]: Repr[Fold[S, S]]
  def andThen_fl[S, A, B](u: Repr[Fold[S, A]],
                         d: Repr[Fold[A, B]]): Repr[Fold[S, B]]
  def nonEmpty[S, A](fl: Repr[Fold[S, A]]): Repr[Getter[S, Boolean]]
  def to_fl[S, A](afl: Repr[AffineFold[S, A]]): Repr[Fold[S, A]]
}

trait OpticaCom[Repr[_]] extends GetterCom[Repr]
    with AffineFoldCom[Repr]
    with FoldCom[Repr] {
  def empty[S, A](fl: Repr[Fold[S, A]]): Repr[Getter[S, Boolean]] =
    fl.nonEmpty.not
  def all[S, A](fl: Repr[Fold[S, A]])(
                p:  Repr[Getter[A, Boolean]]): Repr[Getter[S, Boolean]] =
    (fl >>> filtered(p.not)).empty
  def any[S, A](fl: Repr[Fold[S, A]])(
                p:  Repr[Getter[A, Boolean]]): Repr[Getter[S, Boolean]] =
    fl.all(p.not).not
  def elem[S, A: Base](fl: Repr[Fold[S, A]])(a: A): Repr[Getter[S, Boolean]] =
    fl.any(id_gt === like(a))
}
  \end{lstlisting}
  \caption{OpticaCom symantics (optic combinators).}
  \label{fig:symantics}
\end{figure}

\begin{figure}
  \begin{lstlisting}
trait GetterQuery[Repr[_], Obs[_]] {
  def get[S, A](gt: Repr[Getter[S, A]]): Obs[S => A]
}

trait AffineFoldQuery[Repr[_], Obs[_]] {
  def preview[S, A](af: Repr[AffineFold[S, A]]): Obs[S => Option[A]]
}

trait FoldQuery[Repr[_], Obs[_]] {
  def getAll[S, A](fl: Repr[Fold[S, A]]): Obs[S => List[A]]
}

trait Optica[Repr[_], Obs[_]] extends OpticaCom[Repr]
  with GetterQuery[Repr, Obs]
  with AffineFoldQuery[Repr, Obs]
  with Fold[Repr, Obs]
  \end{lstlisting}
  \caption{Optica symantics (generic combinators and queries).}
  \label{fig:optica}
\end{figure}

\subsection{Domain queries}

In order to write domain queries, we need to extend the syntax and type system
of the Optica language, as we have seen in Sect.~\ref{sub:Optica:Generic}.
Quoting from \cite{oleg-nondet}, ``extensibility is the strong suite of the
tagless-final embedding''; therefore, this task should be easy. Indeed, we
simply need to declare a new type class where we have a component containing an
entry for each domain optic in the model, as shown in
Fig.~\ref{fig:domain-primitives}. The types \lstinline{Couples},
\lstinline{Person}, etc., are immutable data structures which mostly behave as
phantom types and aid in the extension of the type system of the language.

\begin{figure}[hbt]
\centering
\begin{lstlisting}
trait CoupleModel[Repr[_]] {
  def couples: Repr[Fold[Couples, Couple]]
  def fst: Repr[Getter[Couple, Person]]
  def snd: Repr[Getter[Couple, Person]]
  def name: Repr[Getter[Person, String]]
  def age: Repr[Getter[Person, Int]]
}
\end{lstlisting}
\caption{Couple domain symantics}
\label{fig:domain-primitives}
\end{figure}

Once we have the core and domain primitives available, we should be able to
implement generic optic expressions by declaring both dependencies, the
combinators of the Optica API and the domain model (note that observations are
not needed to write pure optic expressions):
\begin{lstlisting}
def differencesFl[Repr[_]](implicit
    O: OpticaCom[Repr],
    M: CoupleModel[Repr]): Repr[Fold[Couples, (String, Int)]] =
  couples >>> filtered((fst >>> age) > (snd >>> age)) >>>
            (fst >>> name) *** ((fst >>> age) - (snd >>> age))
\end{lstlisting}
and generic query expressions (in this occasion, we pass the whole Optica type
class, which includes the queries):
\begin{lstlisting}
def differences[Repr[_], Obs[_]](implicit
    O: Optica[Repr, Obs],
    M: CoupleModel[Repr]): Obs[Couples => List[(String, Int)]] =
  differencesFl.getAll
\end{lstlisting}
As can be seen, the required primitives are injected using the Scala implicit
mechanism. In contrast with Def.~\ref{def:differences}, this version remarks the
aforementioned existence of different representations for optics and queries, as
evidenced by the result types. Scala implicits are also exploited by the library
to omit invocations to casting methods, although the required syntax is not
shown for brevity.

\subsection{Standard semantics}

Type classes in the tagless-final style are commonly named {\em Symantics}, a
portmanteau of `syntax' and `semantics', to emphasise the fact that the same
abstraction serves a double purpose: the type class declaration defines the
syntax and type system of the language, whereas type class instances provide its
semantics. The standard semantics of the language is no exception, and for this
purpose we greatly benefit from having reused the standard semantic domains at
the syntactic level: simply use the identity \emph{type lambda}
\lstinline[columns=fixed]{lambda[x => x]} for both the \lstinline{Repr} and
\lstinline{Obs} parameters, and map each primitive into its concrete
counterpart\footnote{This is the common case in which standard semantic domains
  do not eventually behave as phantom types.}.

We can find the interpretation that supplies the standard semantics of Optica in
Fig.~\ref{fig:scala-standard}. In particular, it is represented by the singleton
object \lstinline{R}, which is also a common name for meta-circular
interpretations in the tagless-final style. We follow the very same pattern to
instantiate the couple domain terms, as we show in Fig.~\ref{fig:scala-domain}.

Now we can use the standard semantics to evaluate generic queries, and to
re-implement, in a modular way, the ad-hoc functions that deal with immutable data
structures. For instance:
\begin{lstlisting}
val differencesR: Couples => List[(String, Int)] = 
  differences[lambda[x => x], lambda[x => x]](R, CoupleModelR)
\end{lstlisting}

As can be seen, we have specified the standard representation types for optics
and queries alongside the associated evidences. Fortunately, they could be
inferred implicitly, as shown in this alternative and preferred version.

\begin{lstlisting}
val differencesR: Couples => List[(String, Int)] = differences
\end{lstlisting}

The resulting function is extensionally equal to \lstinline{differences} from
Sect.~\ref{subsub:couple-example}. The implementation of the rest of
interpretations in this article (XQuery, SQL and T-LINQ) follows the same
principles. Interested readers can find a \emph{README} file in the companion
sources~\cite{habla2019scp19} which briefly describes the library structure and
supplies links to the aforementioned interpreters ---and other relevant modules.

\begin{figure}
  \centering
  \begin{lstlisting}
trait RGetterCom extends GetterCom[lambda[x => x]] {
  def id_gt[S] = Getter.id
  def andThen_gt[S, A, B](u: Getter[S, A], d: Getter[A, B]) = Getter.andThen(u, d)
  def fork_gt[S, A, B](l: Getter[S, A], r: Getter[S, B]) = Getter.fork(l, r)
  def like[S, A: Base](a: A) = Getter.like(a)
  def not[S](b: Getter[S, Boolean]) = Getter.not(b)
  def eq[S, A: Base](x: Getter[S, A], y: Getter[S, A]) = Getter.eq(x, y)
  def gt[S](x: Getter[S, Int], y: Getter[S, Int]) = Getter.gt(x, y)
  def sub[S](x: Getter[S, Int], y: Getter[S, Int]) = Getter.sub(x, y)
}

trait RAffineFoldCom extends AffineFoldCom[lambda[x => x]] {
  def id_af[S] = AffineFold.id
  def andThen_af[S, A, B](u: AffineFold[S, A], d: AffineFold[A, B]) = AffineFold.andThen(u, d)
  def filtered[S](p: Getter[S, Boolean]) = AffineFold.filtered(p)
  def as_af[S, A](gt: Getter[S, A]) = gt
}

trait RFoldCom extends FoldCom[lambda[x => x]] {
  def id_fl[S] = Fold.id
  def andThen_fl[S, A, B](u: Fold[S, A], d: Fold[A, B]) = Fold.andThen(u, d)
  def nonEmpty[S, A](fl: Fold[S, A]) = fl.nonEmpty
  def as_fl[S, A](afl: AffineFold[S, A]) = afl
}

trait RGetterAct extends GetterAct[lambda[x => x], lambda[x => x]] {
  def get[S, A](gt: Getter[S, A]) = gt.get
}

trait RAffineFoldAct extends AffineFoldAct[lambda[x => x], lambda[x => x]] {
  def preview[S, A](af: AffineFold[S, A]) = af.preview
}

trait RFoldAct extends FoldAct[lambda[x => x], lambda[x => x]] {
  def getAll[S, A](fl: Fold[S, A]) = fl.getAll
}

implicit object R extends Optica[lambda[x => x], lambda[x => x]]
  with RGetterCom with RGetterAct
  with RAffineFoldCom with RAffineFoldAct
  with RFoldCom with RFoldAct
  \end{lstlisting}
  \caption{Optica standard semantics.}
  \label{fig:scala-standard}
\end{figure}

\begin{figure}
\begin{lstlisting}
implicit object CoupleExampleR extends CoupleModel[lambda[x => x]] {
  val couples = CoupleModel.couples
  val fst = CoupleModel.fst
  val snd = CoupleModel.snd
  val name = CoupleModel.name
  val age = CoupleModel.age
}
\end{lstlisting}
\caption{Couple domain standard semantics}
\label{fig:scala-domain}
\end{figure}

\section{Discussion}
\label{sec:Discussion}

\subsubsection*{The language of optics.}

One the most prominent sought-after features of optics is {\em modularity},
i.e.\ the capacity of creating optics for compound data structures out of
simpler optics for their parts. This is specially emphasized in the framework of
profunctor optics \cite{pickering2017profunctor}, where optic composition builds
upon plain function composition and enables straightforward combinations of
isos, prisms, lenses, affine traversals, and traversals. The profunctor
representation is particularly convenient to implement (and even reveal) the
compositional structure of the different varieties of optics, but, in essence,
this structure is also enjoyed by concrete optics, van Laarhoven optics, etc.
Modularity is a feature of the {\em language} of optics, rather than of any
particular representation. This paper has shown, albeit for a very restricted
subset of optics (getters, affine folds and folds), that this compositional
structure of optics can be encoded in the type system of a formal language, that
we have named Optica. The denotational semantics of this language was given in
terms of concrete optics but any other isomorphic representation, such as
profunctor optics, may have served as well.

Now, the specification of Optica includes not only the compositional features of
read-only optics but also, and significantly, their characteristic queries.
Taking into account this non-compositional character of optics is essential as
soon as we tackle the extension of Optica with new varieties of optics. For
instance, the major difference between folds and traversals is not found in
their compositional properties, but in the queries that they must support:
besides {\em getAll}, traversals must also support a {\em putAll} query to
replace the content of the elements that they are selecting.

On the implementation side, we have found the typed tagless-final approach
especially suitable in order to encode this separation of concerns between
declarative optic combinators and their intended queries. In essence, it closely
corresponds to the difference between {\em representations} of the DSL and their
{\em observations} or interpreters. Another essential feature from the
tagless-final pattern that we plan to profit from is extensibility. In
particular, new optics will be added to S-Optica through their own type classes
(as we have done for getters, affine folds and folds) so that we can fully reuse
old queries without recompiling sources.

\subsubsection*{Optics versus comprehensions}

Optics can be seamlessly combined with comprehensions, as shown in
Sect.~\ref{sec:T-LINQ}. Indeed, by using the T-LINQ interpreter of Optica we can
freely mix optic expressions with general comprehension queries. In this way,
optics may play within comprehensions a similar role to that which is played by
XPath within XQuery \cite{cheneyEmail}. In the following paragraphs, we discuss
the basic trade-off between expressiveness and modularity of comprehension and
Optica queries, so as to better appreciate their role in the LINQ landscape.

The separation of concerns between declaratively {\em selecting} parts of a data
structure and building a variety of {\em queries} related to those parts is
the cornerstone of optics. In this regard, the LINQ approach based on
comprehensions focuses on the query building side and, commonly, on
constructing queries of a simple kind: retrieval queries denoting a multiset
(the semantic domain for queries on QUE$\Lambda$~\cite{suzuki2016finally},
T-LINQ~\cite{cheney2013practical},
NRC~\cite{DBLP:journals/sigmod/BunemanLSTW94}, etc.). The optics approach is,
hence, potentially more modular. For instance, a representation of traversals
intended for SQL should allow us to generate both \sqlinline{SELECT} and
\sqlinline{UPDATE} statements for the queries {\em getAll} and {\em putAll},
respectively. We plan to deal with this extension and its trade-offs with
expressiveness in future versions of Optica.

We can still claim further modularity advantages of optics over
comprehensions. Basically, these are due to the fact that optics provide a
language which is more akin to relational algebra than the calculus approach that monads
provide for comprehensions \cite{DBLP:journals/pacmpl/GibbonsHHW18}. Arguably,
the support for functional abstraction and intermediate nested data of
comprehensions languages and systems such as Links, T-LINQ or DSH
offer\footnote{DSH, in particular, comes with an extensive catalog of
  list-processing combinators:
  \url{https://github.com/ulricha/dsh/blob/master/src/Database/DSH/Frontend/Externals.hs}.},
leads also to highly compositional queries\footnote{Indeed, our version of the
  expertise query in Sect.~\ref{sec:T-LINQ} is no more simple than the
  equivalent version using nested data in \cite{cheney2013practical}.}. We can
find examples, however, where the difference in style manifests itself. For
instance, this is the query that remains to complete Table~\ref{tab:intro}, in
the style of S-Optica:

\begin{lstlisting}
def under50_d[Repr[_], Obs[_]](implicit 
    O: Optica[Repr, Obs],
    M: CoupleModel[Repr]): Obs[Couples => List[String]] =
  (couples >>> fst >>> filtered (age < 50) >>> name).getAll
\end{lstlisting}
which we compare to an analogous query using the Scala implementation of T-LINQ
\footnote{\lstinline{Tlinq[_[_]]} provides the tagless-final implementation in
  Scala of T-LINQ, that we have used to implement the corresponding interpreter for
  S-Optica. The role of \lstinline{CoupleNested} in the sample query is similar
  to the \lstinline{OrgNested} model in Sect.~\ref{sub:background-tlinq}.}:

\begin{lstlisting}
def under50_e[Repr[_]](
    couples: Repr[Couples])(implicit
    Q: Tlinq[Repr],
    N: CoupleNested[Repr]): Repr[List[String]] =
  foreach(couples)(c => where (c.fst.age < 50) (yields (c.fst.name)))
\end{lstlisting}
As this example shows, in adopting the language of optics, modularity is
improved in several respects. First, as we have mentioned earlier, the query is
actually composed of two major parts: the optic expression, which declares what
to select from, and the query expression, which actually specifies the kind of
query to be executed over the selection. Second, the optic expression is unaware
of variables and builds upon finer grained and reusable modules, such as
\lstinline{couples}, \lstinline{fst}, \lstinline{age} and \lstinline{name}.
This results in pure algebraic queries that are arguably more simple to compose
and maintain. In essence, we are building out of simpler optics in a
purely compositional style, and deriving queries in one shot.

The downside of the optics approach in relation to comprehensions, at least in
the current version of Optica, is its limited expressiveness. Indeed, variables
are fruitfully exploited in comprehensions to express arbitrary joins (e.g.
cyclic) whereas optic queries appear to move only downwards from the root of the
hierarchy. Relational models are more general than nested models, providing the
programmer with better navigation tools~\cite{bachman1973programmer}, and
therefore not every model is expressible in Optica. Take the couple model as an
example. We assume that each person is hanging from a couple, so that we can
find them by diving into the couple fields \lstinline{fst/snd}. However, the
relational model is able to supply more entries for people who do not
necessarily form a couple. To alleviate this problem, we may introduce a new
fold \lstinline{people} besides the existing \lstinline{couples}, sharing a
virtual \emph{root} type as source. The connections between \lstinline{people}
and \lstinline{fst}/\lstinline{snd} would still be unclear in the optic model;
therefore, new mechanisms should be introduced in order to establish the precise
relationship among them. We leave for future work a more precise investigation
of the compared expressiveness of the comprehension and optic languages, as well
as the extension of Optica with already supported features in comprehension
languages like grouping, aggregation and order-by
queries~\cite{suzuki2016finally, DBLP:conf/aplas/KiselyovK17,
DBLP:conf/pepm/KatsushimaK17}.

\subsubsection*{Optics as a general query language.}

The role of optics in LINQ expands beyond its combined use with comprehensions.
By lifting optics into a full-fledged DSL, we have opened the door to
non-standard interpretations that directly translate the language of optics to
data accessors for alternative representations beyond immutable data structures.
For instance, we have provided an interpretation to turn Optica queries into
XQuery expressions where we have seen that the connection among them is
straightforward, leading to a compositional interpreter. The translation ignores
the XQuery FLWOR syntax and basically focuses on XPath features. Indeed, we
understand XPath as a language to select parts from an XML document, which makes
it a perfect example of optic representation. Moreover, since XPath does not
provide the means to update an XML document, it also fits perfectly with
read-only optics such as getters, affine folds and folds.

It might be worth mentioning that synergies among optics and XML are by no means
new. In fact, prominent optic libraries are extended with modules to cope with
XML\footnote{\url{https://hackage.haskell.org/package/xml-lens-0.1.6.3/docs/Text-XML-Lens.html}}
or JSON documents, even packaged as domain-specific query languages, such as
JsonPath~\footnote{\url{https://github.com/julien-truffaut/jsonpath.pres}}. In
these projects, standard optics facilitate the definition of these DSLs for
querying JSON or XML documents. Nevertheless, our approach is radically
different since we provide a general optic language in order to build generic
optics which may be translated over those DSLs (JsonPath, XQuery, etc.).

Our approach also differs from others where the process is reversed and a
translation of XPath expressions into a general query language based on
comprehensions is performed~\cite{cheney2013practical}. In this spirit, we could
also relate this paper with SilkRoute~\cite{fernandez2002silkroute}, where the
database administrator exposes the database application data in terms of a
public XML view, and external components issue XQuery application queries
against it. Then, the framework translates them into one or more SQL statements
and collects the results from the inner database, which are returned as XML
documents. One of the key aspects of SilkRoute is \emph{view forests}, a concept
that is exploited by the framework to separate the XML structure from its
computation. On its part, Optica exposes a hierarchy of domain optics that
external components may use to compose optic expressions, as application
queries. In addition, we could understand view forests as a kind of optic since
they also select parts from the underlying database. However, Optica is more
general considering that the same application queries can be reused against
different targets, and not only SQL.

Nevertheless, SQL is the primary target of classical LINQ with comprehensions,
and we have also provided a non-standard SQL interpreter for Optica. Commonly,
comprehension-based queries need to be flattened in order to guarantee a good
performance: the naive translation to SQL is not optimal since it typically
leads to nested subqueries. Moreover, translations of flat-flat queries to SQL
are guaranteed to be total and to avoid the problem of query avalanche
\cite{DBLP:conf/dbpl/Cooper09}. In systems like Links or T-LINQ, these
guarantees are even statically checked.  In Sect.~\ref{sub:sql-actions-queries},
our translation to SQL attains similar guarantees concerning the type of
generated queries, which are absent of subqueries, beyond those that are
generated by \sqlinline{EXISTS}, which are unavoidable. However, failures in
query generation are signalled at run-time rather than at compile time. We plan
to solve this limitation in future work by using the optimization techniques
that the tagless-final approach offers \cite{suzuki2016finally}. Our translation
process resembles the denotational approach of SQUR
\cite{DBLP:conf/aplas/KiselyovK17} and Links \cite{DBLP:conf/tldi/LindleyC12}
rather than the rewriting approach followed in \cite{DBLP:conf/dbpl/Cooper09,
cheney2013practical, suzuki2016finally}. In particular, we use an intermediate
language {\em TripletFun} to decouple the filtering, selection and collection
aspects of the final SQL query. We differ from SQUR, however, because the
ultimate translation to SQL is performed directly from this non-standard
semantic domain rather than from a normalized optic query. We plan to
incorporate normalization and partial evaluation in future work, which will be
convenient as soon as we extend the language with projections \lstinline{first}
and \lstinline{second}, in correspondence with the \lstinline{fork} combinator.

Given the existing translation to comprehensions from Optica and the established
results concerning the generation of SQL from comprehensions the usefulness of
\lstinline{TripletFun} for this purpose is certainly relative. However, this
demonstrates an instance of optic representation in the relational setting,
which we believe to have the potential of being very useful when we extend our
results for optics with updating capabilities. In this light, we intend to
exploit the very same \lstinline{TripletFun} representation to generate both
\sqlinline{SELECT} and \sqlinline{UPDATE} statements. Moreover, the
\lstinline{TripletFun} interpreter represents an example of complex translation
using an intermediate optic representation, which resembles the denotational
approach of \cite{DBLP:conf/aplas/KiselyovK17} but performed in the algebraic
setting of optics rather than in the relational calculus of comprehensions. This
semantic style may serve as a reference for similar complex interpreters, e.g.
for NoSQL databases such as MongoDB \cite{MongoDB}.

\subsubsection*{Optica versus ORMs and LINQ libraries}

Connections between optics and databases are widespread. As a matter of fact, lenses
emerged in this context~\cite{foster2007combinators} under the umbrella of
\emph{bidirectional programming}. We remark~\cite{horn2018incremental} as a
recent work in this field, where a practical approach to the view update problem
is introduced by means of the so-called \emph{incremental relational lenses}.
Although we still do not know if extending Optica will lead us to contemplate
views in the non-standard SQL semantics, we find this research essential to deal
with updating optics in an effective way.

S-Optica and object-relational mappers (ORMs), like Hibernate, pursue similar
goals: they aim at working with data in persistent stores as if it were plain
in-memory data. However, S-Optica uses the language of optics while ORMs try to
remain as close as possible to the customary object-oriented style. These are
other relevant differences:
\begin{itemize}

\item S-Optica does not stick to relational databases as its preferred target
  infrastructure. In fact, Sect.~\ref{sec:Optica} and Sect.~\ref{sec:XQuery}
  show that in-memory immutable data structures and XML files are also
  potential sources of information.

\item S-Optica is eminently declarative. Indeed, S-Optica queries are simply
  values\footnote{Strictly speaking, values are objects in Scala whereas
    S-Optica queries are polymorphic methods. These may be easily turned into values
    by using a Church encoding representation.} that do not produce side
  effects on their own. This contrasts with ORMs, where it requires a huge
  understanding of the particular ORM to identify which queries are being
  launched at any time. The declarative style of S-Optica enables
  compositionality as well as the possibility to introduce further optimizations.

\item S-Optica queries are expressive and well-typed. Many ORMs introduce
  contrived additional languages to express queries, and their expressions are
  usually presented as plain strings, so that errors are not detected at compile
  time.

\item ORMs usually consider the notion of object as the smallest granularity
  concept to deal with, while S-Optica supports queries that select very specific
  parts from the whole data.

\item ORMs are able to write data back to the store, while this feature is
  future work in Optica.

\end{itemize}

In general, ORMs have been used for a long time and they are consequently very mature, while
Optica is still an experimental and limited library. However, it already solves
many of the problems that are deep-rooted in the ORM approach.

The Scala libraries Quill~\cite{quill19} and Slick~\cite{slick19} are arguably
the most similar frameworks to S-Optica. The former is strongly inspired by
T-LINQ~\cite{cheney2013practical} and it therefore follows the same theoretical
principles. The major benefit with regard to the original P-LINQ, the F\#
implementation of T-LINQ, is the ability of Quill to produce the final queries
at compile-time, exploiting the Scala metaprogramming
facilities~\cite{burmako2013scala}. Unfortunately, although Quill provides a
\lstinline{flatMap} method for the type constructor \lstinline{Query}, it
apparently lacks an implementation for \lstinline{point} which is required to
translate some Optica queries into Quill expressions\footnote{For instance, the
    S-Optica query \lstinline{(like 1).getAll}.}. Slick is similar to Quill, but
    it does not build upon a theoretical language like T-LINQ\footnote{A
        comparison between Quill and Slick (written by Quill's author) is
    provided here
\url{https://github.com/getquill/quill/blob/master/SLICK.md}.}. In any case,
both Quill and Slick map relational models in Scala in a direct way, i.e. as
flat data models, whereas S-Optica works with nested data models and has to
solve a bigger impedance mismatch. On the other hand, although Quill and Slick
support updates and deletes, they do this with ad-hoc languages that escape the
collection-like interface. Optica should be able to supply a standard interface
in order to support writes by introducing additional optics. As a final note, we
want to recall that, as Sect.~\ref{sec:T-LINQ} points out, Optica should not be
seen as a competitor but as a complement for these libraries, since optics and
comprehensions were shown to be compatible.

\section{Conclusions}
\label{sec:Conclusions}

This paper has attempted to demonstrate that optics embrace a much wider range
of representations beyond concrete, van Laarhoven, profunctor optics and other
isomorphic acquaintances. We have shown, for instance, that a restricted subset
of XQuery can be properly understood as an {\em optic representation}, i.e.\ as
an abstraction whose essential purpose is to allow us to select parts from a
data source by using powerful combinators, declaratively, and derive queries
from those selectors. From this standpoint, data sources of optic
representations may range far beyond general immutable structures: they might be
XML documents, as in the case of XQuery, or relational databases. In fact, we
have also shown how to derive SQL queries from {\em TripletFun}, an optic
representation that endorses the separation of concerns between selection,
filtering and collection aspects, which characterizes SQL \sqlinline{SELECT}
statements. Strictly speaking, we may say that SQL is not an optic but a query
language which is translatable from an optic representation. In future work, we
aim at testing the generality of the language of optics through the generation
of other effective, idiomatic translations into a diverse range of querying
infrastructures. We will particularly pay attention to technologies that are
more recent than XQuery, with a clear bias towards nested data models such as
document-oriented NoSQL databases like MongoDB~\cite{MongoDB}~\cite{MongoDB},
and languages like GraphQL~\cite{hartig2017initial}.

We put forward Optica, a full-fledged DSL, to {\em specify} what all these
representations have in common, i.e. the concept of optic itself. Technically
speaking, the type system of Optica encodes the compositional and querying
features of getters, affine folds and folds, independently of any particular
representation; concrete optics provide the semantic domains for its standard
denotational semantics; and XQuery, TripletFun and T-LINQ represent semantic
domains for non-standard optic representations. Currently, Optica only pays
attention to a very restricted set of optics, namely getters, affine folds and
folds. In future work, we will contemplate other optics like lenses,
affine traversals or traversals, as well as additional combinators that populate
de-facto libraries like Haskell lens and Monocle. This will force us to also pay
attention to the laws (e.g.\ the get-set law of lenses) that the
intended queries of optics must comply with. We think optic algebras
\cite{lopez2018towards} will be instrumental in that formalization.

The ultimate goal behind this quest for the language of optics has been to show
that optics can play a significant role in the theory and practice of
language-integrated query. In particular, we have demonstrated how optics can be
used as a high-level language in order to derive comprehension queries, the most
common approach in the LINQ field nowadays. This has the advantage of allowing
programmers to exploit optics, the de-facto standard for dealing with
hierarchical data structures, in their LINQ developments. Additionally, the
XQuery and SQL interpretations have also shown that the language of optics is
general enough to cope with LINQ systems independently from comprehensions.
However, in the case of SQL, this is done at the expense of a more limited
expressiveness since joins are not currently supported. We plan to investigate
possible extensions to Optica based on the compositional encoding of equijoins
in \cite{DBLP:journals/pacmpl/GibbonsHHW18}. We also plan to investigate future
interpretations of Optica into declarative query languages such as
Datalog~\cite{Ceri:1989:YAW:627272.627357} and description
logics~\cite{DL-primer}, as well as its connection with recent developments in
comprehension-related languages based on monoids~\cite{fegaras_2017}. We also
think that Optica in its current shape has a great potential to deal with modern
warehouse technologies aimed at data analytics, where updates are not customary.

Optics show a potential to cope not only with retrieval queries but also with
updates, a kind of query that is commonly unaddressed in theoretical accounts
but patently necessary in practice. This paper lays the foundation to engage
with this issue in future work. On the one hand, extending the syntax and type
system of Optica (and S-Optica) with new optic types and combinators is trivial.
On the other hand, the feasibility of introducing updates in the interpretation
is subject to limitations of the particular infrastructure. For example, XQuery
does not support updates (although there are extensions that deal with
them~\cite{benedikt2009semantics}), and thus the evaluation of optics with
updating capabilities would be partial. As for SQL, it does support updates, but
there is a tradeoff with expressiveness: not all relational queries can be
updatable views, which introduces a new level of partiality. Whether triplets
need to be extended in order to accomodate updates is something that requires
further research. Lastly, we are very optimistic about the potential of updates
in modern technologies based on nested models~\cite{MongoDB,hartig2017initial},
where we have carried out several simple experiments with positive results. 

Finally, we have implemented a proof-of-concept of Optica and its interpreters
in the Scala library S-Optica by using the tagless-final style. Optica is thus
implemented as a type class: the class of optic representations and their
intended queries. Beyond the generic queries that we have used to guide the
explanations, we have tested S-Optica with other queries around the same domains
and with new domains that were extracted from the official documentation of
Monocle, Slick and Quill. These examples are located in an experimental branch
of S-Optica that will be available as soon as the library matures. In this
sense, we intend to profit from the many improvements in the inminent release of
Scala 3.0, particularly in regard to type classes and metaprogramming
facilities\footnote{\url{https://dotty.epfl.ch/docs/reference/metaprogramming/toc.html}},
with a new implementation of Optica in Dotty~\cite{dotty}. Similar
implementations may have also been developed in other languages that support the
tagless-final approach, such as Haskell or OCaml. In any case, the results that
have been obtained are encouraging enough to anticipate the feasibility of
extending existing comprehension-based libraries in these languages for LINQ,
with optic capabilities.


\appendix

\section{Scala Background}
\label{app:scala-cheatsheet}

\begin{table}
\centering
\footnotesize
\begin{tabular}{|p{0.19\linewidth}|l|p{0.36\linewidth}|}
\hline

\multicolumn{1}{|c|}{\textbf{Abstraction}} & 
\multicolumn{1}{c|}{\textbf{Code Example}} & 
\multicolumn{1}{c|}{\textbf{Description}} \\ \hline

\emph{algebraic data type}
&
\begin{lstlisting}
sealed abstract class Option[A]
case class None[A]() extends Option[A]
case class Some[A](a: A) extends Option[A]
\end{lstlisting}
&
ADTs are implemented using object inheritance. The example shows Option, which
is also known as Maybe.
\\ \hline

\emph{case class} 
&  
\begin{lstlisting}
case class Person(
  name: String, 
  age: Int)
\end{lstlisting}
&  
Defines a class with special features, like construction and observation
facilities. 
\\ \hline

\emph{companion object}
&  
\begin{lstlisting}
trait Person
object Person
\end{lstlisting}
&  
Module that serve as a companion to a class or trait with the very same name. We
use it to supply class members and provide implicit definitions, like conversors
and type class instances.
\\ \hline

\emph{for comprehension} 
&  
\begin{lstlisting}
for {
  i <- List(1, 2)
  j <- List(3, 4)
} yield i + j 
// res: List[Int] = List(4,5,5,6)
\end{lstlisting}
&  
Syntactic sugar for \emph{flatMap}, \emph{map}, etc. Analogous for
Haskell's \emph{do-notation}.
\\ \hline

\emph{function type} 
&  
\begin{lstlisting}
val f: Int => Boolean = i => i > 0
f(3) // res: Boolean = true
\end{lstlisting}
&  
Function types are represented with arrows separating domain and codomain.
Lambda expressions follow a similar syntax, where the arrow separates parameter
and function body.
\\ \hline

\emph{implicit resolution} 
&  
\begin{lstlisting}
def isum(
  x: Int)(implicit 
  y: Int): Int = x + y

implicit val i: Int = 3
isum(1) // res: Int = 4
\end{lstlisting}
&  
Family of techniques to let the compiler infer certain parameters automatically.
In the example, \emph{i} is implicitly passed as second argument to \emph{isum}.
\\ \hline

\emph{partial function syntax} 
&  
\begin{lstlisting}
val f: Option[Int] => Boolean = {
  case None => true
  case Some(_) => false
}
\end{lstlisting}
&  
Special syntax for those situations where we want to produce an anonymous
(potentially partial) function that requires pattern matching on its parameter.
\\ \hline

\emph{placeholder syntax} 
&  
\begin{lstlisting}
val inc:  Int => Int = i => i + 1
val inc2: Int => Int = _ + 1
\end{lstlisting}
&  
Syntax for lambda expressions where we refer to the parameter as \emph{'\_'}; consequently,
there is no need to name it. Both \emph{inc} and \emph{inc2} (placeholder
syntax) are equivalent.
\\ \hline

\emph{trait} 
&  
\begin{lstlisting}
trait Person {
  def name: String = "John"
  def age: Int
}
\end{lstlisting}
&
Similar to Java interfaces (as they enable multiple inheritance), but traits
support partial implementation of members.
\\ \hline

\emph{type parameter} 
&  
\begin{lstlisting}
trait List[A]
def nil[A]: List[A]
trait Symantics[Repr[_]]
\end{lstlisting}
&  
Types that are taken as parameters by class or method definitions. It is
required a special notation if we expect higher kinded types, like \emph{Repr}.
\\ \hline

\emph{type lambda} 
&  
\begin{lstlisting}
lambda[x => x]
lambda[x => Int]
lambda[x => Option[x]]
\end{lstlisting}
&  
Notation enabled by the \emph{kind-projector} compiler plugin to produce
anonymous type functions of the kind $* \rightarrow *$.
\\ \hline

\end{tabular}
\caption{Scala Cheat Sheet.}
\label{tab:cheatsheet}
\end{table}

This section aims at providing a brief background of those Scala features that
we use in this paper. First, Table~\ref{tab:cheatsheet} supplies a cheat sheet
where we can find examples and short descriptions of the abstractions and
constructions that we consider to be more relevant in the particular context of this
work. As can be seen, some of them are specific to Scala but there are other
concepts which are widespread in the functional programming community, where we
just want to show how to encode them in this language. Second, we describe the
general pattern to encode type classes~\cite{wadler1989make} in
Scala~\cite{oliveira2010type}. 

\subsection{Encoding Type Classes in Scala}
\label{appsub:encoding-type-classes}

In Scala, we can use \emph{traits} to define new type classes. For instance, we
encode \emph{Functor} as follows:
\begin{lstlisting}
trait Functor[F[_]] {
  def fmap[A, B](fa: F[A])(f: A => B): F[B]
}
\end{lstlisting}
The trait itself is parameterized with a type constructor \lstinline{F[_]};
therefore, this is a type constructor class. It declares the \lstinline{fmap}
method, which is parameterized with concrete type parameters \lstinline{A} and
\lstinline{B}  and value parameters \lstinline{fa} and \lstinline{f} (organized
in two sections\footnote{Scala supports definitions with multiple groups of
    value parameters, delimited by parentheses. In this particular situation,
    the separation turns out to be helpful to improve type inference while
invoking the method.}). As can be seen, function types are represented with the
arrow \lstinline{=>}. Now, we could follow the same pattern to provide other
type classes, like \lstinline{Pointed}:
\begin{lstlisting}
trait Pointed[F[_]] {
  def point[A](a: A): F[A]
}
\end{lstlisting}
or \lstinline{Bind}:
\begin{lstlisting}
trait Bind[F[_]] {
  def bind[A, B](fa: F[A])(f: A => F[B]): F[B]
}
\end{lstlisting}
which follow the very same pattern. The previous definitions form the building
blocks of \lstinline{Monad}.  Thereby, we could compose them to provide the
corresponding type class:
\begin{lstlisting}
trait Monad[F[_]] extends Functor[F] with Pointed[F] with Bind[F] {
  def fmap[A, B](fa: F[A])(f: A => B) = bind(fa)(a => point(f(a)))
}
\end{lstlisting}
Here, we exploit the multiple inheritance mechanism provided by Scala to mix the
involved traits. At this point, we should be able to implement for
\lstinline{fmap} in terms of \lstinline{bind} and \lstinline{point}, once and
for all.
It is common practice to deploy type class instances in the type class
companion object since the Scala compiler will search for instances in this
module, among other places. For example, this is the \lstinline{Monad}
companion object, where we have placed the monad instance for
\lstinline{Option} (\emph{Maybe} in Haskell).
\begin{lstlisting}
object Monad {
  implicit object OptionMonad extends Monad[Option] {
    def point[A](a: A) = Some(a)
    def bind[A, B](fa: Option[A])(f: A => Option[B]) = fa match {
      case None => None
      case Some(a) =de}
\end{lstlisting}
There are several alternatives to supply an instance. In this occasion, we have decided to
implement it as an object \lstinline{OptionMonad} which is declared with the
\emph{implicit} modifier, so that the compiler could find it \emph{implicitly} if
necessary. The implementation of \lstinline{point} and \lstinline{bind} turns
out to be trivial.
Once we have defined a type class, we could implement derived functionality. For
instance, we could define the typical \lstinline{join} method. 
\begin{lstlisting}
def join[F[_], A](ffa: F[F[A]])(implicit M: Monad[F]): F[A] =
  M.bind(mma)(identity)
\end{lstlisting}
This method requires an implicit evidence of \lstinline{Monad} for
\lstinline{F}, which is used in the implementation to invoke \lstinline{bind}.
Now, we could use the \lstinline{Option} instance to flatten a nested optional
value by means of \lstinline{join}.
\begin{lstlisting}
join[Option, Int](Option(Option(3)))(Monad.OptionMonad)
// res: Option[Int] = Some(3)
\end{lstlisting}
Here, we manually supply the type parameters and the monad evidence.
Fortunately, the Scala compiler is able to infer them; therefore, the next version is
preferred.
\begin{lstlisting}
join(Option(Option(3)))
// res: Option[Int] = Some(3)
\end{lstlisting}
As a final remark, note that a monad instance subsumes instances for the rest of
type classes that form it, e.g. \lstinline{OptionMonad} is also an
\lstinline{Option} instance for \lstinline{Pointed}.

\section{XML Schemas}

\subsection{Couple XSD}
\label{app:XML:Couple}

\begin{lstlisting}[language=XML, basicstyle=\scriptsize\ttfamily]
<?xml version="1.0" encoding="UTF-8"?>
<xs:schema xmlns:xs="http://www.w3.org/2001/XMLSchema">
    <xs:element name="xml">
        <xs:complexType>
            <xs:sequence>
                <xs:element name="couple" minOccurs="0" maxOccurs="unbounded">
                    <xs:complexType>
                        <xs:sequence>
                            <xs:element name="fst">
                                <xs:complexType>
                                    <xs:sequence>
                                        <xs:element name="name" type="xs:string"/>
                                        <xs:element name="age" type="xs:positiveInteger"/>
                                    </xs:sequence>
                                </xs:complexType>
                            </xs:element>
                            <xs:element name="snd">
                                <xs:complexType>
                                    <xs:sequence>
                                        <xs:element name="name" type="xs:string"/>
                                        <xs:element name="age" type="xs:positiveInteger"/>
                                    </xs:sequence>
                                </xs:complexType>
                            </xs:element>
                        </xs:sequence>
                    </xs:complexType>
                </xs:element>
            </xs:sequence>
        </xs:complexType>
    </xs:element>
</xs:schema>
\end{lstlisting}

\subsection{Organization XSD}
\label{app:XML:Organization}

\begin{lstlisting}[language=XML, basicstyle=\scriptsize\ttfamily]
<?xml version="1.0" encoding="UTF-8"?>
<xs:schema xmlns:xs="http://www.w3.org/2001/XMLSchema">
    <xs:element name="xml">
        <xs:complexType>
            <xs:sequence>
                <xs:element name="department" minOccurs="0" maxOccurs="unbounded">
                    <xs:complexType>
                        <xs:sequence>
                            <xs:element name="dpt" type="xs:string"/>
                            <xs:element name="employee" minOccurs="0" maxOccurs="unbounded">
                                <xs:complexType>
                                    <xs:sequence>
                                        <xs:element name="emp" type="xs:string"/>
                                        <xs:element name="task" minOccurs="0" maxOccurs="unbounded" 
                                                    type="xs:string"/>
                                    </xs:sequence>
                                </xs:complexType>
                            </xs:element>
                        </xs:sequence>
                    </xs:complexType>
                </xs:element>
            </xs:sequence>
        </xs:complexType>
    </xs:element>
</xs:schema>
\end{lstlisting}

\section*{Acknowledgements}

We would like to thank James Cheney, Oleg Kiselyov, Eric Torreborre and the
anonymous reviewers for their helpful comments and corrections to a previous
version of this paper. In particular, we give Cheney credit for
Sect.~\ref{sec:T-LINQ}, since he showed us how to translate into comprehensions
the optic-based organization example, using the Links programming language.
 
\bibliographystyle{elsarticle-num} 
\bibliography{opticlinq}




 
\end{document}
\endinput